\documentclass[traditabstract]{aa} 

\usepackage{graphicx}
\usepackage{txfonts}
\usepackage{longtable}
\usepackage{lscape}
\begin{document}

   \title{A Virgo Environmental Survey Tracing Ionised Gas Emission (VESTIGE).XII. Ionised gas emission in the inner regions of lenticular galaxies\thanks{Based on observations obtained with
   MegaPrime/MegaCam, a joint project of CFHT and CEA/DAPNIA, at the Canada-French-Hawaii Telescope
   (CFHT) which is operated by the National Research Council (NRC) of Canada, the Institut National
   des Sciences de l'Univers of the Centre National de la Recherche Scientifique (CNRS) of France and
   the University of Hawaii. Based on observations collected at the European Southern Observatory under ESO programmes 097.D-0408 and 099.B-0384.
   Based in part on observations obtained at the Southern Astrophysical Research (SOAR) telescope, which is a joint project of the 
   Minist\'{e}rio da Ci\^{e}ncia, Tecnologia e Inova\c{c}\~{o}es (MCTI/LNA) do Brasil, the US National Science Foundation, NOIRLab, the University of North Carolina at Chapel Hill (UNC), 
and Michigan State University (MSU).}
      }
   \subtitle{}
  \author{A. Boselli\inst{1},  
 	  M. Fossati\inst{2},
          A. Longobardi\inst{1,2},
	  K. Kianfar\inst{3,4},
	  N.Z. Dametto\inst{5},
	  P. Amram\inst{1},
	  J.P. Anderson\inst{6},
	  P. Andreani\inst{4}, 
	  S. Boissier\inst{1}, 
	  M. Boquien\inst{5},
	  V. Buat\inst{1},
       	  G. Consolandi\inst{2},
	  L. Cortese\inst{7,8},
          P. C{\^o}t{\'e}\inst{9},
          J.C. Cuillandre\inst{10},
          L. Ferrarese\inst{9},
	  L. Galbany\inst{11,12},
	  G. Gavazzi\inst{2},
          S. Gwyn\inst{9}, 
	  G. Hensler\inst{13},
	  J. Hutchings\inst{9},
	  E.W. Peng\inst{14},
	  J. Postma\inst{15},
	  J. Roediger\inst{9},
	  Y. Roehlly\inst{1},
	  P. Serra\inst{16},
	  G. Trinchieri\inst{17}  
       }

\institute{     
                Aix Marseille Univ, CNRS, CNES, LAM, Marseille, France
                \email{alessandro.boselli@lam.fr}
        \and
		Universit\'a di Milano-Bicocca, piazza della scienza 3, 20100 Milano, Italy
	\and
		Instituto Tecnol\'ogico de Aeron\'autica (ITA),  S\~ao Jos\'e dos Campos, Sao Paolo, Brazil
 	\and
 		European Southern Observatory, Karl-Schwarzschild-Strasse 2, 85748, Garching, Germany 
	\and
		Centro de Astronomi\'a (CITEVA), Universidad de Antofagasta, Avenida Angamos 601, Antofagasta, Chile
	\and
		European Southern Observatory, Alonso de C\'ordova 3107, Casilla 19, Santiago, Chile
	\and
		International Centre for Radio Astronomy Research, The University of Western Australia, 35 Stirling Hw, 6009 Crawley, Australia
	\and
		ARC Centre of Excellence for All Sky Astrophysics in 3 Dimensions (ASTRO 3D), Australia
        \and  
                National Research Council of Canada, Herzberg Astronomy and Astrophysics, 5071 West Saanich Road, Victoria, BC, V9E 2E7, Canada
        \and        
		AIM, CEA, CNRS, Universit\'e Paris-Saclay, Universit\'e Paris Diderot, Sorbonne Paris Cit\'e, Observatoire de Paris, PSL University, F-91191 Gif-sur-Yvette Cedex, France
	\and
		Institute of Space Sciences (ICE, CSIC), Campus UAB, Carrer de Can Magrans, s/n, E-08193 Barcelona, Spain
	\and
		Institut d'Estudis Espacials de Catalunya (IEEC), E-08034 Barcelona, Spain
	\and
		Department of Astrophysics, University of Vienna, T\"urkenschanzstrasse 17, 1180 Vienna, Austria
	\and
		Department of Astronomy, Peking University, Beijing 100871, PR China
	\and
		University of Calgary, 2500 University Drive NW, Calgary, Alberta, Canada
	\and
		Osservatorio Astronomico di Cagliari, via della scienza 5, 09047 Selargius, Cagliari, Italy
	\and
		INAF - Osservatorio Astronomico di Brera, via Brera 28, 20159 Milano, Italy
                 }

\authorrunning{Boselli et al.}
\titlerunning{VESTIGE}

   \date{}

 
  \abstract  
{As part of the Virgo Cluster Survey Tracing Ionised Gas Emission (VESTIGE), a blind narrow-band H$\alpha$+[NII] imaging survey of the Virgo cluster 
carried out with MegaCam at the CFHT, we discovered eight massive (10$^{10}$ $\lesssim$ $M_{star}$ $\lesssim$ 10$^{11}$ M$_{\odot}$) lenticular galaxies with 
prominent ionised gas emission features in their inner (few kpc) regions. These features are either ionised gas filaments similar to those observed in cooling flows (2 galaxies), 
or thin discs with sizes 0.7 $\lesssim$ $R(H\alpha)$ $\lesssim$ 2.0 kpc (6 galaxies), thus significantly smaller than those of the stellar disc ($R(H\alpha)$ $\simeq$ 7-22 \% $R_{iso}(r)$).
These discs have morphological properties similar to those of the dust seen in absorption in high-resolution HST images.
Using a unique set of multifrequency data, including new or archival ASTROSAT/UVIT, GALEX, HST, CFHT, \textit{Spitzer}, and \textit{Herschel} imaging data, 
combined with IFU (MUSE, ALMA) and long-slit (SOAR) spectroscopy, we show that while the gas located within these inner discs is photoionised by young stars,
signaling ongoing star formation, the gas in the filamentary structures is shock-ionised. These discs have a star formation surface brightness similar to 
those observed in late-type galaxies. Because of their reduced size, however, these lenticular galaxies are located below the main sequence of 
unperturbed or cluster star-forming systems. By comparing the dust masses measured from absorption maps in optical images, from the Balmer decrement,
or estimated by fitting the UV-to-far-IR spectral energy distribution of the target galaxies, we confirm that those derived from optical attenuation maps
are heavily underestimated because of geometrical effects due to the relative distribution of the absorbing dust and the emitting stars.
We have also shown that these galaxies have gas-to-dust ratios of $G/D$ $\simeq$ 80$^{320}_{30}$, and that the star formation within these discs follows
the Schmidt relation, albeit with an efficiency reduced by a factor of $\sim$ 2.5. 
Using our unique set of multifrequency data, we discuss the possible origin of the ionised gas in these objects, which suggests
multiple and complex formation scenarios for massive lenticular galaxies in clusters.
 }
   {}
   {}
   {}
   {}
   {}


   \maketitle
%

\section{Introduction}

Lenticular galaxies are an intermediate class of objects with their physical, structural, and kinematical properties being between 
quiescent ellipticals and star-forming, rotating discs (Hubble 1936, van den Bergh 1976). They are generally rotationally-supported 
systems (Dressler \& Sandage 1983, Emsellem et al. 2011) composed of a prominent bulge with spectro-photometric properties  
similar to those of ellipticals and an extended disc similar 
to that of early-type spirals (Burstein 1979, Laurikainen et al. 2005, Erwin et al. 2015). They are composed of evolved stellar populations such as those dominating in 
early-types (Sandage \& Visvanathan 1978a, 1978b, Visvanathan \& Sandage 1977), but can contain a significant quantity of 
cold atomic (van Driel \& van Woerden 1991, Morganti et al. 2006, Serra et al. 2012) and molecular (Thronson et al. 1989, Sage \& Wrobel 1989, Welch \& Sage 2003, Davis et al. 2011, 2013, 
Young et al. 2011, 2014) gas. They are preferentially located in dense regions such as groups and clusters (Dressler 1980, Whitemore et al. 1993, Dressler et al. 1997), 
suggesting that the environment might have played a major role in their evolution, but are also present 
in the field (Sandage \& Visvanathan 1978b, Bamford et al. 2009). Their physical, structural, and kinematical properties
seem also to vary as a function of their total mass, indicating that the evolutionary paths which gave birth to these objects can be 
complex and multiple (van den Bergh 1990, Erwin et al. 2012, Fraser-McKelvie et al. 2018).

Different scenarios have been proposed in the literature to explain the
origin of lenticulars. These include internal secular evolution (e.g. Friedli \& Martinet 1993, Laurikainen et al. 2006, Masters et al. 2010, Fraser-McKelvie et al. 2018, Rizzo et al. 2018), 
with an efficient gas ejection due to the feedback of an AGN (van den Bergh 2009), disc instabilities (Bournaud et al. 2007, 2011, Davis et al. 2014, Bournaud 2016), 
the fading of the star formation activity
of late-type galaxies due to the lack of fresh infalling gas once they become satellites of a larger halo (starvation, 
Larson et al. 1980, Bekki et al. 2002), the quenching of the star formation activity 
of spiral galaxies in rich clusters once their gas content has been
removed during the interaction with the surrounding hostile environment (e.g. Quilis et al. 2000, Burstein et al. 2005, Boselli \& Gavazzi 2006, 2014, Boselli et al. 2021a), or 
more violent gravitational interactions (e.g. Byrd \& Valtonen 1990, Dressler 2004, Bekki \& Couch 2011) including galaxy harassment, minor and major merging events 
(e.g. Spitzer \& Baade 1951, Moore et al. 1998, Bekki 1998, Poggianti et al. 2009, Wilman et al. 2009, Tapia et al. 2017, Diaz et al. 2018, Eliche-Moral et al. 2018, Mendez-Abreu et al. 2018, Davis et al. 2019).
Finally, they can be the descendants of double peak emission line galaxies, a peculiar population of objects observed at different redshift
whose interest is significantly growing in these last years (Dominguez-Sanchez et al. 2018, Maschmann et al. 2020), of dusty starburst galaxies 
observed in intermediate redshift clusters (e.g. Geach et al. 2009), or of E+A and post-starburst galaxies in clusters and in the field (Zabludoff et al. 1996, 
Poggianti et al. 1999, 2004, Yang et al. 2008).

Since a high fraction of lenticulars are located in local high-density environments but were much less frequent in similar structures at earlier epochs
($z$ $\sim$ 0.5, Dressler et al. 1997), they could have been formed only recently. Local systems should thus still keep the inprints of their 
recent transformation in their physical, structural, and kinematical properties. Nearby clusters, where most of these systems are now located (e.g. Dressler et al. 1997), 
are thus a unique laboratory to study in detail the formation and evolution of these intriguing objects. Their proximity makes possible a detailed analysis of the different 
galaxy components (stars, gas, dust) at exquisite sensitivity and angular resolution on large, statistically significant samples spanning a wide range 
in stellar mass and galaxy density.

The Virgo Environmental Survey Tracing Ionised Gas Emission (VESTIGE) is a deep H$\alpha$ narrow-band blind imaging survey of the Virgo cluster (Boselli et al. 2018a).
This survey has been designed to identify galaxies undergoing a perturbation with the hostile surrounding environment (NGC 4254, NGC 4424, NGC 4569, and IC 3476,
Boselli et al. 2016a, 2018b,c, 2021b; NGC 4330, Fossati et al. 2018, Vollmer et al. 2021, Sardaneta et al. 2021) and understand its effects on the star formation process down to scales 
of individual HII regions (Boselli et al. 2020, 2021b). Given its untargeted nature, the spectacular quality of the data in terms of sensitivity and angular resolution 
turned out to be an excellent tool for discovering objects with peculiar features in their ionised gas distribution (e.g. ionised gas filaments in M87, Boselli et al. 2019;  
dust within the tails of ram pressure stripped galaxies, Longobardi et al. 2020; ionised gas associated to an almost dark galaxy, Junais et al. 2021),
opening new exciting perspectives in the study of galaxy evolution in rich environments.

The VESTIGE survey, combined with the multifrequency data covering the entire electromagnetic spectrum available for the Virgo cluster region (GUViCS - Boselli et al. 2011; 
NGVS - Ferrarese et al. 2012; HeViCS - Davies et al. 2010; ALFALFA - Giovanelli et al. 2005), is providing 
us with a unique opportunity to study with exquisite detail a complete sample of lenticular galaxies in the closest cluster of galaxies. At the distance of the cluster (16.5 Mpc,
Gavazzi et al. 1999, Mei et al. 2007), 1 arcsec, corresponds to 80 pc, a sufficiently small value 
for dissecting the different galaxy components (stars, gas, dust) down to small scales unreachable elsewhere. The cluster, which is unrelaxed and thus 
includes objects still undergoing a transformation, is ideal for a detailed study of lenticular galaxies, which are numerous (78 catalogued in the VCC 
as Virgo cluster members by Binggeli et al. 1985) and span a wide range in luminosity and stellar mass (e.g. Sandage et al. 1985).

During the survey we discovered a number of these objects with peculiar features witnessing the presence of ionised gas in their inner regions.
In this paper, we study the properties of a sample of eight massive (10$^{10}$ $\lesssim$ $M_{star}$ $\lesssim$ 10$^{11}$ M$_{\odot}$) 
lenticular galaxies with interesting features in the ionised gas distribution such as discs or filaments.
These kinds of features have been already identified in the literature (e.g. Goudfrooij et al. 1994, Macchetto et al. 1996, Finkelman et al. 2010), 
but the quality of the data in terms of sensitivity and angular resolution, combined with 
an intense stellar continuum emission, hampered a detailed analysis of the properties of the emitting gas, the identification of the dominant ionising source, the study of its
possible relation with star formation, and the ultimate understanding of their origin. Indeed, different questions related to the nature of
these ionised gas features remain unanswered:\\
1) Which is the ionisation source of the gas: a central AGN generally present in massive systems, stellar photoionisation due to young stars, cooling of the hot gas corona,
or the general interstellar radiation field (ISRF) dominant in these evolved systems (e.g. Annibali et al. 2010, Panuzzo et al. 2011, Lagos et al. 2014)?\\
2) If the source of ionisation is young stars (SF), does the gas form stars at a similar rate as expected from the Kennicutt-Schmidt relation 
observed in normal star-forming late-type systems (e.g. Davis et al. 2014)?\\
3) Are the properties of the gas (gas-to-dust ratio, metallicity...) similar to those of the gaseous discs of normal star-forming late-type galaxies
of comparable stellar mass (e.g. Davis et al. 2015)?\\
4) What is the origin of this gas? Recently accreted gas, residual of a gas disc radially stripped outside-in by a recent episode of ram pressure, gas produced by the mass loss
of evolved stars, or gas supplied by the radiative cooling of the galaxy hot haloes (e.g. Lagos et al. 2014)?\\

A detailed analysis of the physical properties of the ionised gas emission of a statistically complete sample of lenticular galaxies in the Virgo cluster will be possible
only once the VESTIGE survey will be completed. Here we limit the analysis to a sample of eight objects with excellent data 
gathered at different frequencies, including medium resolution IFU spectroscopy (MUSE), high angular resolution HST imaging, UV (GALEX and ASTROSAT/UVIT)
and far-infrared (\textit{Spitzer} and \textit{Herschel}) imaging, and millimetric interferometry (ALMA, CARMA). The paper is structured as follows:
the sample is described in Sec. 2, the narrow-band VESTIGE observations, along with the new MUSE IFU and SOAR long-slit spectroscopy, the ASTROSAT/UVIT 
and the ALMA observations and data reduction and the other published multifrequency data analysed in this work are described in Sec. 3. In Sec.4  we analyse the physical properties
of the ionised gas and of the other galaxy components and discuss them in the framework of galaxy evolution in Sec. 5, which also includes our conclusion. 
A dedicated section describing each single object is given in Appendix A.

\section{The sample}

The sample analysed in this work is composed of eight massive (10$^{10}$ $\lesssim$ $M_{star}$ $\lesssim$ 10$^{11}$ M$_{\odot}$) 
S0-S0/a galaxies located within the inner regions of the Virgo cluster, 
where the VESTIGE survey has already reached a sufficient depth to detect low surface brightness features in the ionised gas component. 
For this reason, the sample is by no means complete. We recall, however, that out of the 78 lenticular galaxies (S0-S0/a)
identified as Virgo cluster members by Binggeli et al. (1985) in the VCC, only 33 have a stellar mass $M_{star}$ $>$ 10$^{10}$ M$_{\odot}$.
The eight selected galaxies with ionised gas features analysed in this work correspond to $\sim$ 25\% of the massive lenticulars
in Virgo. The VESTIGE survey, which at the time of writing is still incomplete, covered only $\sim$ 60\%\ of the VCC, with full depth reached only over 25\%\ 
of the cluster. The estimated detection rate of $\sim$ 25\%\ should thus be considered as a lower limit, making these objects a representative population  
of the massive lenticulars of the cluster. These eight galaxies have been selected because they show 
some evident structured ionised gas emission in the deep continuum-subtracted images. These ionised gas features can be either
low surface brightness filaments extending from the nucleus to the outer regions (NGC 4262 and 4552), discs (NGC 4429, 4459, 4476, 4477, 4526),
or both (NGC 4469). The main parameters of the observed galaxies, including those derived in this work, are given in Table \ref{gal}.

\begin{landscape}
\begin{table}
\caption{Physical properties of the target galaxies}
\label{gal}
{\tiny
\[
\begin{tabular}{cccccccccc}
\hline
\noalign{\smallskip}
\hline
Variable                	& Units					& NGC 4262		& Ref	& NGC 4429               & Ref.    & NGC 4459	 	& Ref   & NGC  4469	& Ref	  \\	   
\hline
Type	          		& 					& SB(s)0-?		& 1	& SA(r)0+;LINER, HII     &  1      & SA(r)0+;HII, LINER	& 1	&SB(s)0/a?sp;LINER& 1	  \\
Spec. ~ class.			&					& Passive,Passive	& 2	& - ,wAGN		 &  2	   & -, Retired		& 2	&Transition, wAGN & 2	  \\
$cz$    			& km s$^{-1}$   			& 1359			& 1	& 1104		         &  1      &	1192	  	& 1 	& 539		& 1	  \\
Distance	    		& Mpc                           	& 16.5			& 4,5	& 16.5          	 &  4,5    &	16.5	  	& 4,5	&     23.0	& 5	  \\
Membership			&					& Cluster A		& 5	& Cluster A		 & 5	   & Cluster A		& 5	& Cluster B	& 5	  \\
Proj.~ distance~from~cluster~core& kpc                           	& 1100         		& T.W.	& 440			 & T.W.    & 480	  	& T.W.	& 300$^a$	& T.W.	  \\
$R_{eff}(i)$			& kpc					& 0.60			& 6	& 3.21			 & 6	   &	2.86		& 6	& 2.87		& 6	  \\
Fast/slow~rotator		&					& F			& 7	& F			 & 7	   &	F	  	& 7	& -		&	  \\
log $M_{star}^b$   		& M$_{\odot}$				& 10.23$\pm$0.03	& T.W.	& 10.90$\pm$0.02 	 & T.W     &	10.81$\pm$0.02  & T.W.	& 10.65$\pm$0.02& T.W.	  \\
log $M_{dust}$			& M$_{\odot}$                   	& 6.57$\pm$0.87		& T.W.	& 6.26$\pm$0.07   	 & T.W.    & 6.18$\pm$0.03  	& T.W.	& 6.88$\pm$0.02	& T.W.	  \\
log $M(HI)$         		& M$_{\odot}$   			& 8.75			& 8	& $<$7.12		 & 8       & $<$6.93	  	& 8	& 7.64		& 9	  \\
log $M(H_2)^c$        		& M$_{\odot}$                   	& 8.47$\pm$0.30		& T.W.	& 8.62$\pm$0.20		 & T.W.    & 8.68$\pm$0.10  	& T.W.	& 8.79$\pm$0.30	& T.W.	  \\
\hline
	
log $f(H\alpha+[NII])$           & erg s$^{-1}$ cm$^{-2}$    		& -12.61$\pm$0.30	& T.W.	& -12.67$\pm$0.01	 & T.W.    & -12.79$\pm$0.01  	& T.W.	&-12.62$\pm$0.05& T.W.	  \\
log $L(H\alpha)$      	        & erg s$^{-1}$ 		    		& 39.73$\pm$0.30	& T.W.	& 40.07$\pm$0.04	 & T.W.    & 39.90$\pm$0.21 	& T.W.	& 40.49$\pm$0.05& T.W.	  \\
$R(H\alpha)$		        & kpc					& 10.73			& T.W.	& 1.22			 & T.W.	   & 1.05		& T.W.	& 0.99$^f$	& T.W.	  \\	  
$SFR$	         		& M$_{\odot}$yr$^{-1}$         		& 0.027$\pm$0.019	& T.W.	& 0.059$\pm$0.005        & T.W.    & 0.040$\pm$0.019  	& T.W.	&0.155$\pm$0.019& T.W.	  \\
log $\Sigma(SFR)$                & M$_{\odot}$yr$^{-1}$kpc$^{-2}$	& -4.12$\pm$0.30	& T.W.	& -1.90$\pm$0.04	 & T.W.	   & -1.93$\pm$0.21	& T.W.	&-1.30$\pm$0.05	& T.W.	  \\
\noalign{\smallskip}
\hline
\noalign{\bigskip}

\hline
\noalign{\smallskip}
\hline
Variable                	& Units					& NGC 4476	& Ref   & NGC 4477	& Ref  & NGC4526	       & Ref  & NGC4552 	      & Ref   \\       
\hline
Type	          		& 					& SA(r)0-	& 1	& SB(s)0:?, Sy2	& 1    & SAB(s)0:	       & 1    &     E;LINER,HII,Sy2   & 1     \\
Spec. ~ class.			&					& Passive,-	& 3	& AGN,wAGN	& 2    & HII,wAGN	       & 2    & -,Retired    	      & 2     \\
$cz$    			& km s$^{-1}$   			&     1968	& 1	&     1338	& 1    &     448	       & 1    &     340 	      & 1     \\
Distance	    		& Mpc                           	&     16.5	& 4,5	&     16.5	& 4,5  &     16.5	       & 4,5  &     16.5	      & 4,5   \\
Membership			&					& Cluster A	& 5	& Cluster A	& 5    & Virgo S. Ext.	       & 5    & Cluster A	      & 5     \\
Proj.~ distance~from~cluster~core& kpc                           	& 60		& T.W.	& 360		& T.W. & 1350		       & T.W. & 350		      & T.W.  \\
$R_{eff}(i)$			& kpc					& 1.26		& 6	& 2.56		& 6    & 4.73		       & 6    & 3.32		      & 6     \\
Fast/slow~rotator		&					&	S	& 7	&	F	& 7    &       F	       & 7    &       S 	      & 7     \\
log $M_{star}^b$   		& M$_{\odot}$				& 10.00$\pm$0.02& T.W.	& 10.71$\pm$0.03& T.W. & 11.07$\pm$0.03        & T.W. &     10.99$\pm$0.02    & T.W.  \\
log $M_{dust}$			& M$_{\odot}$                   	& 6.03$\pm$0.06	& T.W.	& 5.76$\pm$0.09	& T.W. & 7.00$\pm$0.07         & T.W. & 6.78$\pm$0.28	      & T.W.  \\
log $M(HI)$         		& M$_{\odot}$   			& -		& 	& $<$6.95	& 8    & 7.11	       	       & 10   & $<$6.91 	      & 8     \\
log $M(H_2)^c$        		& M$_{\odot}$                   	& 8.17$\pm$0.10	& T.W.	& 7.25$\pm$0.20	& T.W. & 8.50$\pm$0.10	       & T.W. & 8.68$\pm$0.30 	      & T.W.  \\
\hline
	
log $f(H\alpha+[NII])$          & erg s$^{-1}$ cm$^{-2}$    		& -13.10$\pm$0.01& T.W.	& -12.45$\pm$0.01& T.W.& -12.63$\pm$0.01       & T.W. & -12.56$\pm$0.01       & T.W.  \\
log $L(H\alpha)^c$      	& erg s$^{-1}$ 		    		& 39.69$\pm$0.04& T.W.	& 39.93$\pm$0.06& T.W. & 40.47$\pm$0.01        & T.W. & 40.05$\pm$0.02	      & T.W.  \\
$R(H\alpha)$		        & kpc					& 0.73		& T.W.	& 2.04		& T.W. & 1.45		       & T.W. & 1.78		      & T.W.  \\      
$SFR^b$         		& M$_{\odot}$yr$^{-1}$         		& 0.025$\pm$0.002& T.W.	& 0.042$\pm$0.006& T.W.& 0.148$\pm$0.001       & T.W. & 0.057$\pm$0.003	      & T.W.  \\
log $\Sigma(SFR)^d$             & M$_{\odot}$yr$^{-1}$kpc$^{-2}$	& -1.83$\pm$0.04& T.W.	& -2.49$\pm$0.06& T.W. & -1.65$\pm$0.01	       & T.W. & -2.23$\pm$0.02	      & T.W.  \\
\noalign{\smallskip}
\hline
\end{tabular}
\]
References: 1) NED, 2) Gavazzi et al. (2018a) (BTP and WHAN classification), 3) derived from the SDSS spectrum, 4) Mei et al. (2007), 5) Gavazzi et al. (1999), 6) NGVS, 
7) Emsellem et al. (2011), 8) Serra et al. (2012), 9) Boselli et al. (2014a), 10) Haynes et al. (2018)\\
Notes: a) for NGC 4469, distance from M49. b) $M_{star}$ and $SFR$ are derived assuming a Chabrier (2003) IMF and the Kennicutt (1998a) calibration. 
c) whenever available, in order of priority: ALMA 12m, CARMA, from dust mass using a constant gas-to-dust ratio $G/D$=80, and assuming an uncertainty of 0.1, 0.2, and 0.3 dex, respectively.

}
\end{table}
\end{landscape}


\begin{table*}[hbt!]
\caption{VESTIGE observational properties of the target galaxies}
\label{obs}
{
\[
\begin{tabular}{cccccc}
\hline
\noalign{\smallskip}
\hline
Name        		& Units	& NGC 4262	& NGC 4429      & NGC 4459	& NGC 4469	      \\       
\hline
N. exposures		&	& 7		& 10		& 12		& 11		      \\
Seeing			& arcsec& 0.65		& 0.79		& 0.84		& 0.69		      \\
\noalign{\smallskip}
\hline
\noalign{\bigskip}
Name        		& Units	& NGC 4476	 & NGC 4477	 & NGC4526	 & NGC4552	 \\	  
\hline
N. exposures		&	&  12		 & 12		 & 8		 & 12		 \\
Seeing			& arcsec&  0.86 	 & 0.84 	 & 0.69 	 & 0.84 	 \\
\noalign{\smallskip}
\hline
\end{tabular}
\]
Notes: each single exposure is 10 minutes in the H$\alpha$ NB filter, 1 minute in the broad-band $r$ filter.
}
\end{table*}

\begin{table*}
\caption{Multiwavelength data of the target galaxies}
\label{data}
{\tiny
\[
\begin{tabular}{ccccccc}
\hline
\noalign{\smallskip}
\hline
Galaxy	&X-rays			&Ref	&UV		&Ref	&Visible	&Ref	  \\	
\hline
4262	&\textit{Chandra}	&1	&GALEX, UVIT	&8,9,TW	&CFHT, HST	&10,11,TW	  \\
4429	&\textit{Chandra}	&1	&GALEX, UVIT	&9,TW	&CFHT, HST	&11,12,TW  \\
4459	&\textit{Chandra}, XMM	&1,2	&GALEX		&9	&CFHT, HST	&10,11,TW	  \\
4469	&\textit{Chandra}	&1	&GALEX		&9	&CFHT, HST	&11,TW	  \\
4476	&\textit{Chandra}, XMM	&1,2	&GALEX, UVIT	&9,TW	&CFHT, HST	&10,11,TW	  \\
4477	&\textit{Chandra}, XMM	&2,3,4	&GALEX		&9	&CFHT, HST	&11,TW	  \\
4526	&\textit{Chandra}, XMM	&2,4	&GALEX		&9	&CFHT, HST	&10,11,TW	  \\
4552	&\textit{Chandra}, XMM	&4,5,6,7&GALEX		&9	&CFHT, HST	&10,11,TW	  \\
\noalign{\smallskip}
\hline

\noalign{\smallskip}
\hline
Galaxy	&IR			  &Ref		  &CO		  &Ref	  	&HI	  &Ref    \\	
\hline
4262	&WISE, \textit{Spitzer}, \textit{Herschel}    &13,13,15,16	  &IRAM		  &18	  	&WSRT     &24,25  \\
4429	&WISE, \textit{Spitzer}, \textit{Herschel}    &13,13,15,16	  &CARMA, ALMA	  &12,19  	&WSRT	  &26	  \\
4459	&WISE, \textit{Spitzer}, \textit{Herschel}    &13,13,15,16	  &CARMA, ALMA, IRAM&18,21,TW  	&WSRT	  &26	  \\
4469	&WISE, \textit{Spitzer}, \textit{Herschel}    &13,13,15,16	  &		  &	  	&Arecibo  &27	  \\
4476	&WISE, \textit{Spitzer}, \textit{Herschel}    &13,17,TW		  &CARMA, ALMA 	  &21,TW  	&WSRT	  &26	  \\
4477	&WISE, \textit{Spitzer}, \textit{Herschel}    &13,13,15,16	  &CARMA, PdBI, IRAM&18,22  	&WSRT	  &26	  \\
4526	&WISE, \textit{Spitzer}, \textit{Herschel}    &13,13,15,16	  &CARMA, ALMA, IRAM&18,23,TW	&Arecibo  &27	  \\
4552	&WISE, \textit{Spitzer}, \textit{Herschel}    &13,13,15,16	  &IRAM		  &18	  	&WSRT	  &26	  \\
\noalign{\smallskip}
\hline
\end{tabular}
\]
References: 
1) \textit{Chandra} data archive (P.Is.: Treu, Soria, Ptak, Gultekin); 2) XMM data archive (P.Is.: David, Forman, Sanders, Jansen, Boehringer, Turner, Aschenbach, Irwin); 
3) Li et al. (2018); 4) Kim et al. 2019; 5) Machaceck et al. (2006a); 6) Machacek et al. (2006b); 7) Kraft et al. (2017);  
8) Bettoni et al. (2010); 9) Boselli et al. (2011); 10) Cote et al. (2004); 11) Ferrarese et al. (2012); 12) Davis et al. (2018);
13) Wright et al. (2010); 14) Bendo et al. (2012a); 15) Cortese et al. (2014); 16) Ciesla et al. (2012); 17) Auld et al. (2013); 18) Combes et al. (2007);
19) Alatalo et al. (2013);  20) Young et al. (2008);
21) Young (2002); 22) Crocker et al. (2011); 23) Utomo et al. (2015); 24) Krumm et al. (1985); 25) Oosterloo et al. (2010); 26) Serra et al. (2012); 27) Haynes et al. (2018).
}
\end{table*}

\section{Observations and data reduction}

\subsection{VESTIGE narrow-band imaging}

Narrow-band (NB) H$\alpha$ imaging observations were carried out during the VESTIGE survey (Boselli et al. 2018a). The data were obtained using MegaCam at the CFHT in 
the NB filter MP9603 ($\lambda_c$ = 6590 \AA; $\Delta\lambda$ = 104 \AA). Given the redshift of the galaxies, this filter includes the emission of the 
Balmer H$\alpha$ line and of the two [NII] lines ($\lambda$ = 6548, 6583 \AA)\footnote{Hereafter we refer to the H$\alpha$+[NII] band simply as H$\alpha$, 
unless otherwise stated.}. Galaxies were observed with typically 1-2 hours of integration 
in the NB filter, and 6-12 min in the broad-band $r$ filter, the latter being necessary for the subtraction of the stellar continuum (see Table \ref{obs}).
A detailed description of the observing strategy and of the data reduction is given in Boselli et al. (2018a). Briefly, MegaCam is composed of 40 CCDs 
with a pixel scale of 0.187 arcsec pixel$^{-1}$. The observations were carried out following a specific observing sequence to optimise the 
determination of the sky background necessary to detect low surface brightness extended features. The observations were taken during excellent
sky conditions, with a typical seeing ranging from 0.65 to 0.86 arcsec (see Table \ref{obs}). The sensitivity of the survey at full depth (2h in the NB filter, 12 min in the broad
$r$-band filter) is of $f(H\alpha)$ $\simeq$ 4$\times$10$^{-17}$ erg~s$^{-1}$~cm$^{-2}$ (5$\sigma$) for point sources and $\Sigma(H\alpha)$ $\simeq$ 2$\times$10$^{-18}$ 
erg~s$^{-1}$~cm$^{-2}$~arcsec$^{-2}$ (1$\sigma$ after smoothing the data to $\sim$ 3\arcsec\ resolution) for extended sources.

As for all the VESTIGE data, the images were reduced using Elixir-LSB (Ferrarese et al. 2012), a data reduction pipeline designed to detect extended and 
low surface brightness features. The photometric calibration in both filters and the astrometric corrections were then secured using the standard MegaCam 
procedures as described in Gwyn (2008). The typical photometric uncertainty in both bands is $\lesssim$ 0.02-0.03 mag. 

\begin{figure*}
   \centering
   \includegraphics[width=0.8\textwidth]{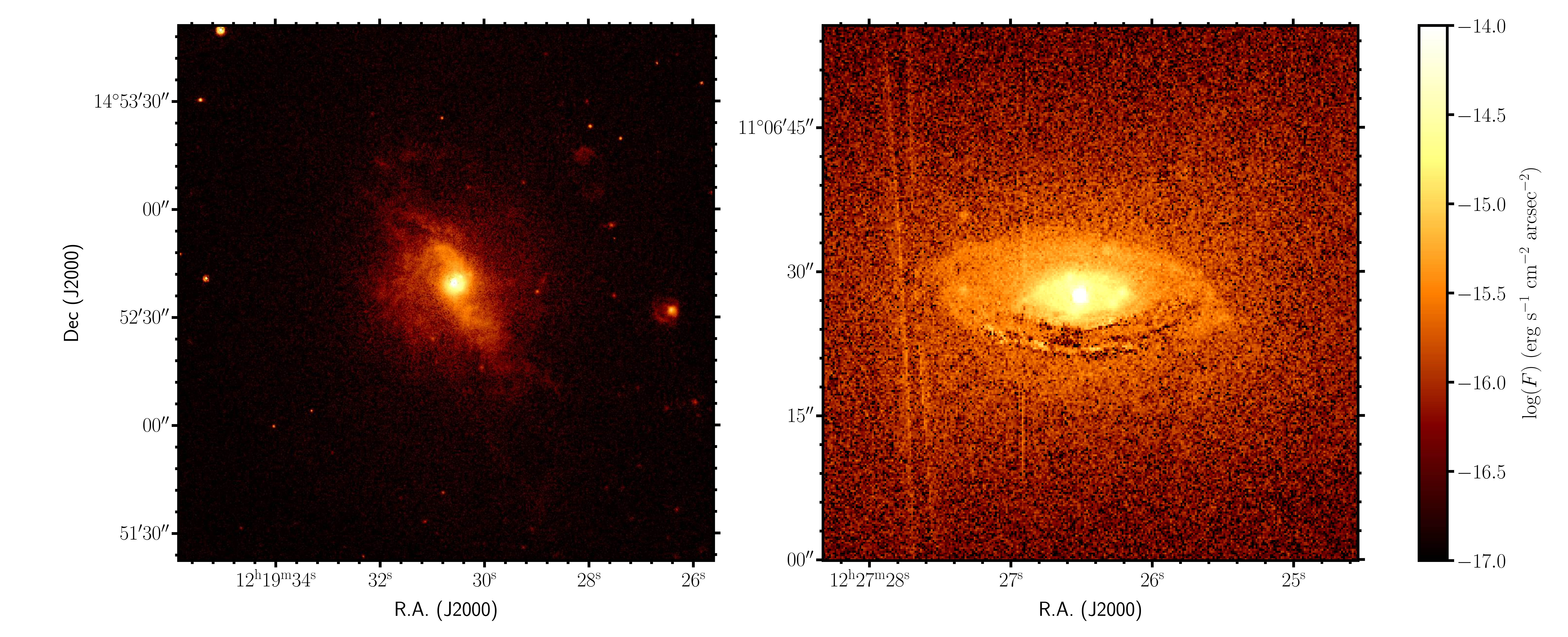}\\
   \includegraphics[width=0.8\textwidth]{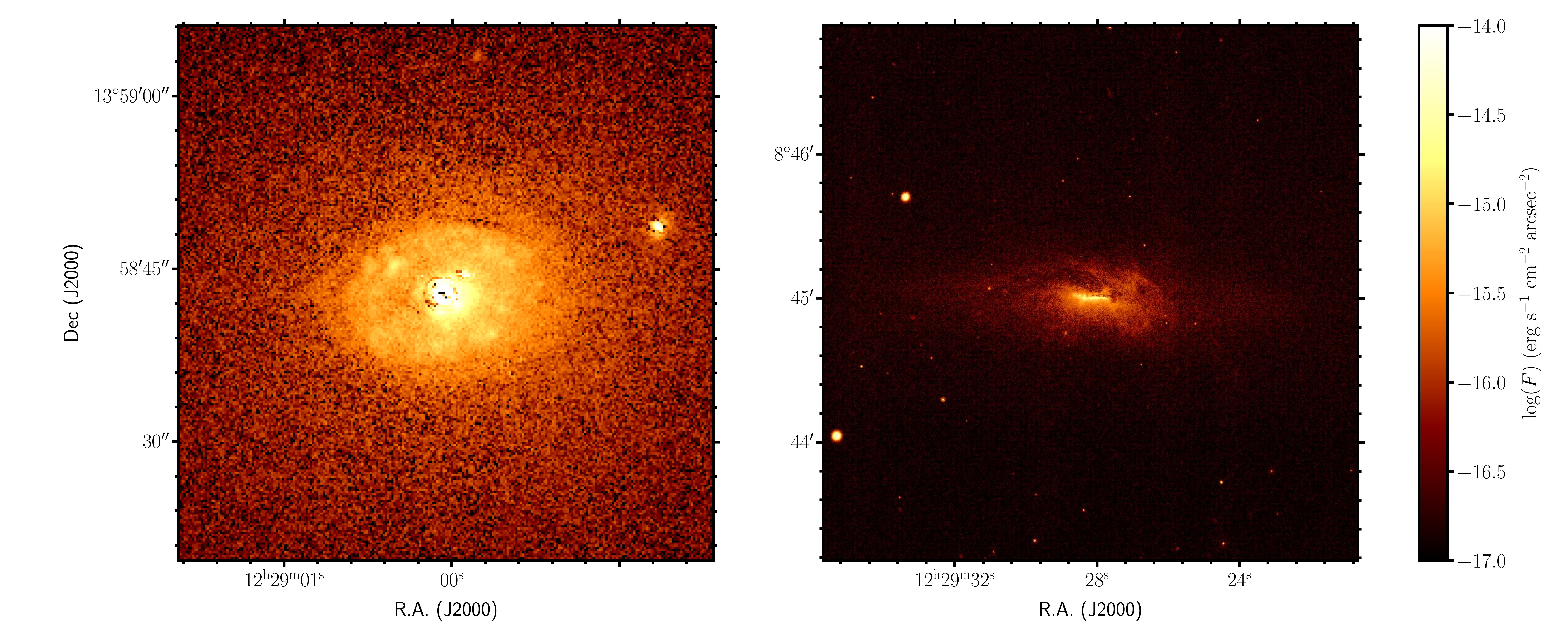}\\
   \includegraphics[width=0.8\textwidth]{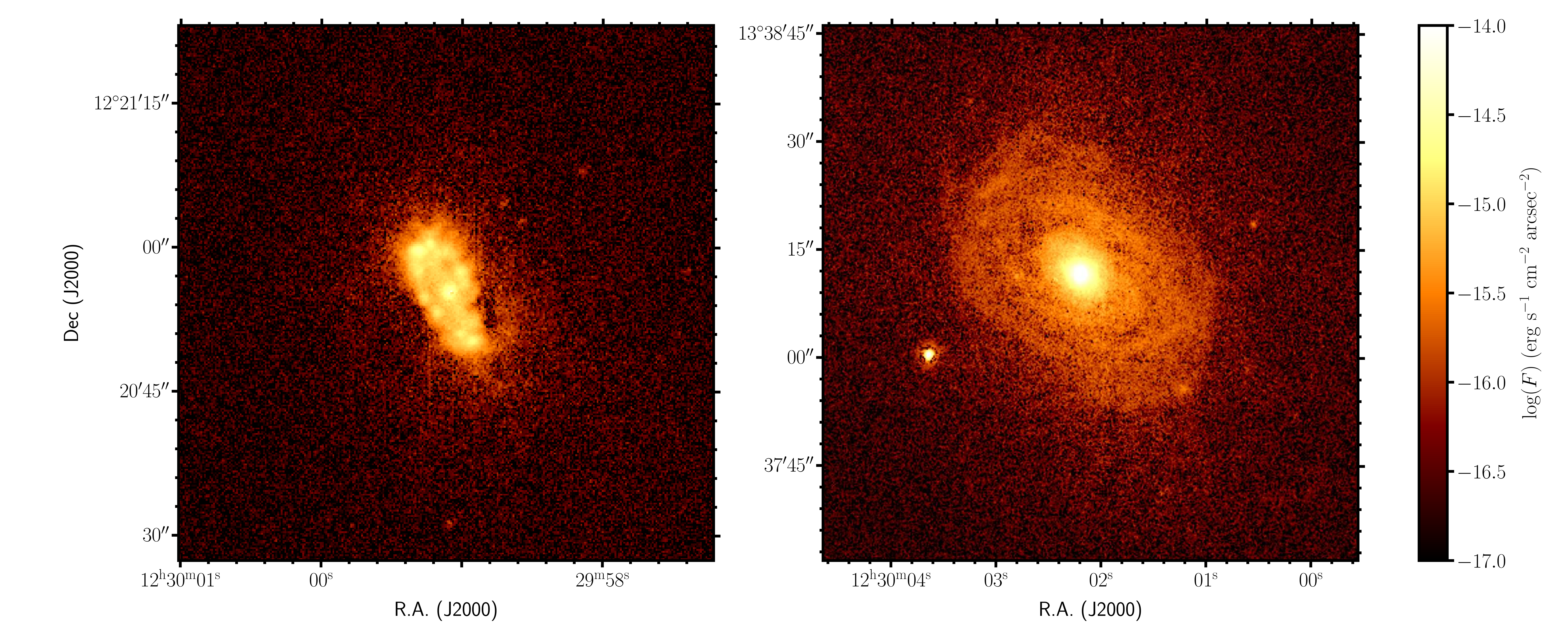}\\
   \includegraphics[width=0.8\textwidth]{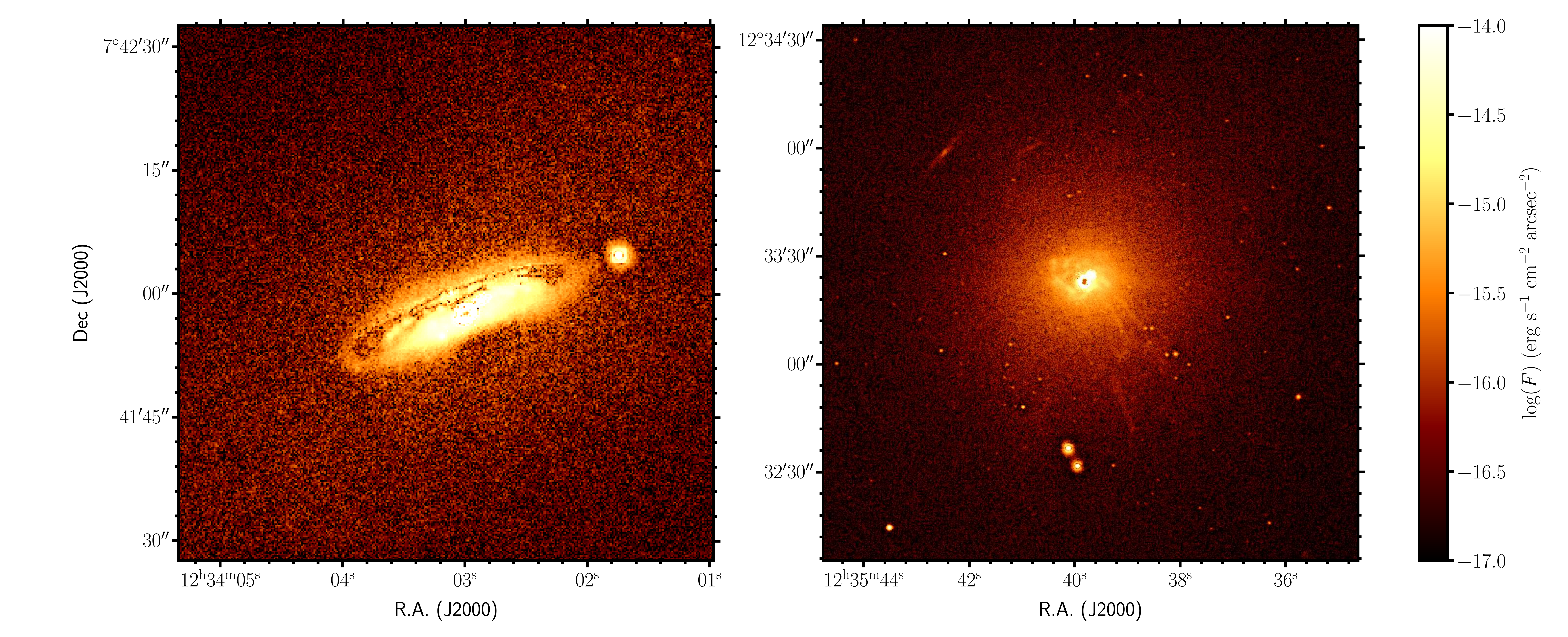}\\
   \caption{Continuum-subtracted H$\alpha$ images of the galaxies NGC 4262 (first row, left), 4429 (first row, right), 4459 (second row, left), 
   4469 (second row, right), 4476 (third row, left), 4477 (third row, right), 4526 (last row, left), and 4552 (last row, right). As in all the following
   figures, North is up, East is left.
 }
   \label{Ha_image}%
   \end{figure*}

As extensively described in Boselli et al. (2019), the subtraction of the stellar continuum is particularly critical in early-type galaxies where
the emission of the ionised gas is marginal with respect to that of the stars. We thus followed the same procedure used for M87, where the stellar continuum was
derived using the $r$-band frame combined with a $g-r$ colour map, the latter constructed using the NGVS $g$-band frame. For this purpose, 
we used a particular $g$-band frame produced by the NGVS survey, where saturated pixels in the core of these bright galaxies were accurately replaced 
with those gathered during short exposures (Ferrarese et al. 2012). The continuum-subtracted H$\alpha$ images of the eight galaxies
analysed in this work are shown in Fig. \ref{Ha_image}. 
Table \ref{gal} gives the total fluxes (in units of erg s$^{-1}$ cm$^{-2}$) and uncertainties of the H$\alpha$ emission line of the eight target galaxies.
As for the other VESTIGE papers (Fossati et al. 2018, Boselli et al. 2018b, 2018c), fluxes and uncertainties were derived by measuring both the galaxy emission and the sky background within the same elliptical aperture, 
randomly displaced on the sky after masking other contaminating sources. To minimise the uncertainty on the sky determination, this procedure was run 1000 times.
In a few galaxies (NGC 4262, 4459, 4526, 4552) the $r$-band images are slightly saturated in the nucleus. For these objects the saturated central region has been masked.
The uncertainties on the fluxes are obtained as the quadratic sum of
the uncertainties on the flux counts and the uncertainties on the
background (rms of the bootstrap iterations). The uncertainties
on the flux counts are derived assuming a Poissonian distribution
for the source photo-electrons.
The extracted fluxes give thus integrated values. With the exception of the two galaxies NGC 4262 and NGC 4552, where the H$\alpha$ emission is 
peaked in the centre, the extracted fluxes are representative of the discs of ionised gas.

\subsection{ASTROSAT/UVIT imaging}

\begin{figure}
   \centering
   \includegraphics[width=0.49\textwidth]{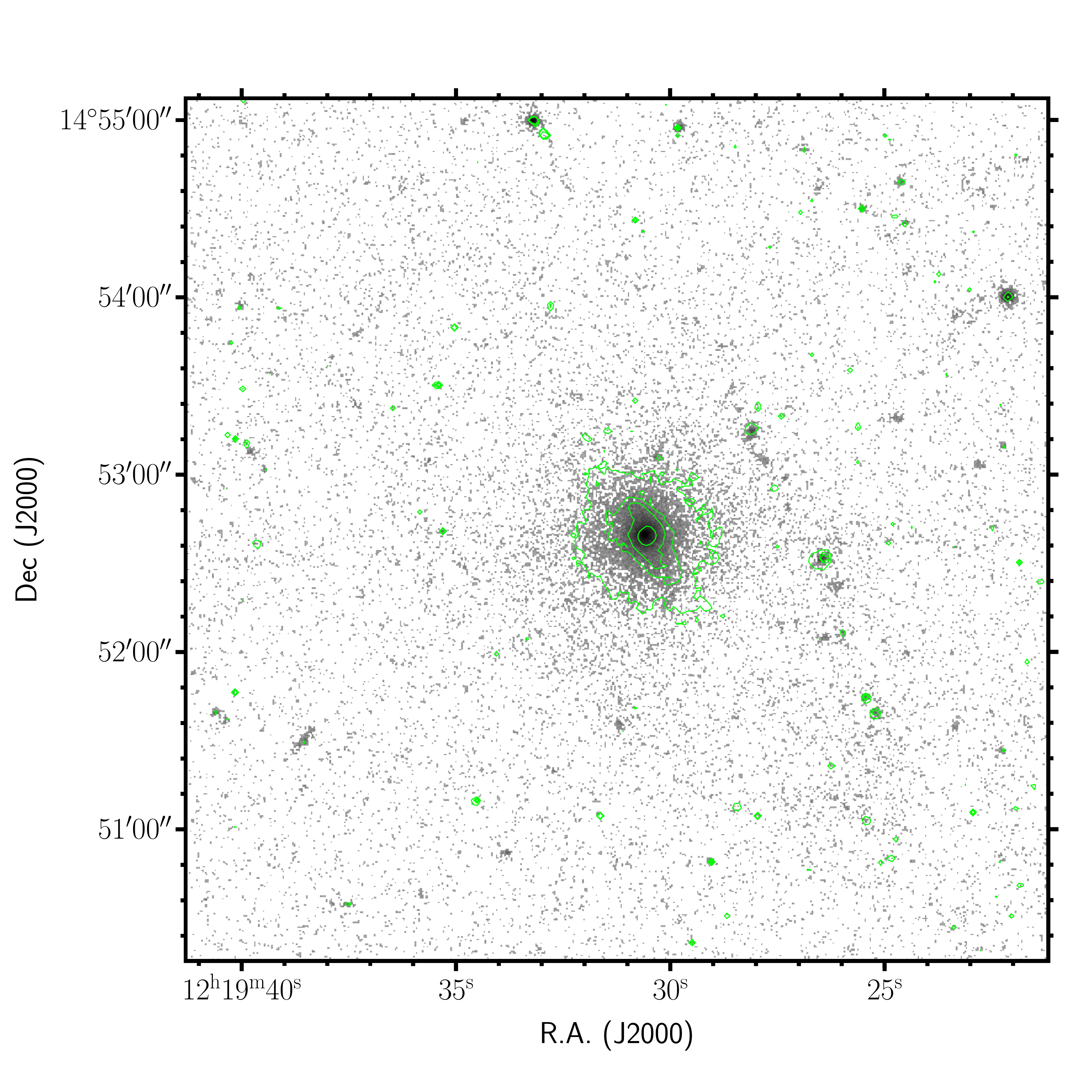}\\
   \caption{H$\alpha$+[NII] contours (at a surface brightness level of $\Sigma(H\alpha+[NII])$ = 3 $\times$ 10$^{-17}$, 1 $\times$ 10$^{-16}$, 2 $\times$ 10$^{-16}$, 1 $\times$ 
   10$^{-15}$ erg s$^{-1}$ cm$^{-2}$ arcsec$^{-2}$) 
   plotted over the UVIT NUV image in the Silica15 band of the galaxy NGC4262 (grey colours).
 }
   \label{UVIT4262}%
   \end{figure}

\begin{figure}
   \centering
   \includegraphics[width=0.49\textwidth]{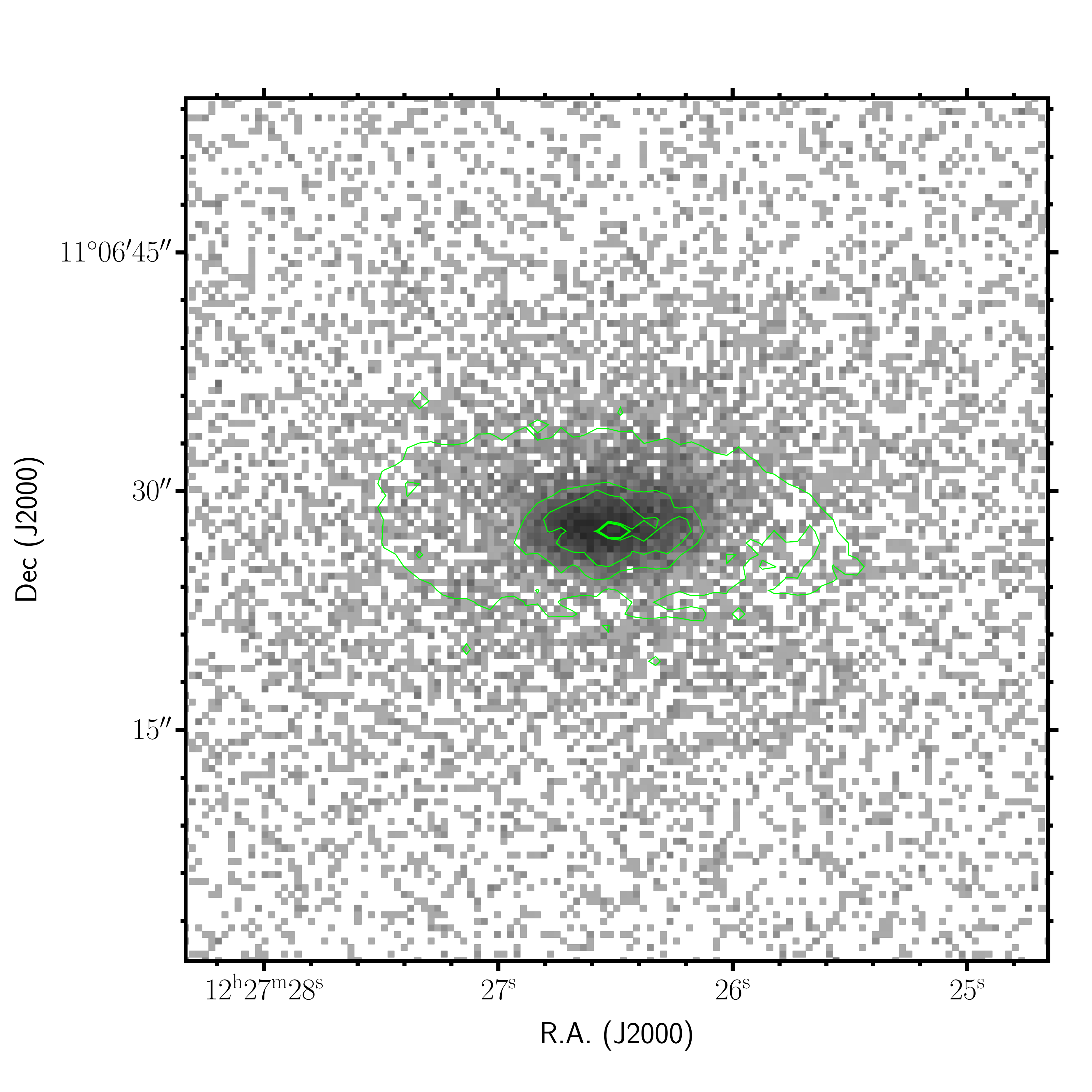}\\
   \caption{H$\alpha$+[NII] contours (at a surface brightness level of $\Sigma(H\alpha+[NII])$ = 1.5 $\times$ 10$^{-16}$, 6 $\times$ 10$^{-16}$, 1.2 $\times$ 
   10$^{-15}$ erg s$^{-1}$ cm$^{-2}$ arcsec$^{-2}$) 
   plotted over the UVIT FUV image in the BaF2 band of the galaxy NGC4429 (grey colours).
 }
   \label{UVIT4429}%
   \end{figure}

\begin{figure}
   \centering
   \includegraphics[width=0.49\textwidth]{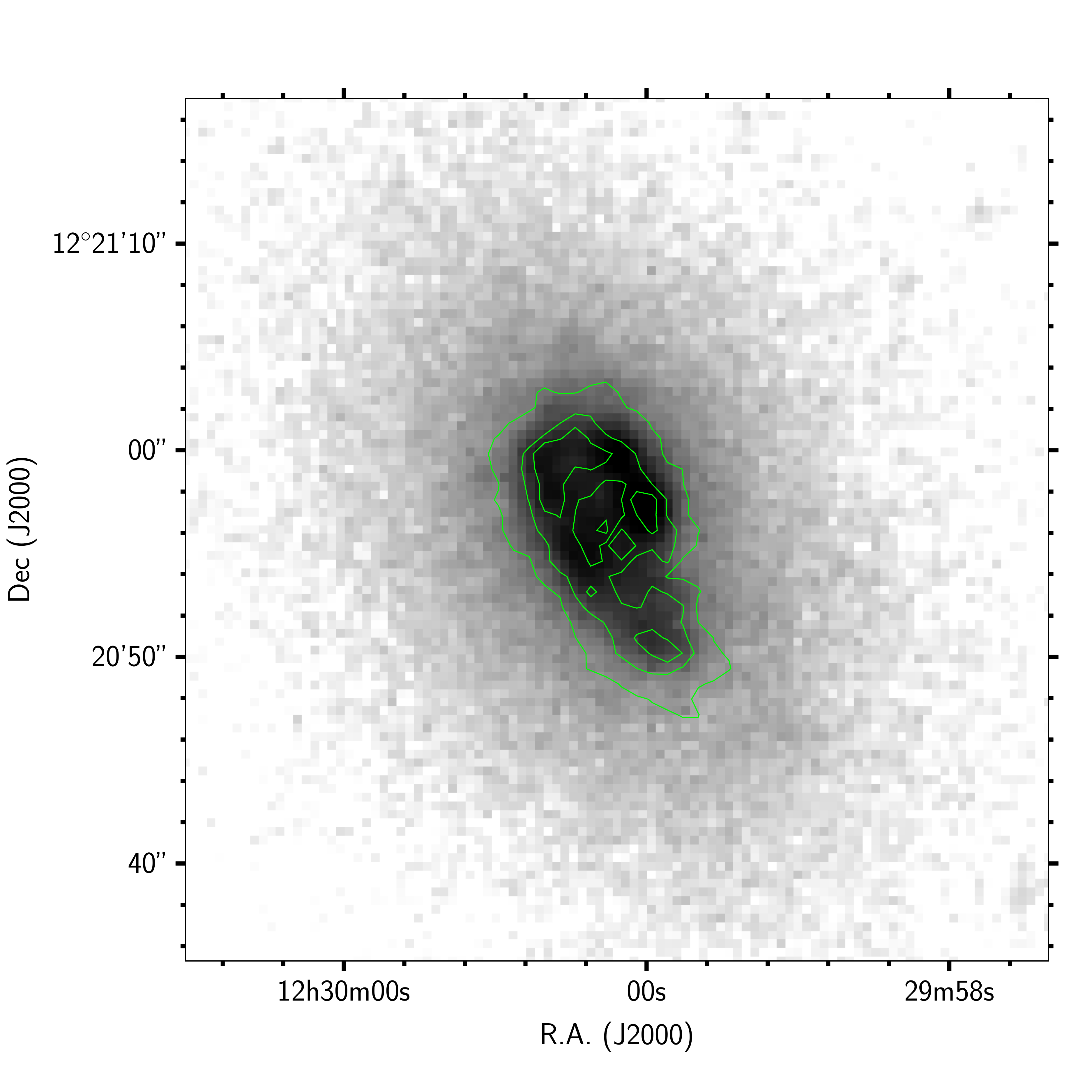}\\
   \caption{H$\alpha$+[NII] contours (at a surface brightness level of $\Sigma(H\alpha+[NII)]$ = 10$^{-16}$, 5 $\times$ 10$^{-16}$, 10$^{-15}$ erg s$^{-1}$ cm$^{-2}$ arcsec$^{-2}$) 
   plotted over the UVIT FUV image in the BaF2 band of the galaxy NGC4476 (grey colours).
 }
   \label{UVIT4476}%
   \end{figure}

Three of the target galaxies were observed with ASTROSAT/UVIT (Agrawal et al. 2006; Tandon et al. 2020). One of them, NGC 4429,
was observed in May 2020 using the far-UV filter BaF2 ($\lambda_c$ = 1541 \AA; $\Delta \lambda$ = 380 \AA) during a run dedicated to the observations of 
a representative sample of Virgo cluster galaxies (PI. J. Hutchings). The other two galaxies, NGC~4262 and NGC~4476, were observed in the far-UV filter BaF2 
and in the near-UV filter Silica15 ($\lambda_c$ = 2418 \AA; $\Delta \lambda$ = 785 \AA) and have data available in the archives, as specified in Table \ref{UVIT}. 
The field of view of the instrument has a diameter of $\simeq$ 28\arcmin\ and 
an angular resolution of $\simeq$ 1.5\arcsec. All the data were reduced following the prescriptions given in Tandon et al. (2020) 
using a zero point of $z_p$ = 17.771 mag for the FUV BaF2 filter and $z_p$ = 19.763 for the NUV Silca15 filter. The
astrometry was checked against the accurate NGVS data (Ferrarese et al. 2012, see below). The ASTROSAT/UVIT images of the three galaxies are shown in 
Fig. \ref{UVIT4262}, \ref{UVIT4429}, and \ref{UVIT4476}.

\begin{table}
\caption{ASTROSAT/UVIT data}
\label{UVIT}
{
\[
\begin{tabular}{cccc}
\hline
\noalign{\smallskip}
\hline
NGC	        & Filter	& exp.time	& limiting S.B.	\\
Units		&		& sec		& mag arcsec$^{-2}$\\ 
\hline
4262		& BaF2		& 2660		& 25.7	\\
		& Silica15	& 2747		& 26.5	\\
4429		& BaF2		& 7142		& 26.2	\\
4476		& BaF2		& 15896		& 27.1	\\
		& Silica15	& 15077		& 28.3	\\	
\noalign{\smallskip}
\hline
\end{tabular}
\]
}
\end{table}

\subsection{MUSE spectroscopy}

Two of the galaxies, NGC 4469 and NGC 4526, were observed with MUSE during the 099.B-0384 (PI O. Gonzalez) 
and 097.D-0408 (PI J. Anderson) programs, the former designed to study 
the kinematics of boxy-shaped early-type discs, the latter the properties of host galaxies of core-collapse supernovae
(the AMUSING survey, Galbany et al. 2016). The MUSE data were acquired in May 2017 and in May 2016 in the Wide Field Mode under medium ($FWHM$ = 1.19\arcsec) and 
good seeing conditions ($FWHM$ = 0.91\arcsec) for NGC 4469 and NGC 4526, respectively. The data cover
the spectral range 4800-9300 \AA\ with a spectral resolution $R$ $\sim$ 2600 (1.25 \AA\ sampling per pixel), corresponding to a limiting velocity dispersion $\sigma$
$\sim$ 50 km s$^{-1}$ (see Boselli et al. 2021b). The data were acquired with an integration time of 2448 s (NGC4469) and 2360 s (NGC 4526), 
able to reach a typical sensitivity to low surface
brightness features at H$\alpha$ of $\Sigma(H\alpha)$ $\simeq$ 4 $\times$ 10$^{-18}$ erg s$^{-1}$ cm$^{-2}$ arcsec$^{-2}$. The data were reduced using ad-hoc procedures
developed within the team and presented in Fossati et al. (2016), Consolandi et al. (2017), and Boselli et al. (2018c, 2021b). The astrometry of the data cubes was
registered on the VESTIGE image using point sources in the field. A comparison of the flux of the H$\alpha$+[NII] emission line extracted from the MUSE data cubes with
the one obtained from the VESTIGE NB image gives $f(H\alpha +[NII])_{VESTIGE}/f(H\alpha +[NII])_{MUSE}$ = 1.05 for NGC 4469 and 
$f(H\alpha +[NII])_{VESTIGE}/f(H\alpha +[NII])_{MUSE}$ = 1.18 for NGC 4526. This difference, which is significantly larger than that observed in star-forming 
galaxies such as NGC 4424 ($\lesssim$ 3.5\%; Boselli et al. 2018c) and IC 3476 ($\lesssim$ 1\%; Boselli et al. 2021b), is probably due to: 
a) the H$\alpha$ emission is coming principally from the inner regions, where the total flux in 
the NB filter is highly dominated by uncertainties in the stellar continuum emission; 
b) the H$\alpha$ flux derived from the MUSE data has been determined after subtraction of the stellar continuum emission using the GANDALF code. In this way, 
the H$\alpha$ line in MUSE is corrected for any possible contamination due to a strong underlying Balmer absorption, not taken into account in the NB imaging data of VESTIGE. 
The 5-18\%\ differences (0.02-0.07 dex) observed in NGC 4469 and NGC 4526 can thus be considered as the typical uncertainty on 
the H$\alpha$ flux determination on the NB VESTIGE data in these bright early-type systems. We recall, however, that 18\%\ should be taken as an upper limit for 
the other galaxies, where the emission in their centre and the dust attenuation along the disc are less extreme than in NGC 4526.

\subsection{SOAR spectroscopy}

SOAR observations were carried out with the Goodman spectrograph (Clemens et al. 2004) on the 4.1 m Southern Astrophysical
Research (SOAR) Telescope at Cerro Pach\'on on April 14th, 2021. 
The spectra were taken with a 600 lines mm$^{-1}$ grating and a 1.5\arcsec wide slit, using a 2$\times$2 binning, resulting in a spectral and 
spatial sampling of 1.3\AA\ pixel$^{-1}$ and 0.3\arcsec pixel$^{-1}$, respectively. The resolving power of Goodman in the wavelength range 4350-7020\AA\ 
covered during the observations (MID configuration) is $R$ $\simeq$ 2800. The exposure time was 2.5 hrs (divided into five exposures of 1800 s) for NGC 4429 
and 3 hrs (divided into six exposures of 1800 s) for NGC 4476. Two spectrophotometric standard stars were observed, one at the beginning and one 
at the end of the night run, to secure the flux calibration.

The data were reduced with basic calibrations including bias, flat, cosmic ray removal and wavelength calibration obtained using a combination of CuHeAr and HgArNe lamps. 
The standard star observations were reduced and the stellar spectrum was extracted and compared to the reference spectrum from the X-SHOOTER 
calibration database, to derive the instrument response function. The function was fit with a 4$^{th}$ degree polynomial and applied to calibrate the science data which was 
subsequently shifted and combined with mean statistics. To trace the properties of the H$\alpha$ emitting disc
and avoid any possible contamination from the nucleus, a pair of 1D spectra were extracted on both sides of the spectral slit at radial distances from the galaxy centres 
of $3.3"-7.5"$ and $3.3"-7.8"$ for NGC4429 and NGC4476, respectively. Each spectrum was fit with the {\sc GANDALF} code (Sarzi et al. 2006) using the MILES 
stellar spectral library (Vazdekis et al. 2010) to model the stellar continuum and the nebular line emission. Emission line fluxes, reported in Table \ref{SOAR}, 
were averaged from the two spectral extractions when used in the following analysis.

\begin{table*}
\caption{SOAR data}
\label{SOAR}
{
\[
\begin{tabular}{ccccccccccc}
\hline
\noalign{\smallskip}
\hline
NGC	& position& H$\beta$ & [OIII]$\lambda$5007 & H$\alpha$ & [NII]$\lambda$6583 & [SII]$\lambda$6716 & [SII]$\lambda$6731 & S/N(H$\alpha$) & S/N(H$\beta$) & C(H$\beta$) \\
\hline
4429    &  W	  & 0.16     &	0.05	           &  1.00    &  0.84              & 0.20	        & 0.21	             & 20.19          & 2.19          & 1.16$\pm$0.67 \\
4429    &  E	  & 0.01     &	0.00	           &  1.00    &  1.06              & 0.24	        & 0.20	             & 16.82          & 0.60          & 4.78$\pm$2.45 \\
4476    &  NW	  & 0.16     &	0.03	           &  1.00    &  0.47              & 0.16	        & 0.11	             & 55.65          & 5.92          & 1.11$\pm$0.25 \\
4476    &  SE	  & 0.11     &	0.02	           &  1.00    &  0.53              & 0.19	        & 0.15	             & 53.66          & 3.31          & 1.72$\pm$0.44 \\

\noalign{\smallskip}
\hline
\end{tabular}
\]
Notes: fluxes are normalised to H$\alpha$; the orientation of the slit was E-W for NGC 4429 and SE-NW (31$^o$ from North, counter clockwise) for NGC 4476.
}
\end{table*}

\subsection{ALMA}

High quality millimetric data necessary for the determination of the molecular gas content and distribution were gathered using the
Atacama Large Millimeter/submillimeter Array (ALMA) and are available in the archives for the galaxies NGC 4429, 4459, 4476, and 4526
at different frequencies, as indicated in Table \ref{ALMA}.

NGC4429 was observed during ALMA Cycle 1 (Project 2013.1.00493.S, PI M. Bureau) with 42 antennae in band-7 tuned between 343.5 GHz to 345.5 GHz, 
targeting the $^{12}$CO(3-2) line. The spatial resolution is 0.19$\times$0.16 arcsec. The observations were taken on 26$^{th}$ and 27$^{th}$ June 2015 with 
a total integration time of 4233 seconds. 
The observations for NGC 4459 and NGC 4476 were carried out in Cycle 5 on
September 2017 with a total integration time of $\sim$ 25200 seconds using the 12~m and 7~m arrays (PI: M. Chevance; Programme ID: 2017.1.00766.S).
The spectral setup was optimised for the $^{12}$CO(2-1) transition line in band-6, at 229.68 GHz rest frequency.
At this frequency, the beam size for NGC 4459 is 0.58\arcsec$\times$0.50\arcsec for the 12~m array and 6.64\arcsec$\times$4.99\arcsec for 7~m array, while for NGC 4476
0.744\arcsec$\times$0.555\arcsec and 7.08\arcsec$\times$4.77\arcsec 
for the 12~m and 7~m arrays, respectively.
The observations for NGC 4526 were carried out in Cycle 6 on April 2019
with an integration time of 2580 seconds using the 12~m array (PI: S. Raimundo; Programme ID: 2018.1.01599.S).
The spectral setup was optimised for the $^{12}$CO(1-0) transition line in
band-6, at 114.96 GHz rest frequency, with a beam size of 1.14\arcsec$\times$0.76\arcsec at a position angle of $PA$ = -78.828$^o$.

The original raw data were recalibrated with the Common Astronomy Software Applications package (CASA, McMullin et al. 2007), version  4.2.0, and re-imaged with the 5.8 version. The image reconstruction was
performed using the standard \textsc{TCLEAN} deconvolution using the Briggs weighting algorithm (robust parameter 0.5). The moments were computed in CASA using a mask to cover all pixels under a 
given threshold to shrink the noise contribution to the final maps, using the CASA task \textsc{immoment}.

\begin{table*}
\caption{CO data}
\label{ALMA}
{
\[
\begin{tabular}{ccccccccc}
\hline
\noalign{\smallskip}
\hline
	        & ALMA		& 				& 		& 			& CARMA		&				&		&		\\
NGC		& Transition	& Beam				& I(CO)		& $M(H_2)$		& Transition	& Beam				& I(CO)		& $M(H_2)$	\\ 
Units		&		& \arcsec $\times$ \arcsec	& Jy km s$^{-1}$& M$_{\odot}$		&		& \arcsec $\times$ \arcsec	& Jy km s$^{-1}$& M$_{\odot}$	\\
\hline
4429		& CO(3-2)	& 0.18$\times$0.15		& 52.70		& 4.15$\times$10$^8$ 	& CO(1-0)	& 4.69$\times$3.73		& 64.66 	& 1.58$\times$10$^8$\\
4459		& CO(2-1)	& 0.58$\times$0.50		& 127.00	& 4.77$\times$10$^8$	& CO(1-0)	& 9.01$\times$5.53		& 55.52		& 1.36$\times$10$^8$\\
4459		& CO(2-1)	& 6.64$\times$4.99		& 103.90	& 3.90$\times$10$^8$	& CO(1-0)	& 				& 		& \\
4476		& CO(2-1)	& 0.74$\times$0.55		& 39.15		& 1.47$\times$10$^8$ 	& CO(1-0)	& 8.29$\times$5.66		& 29.25		& 7.14$\times$10$^7$\\
4476		& CO(2-1)	& 7.08$\times$4.77		& 40.17		& 1.51$\times$10$^8$ 	& CO(1-0)	& 				& 		& \\
4477		&		&				&		& 			& CO(1-0)	& 3.30$\times$2.62		& 7.34		& 1.79$\times$10$^7$\\
4526		& CO(1-0)	& 1.14$\times$0.76		& 128.60	& 3.14$\times$10$^8$	& CO(1-0)	& 4.99$\times$3.84		& 166.10	& 4.06$\times$10$^8$\\
\noalign{\smallskip}
\hline
\end{tabular}
\]
Notes: $M(H_2)$ are derived assuming a standard CO-to-H$_2$ conversion factor $X_{CO}$ = 2.3 $\times$ 10$^{20}$ cm$^{-2}$/(K km s$^{-1}$) (Strong et al. 1988) and an 
intensity ratio CO(2-1)/CO(1-0) = $R_{21}$ = 0.65 and CO(3-2)/CO(1-0) = $R_{31}$ = 0.31 (Leroy et al. 2021).
}
\end{table*}

\subsection{Multifrequency data}

\begin{figure*}
   \centering
   \includegraphics[width=0.39\textwidth]{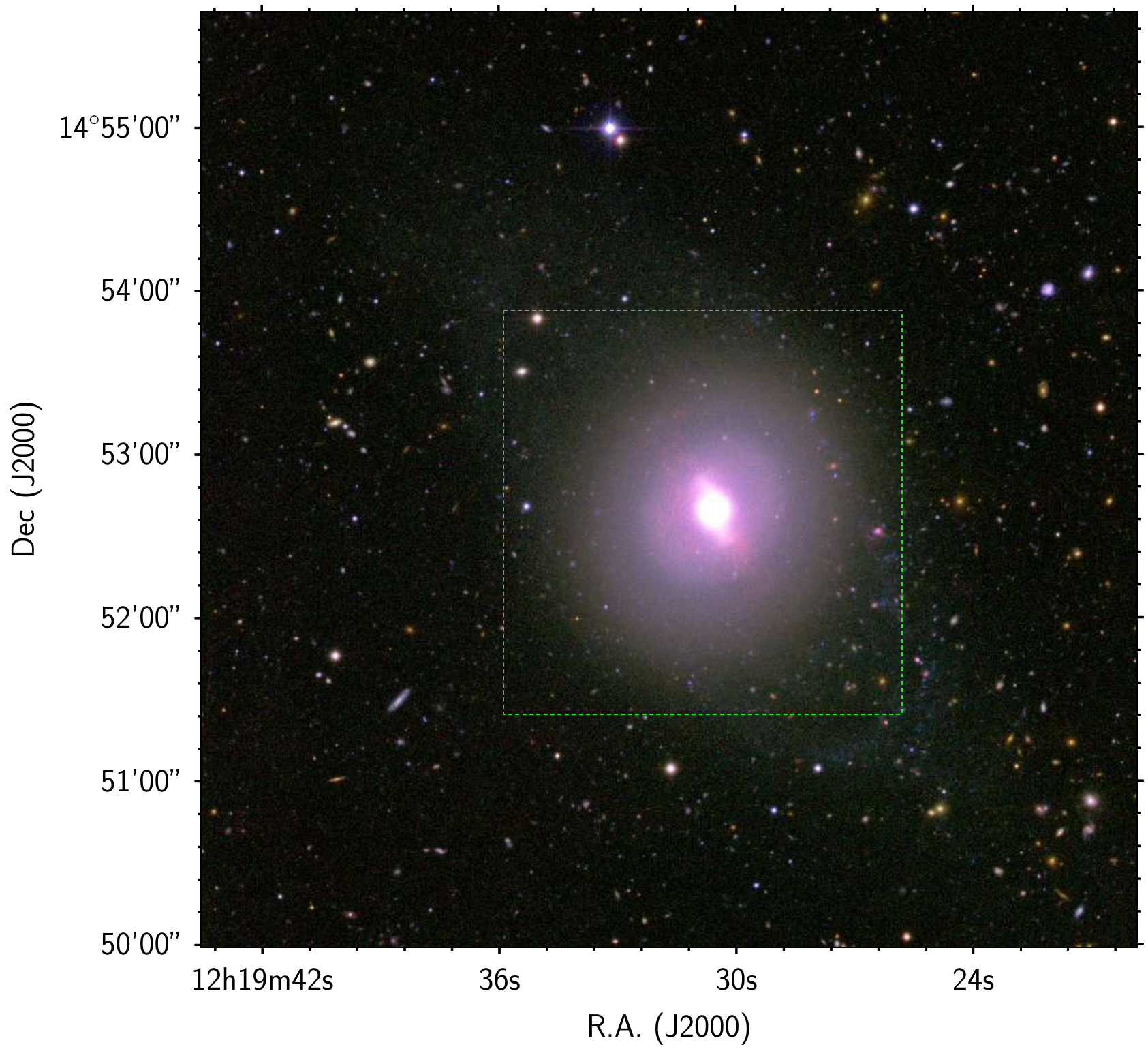}
   \includegraphics[width=0.45\textwidth]{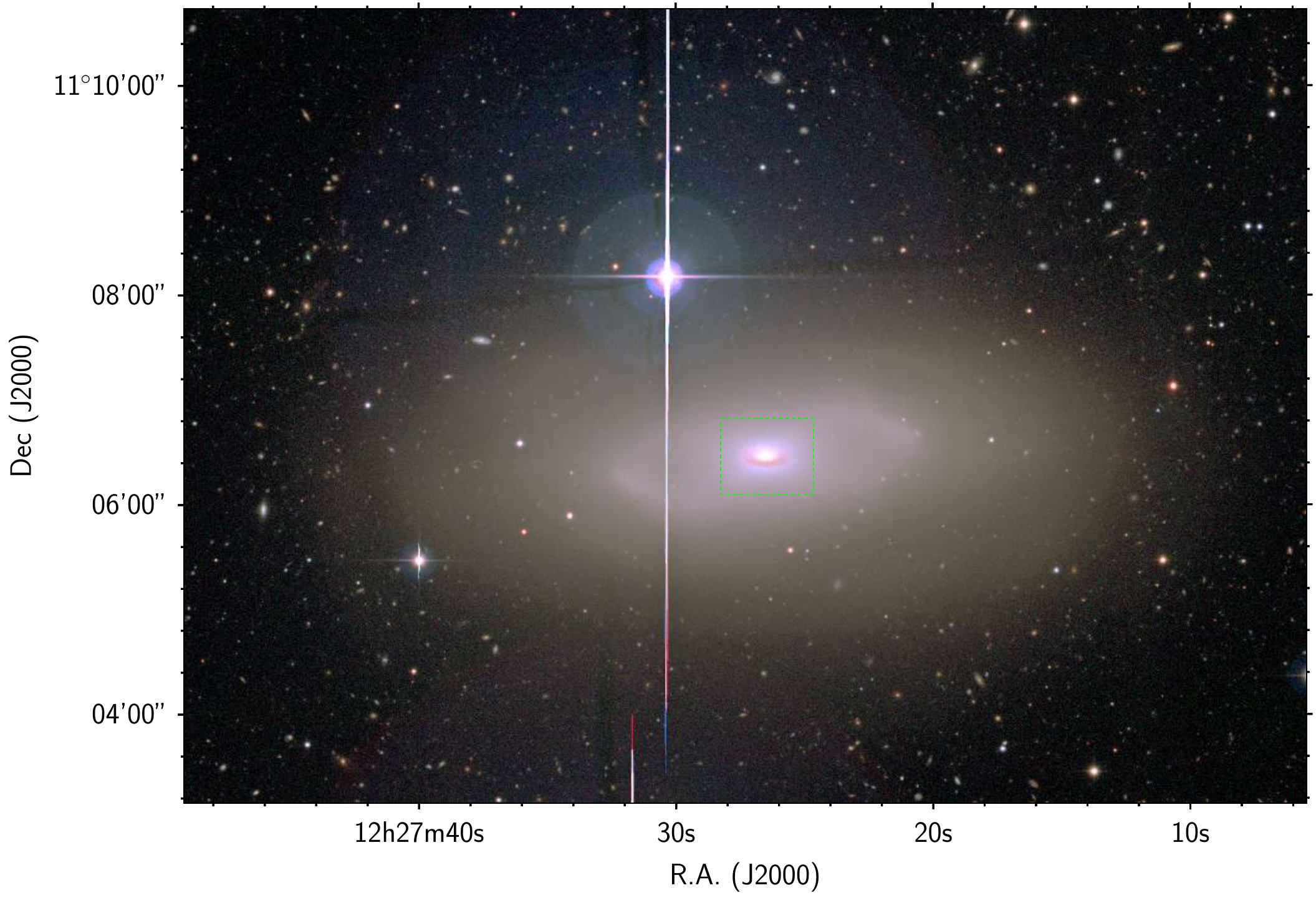}\\
   \includegraphics[width=0.39\textwidth]{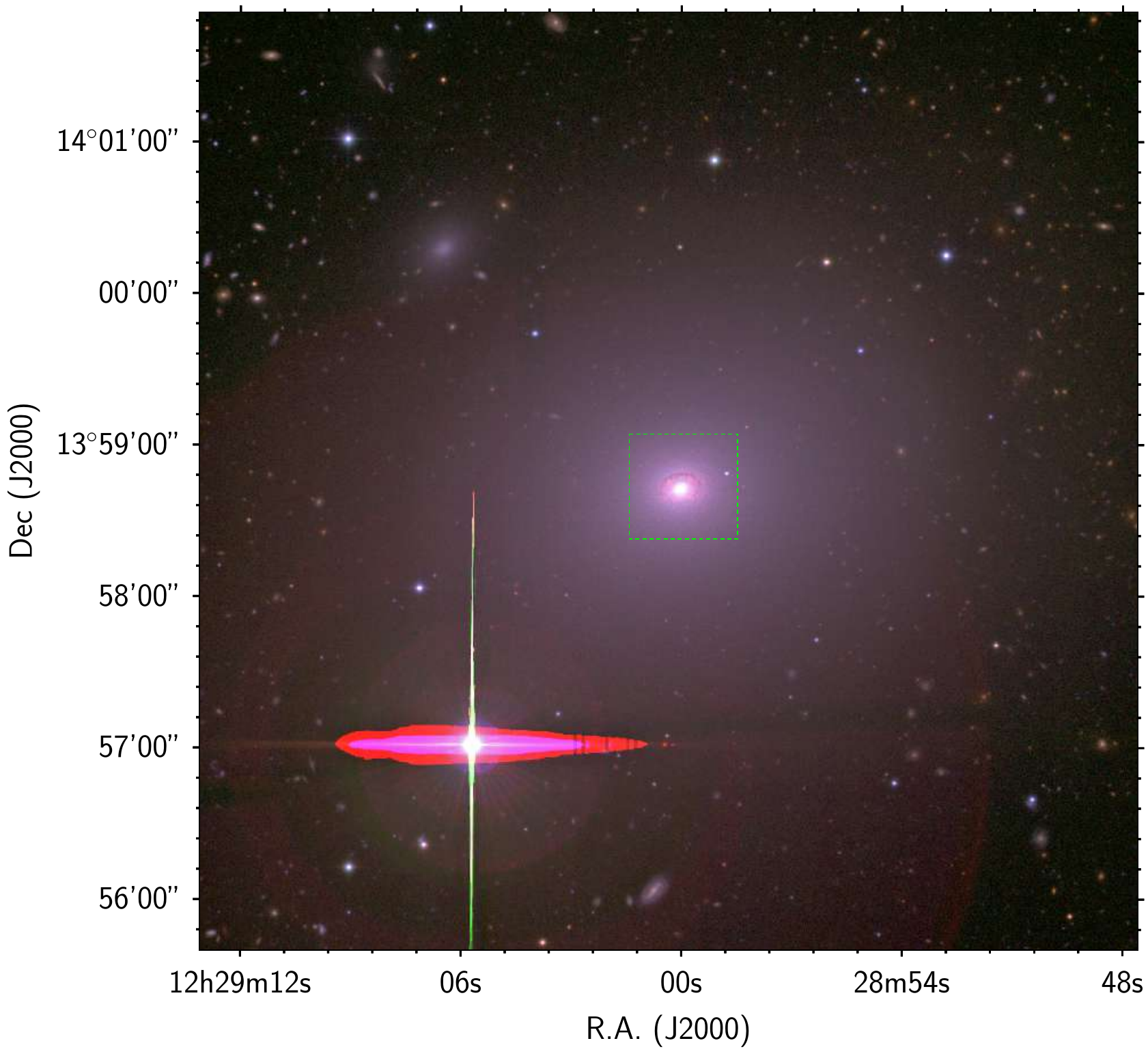}
   \includegraphics[width=0.39\textwidth]{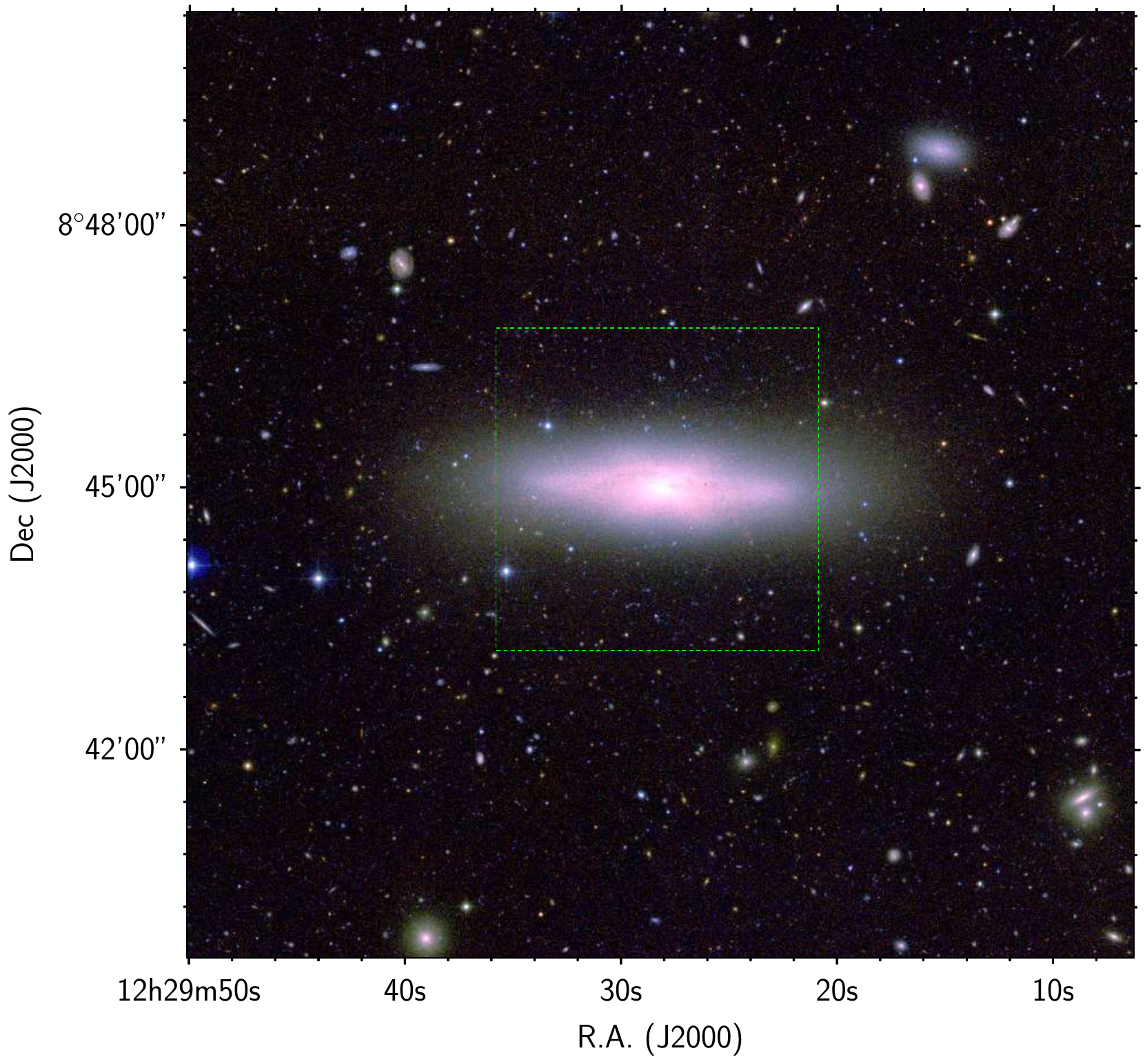}\\
   \includegraphics[width=0.39\textwidth]{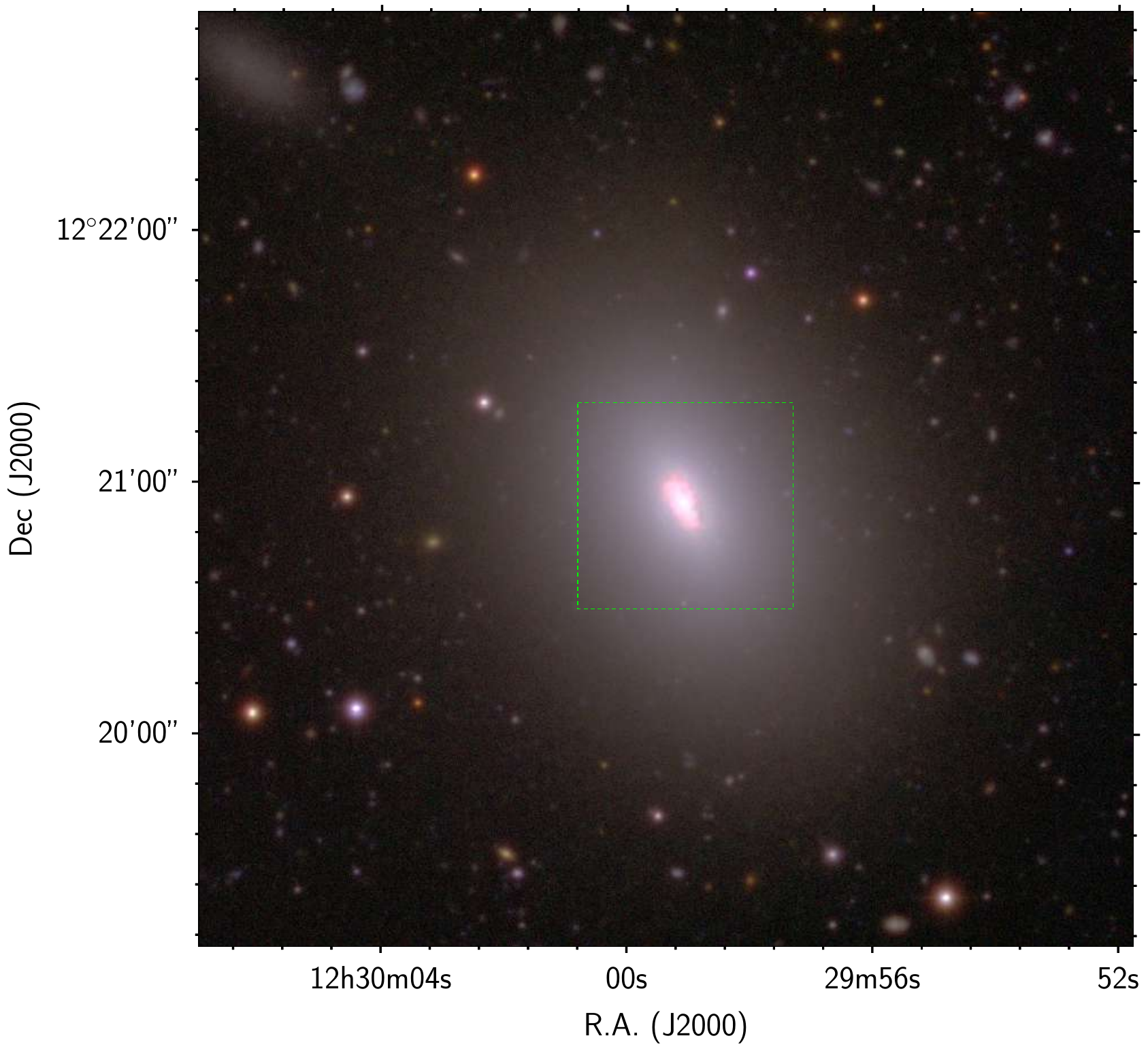}
   \includegraphics[width=0.39\textwidth]{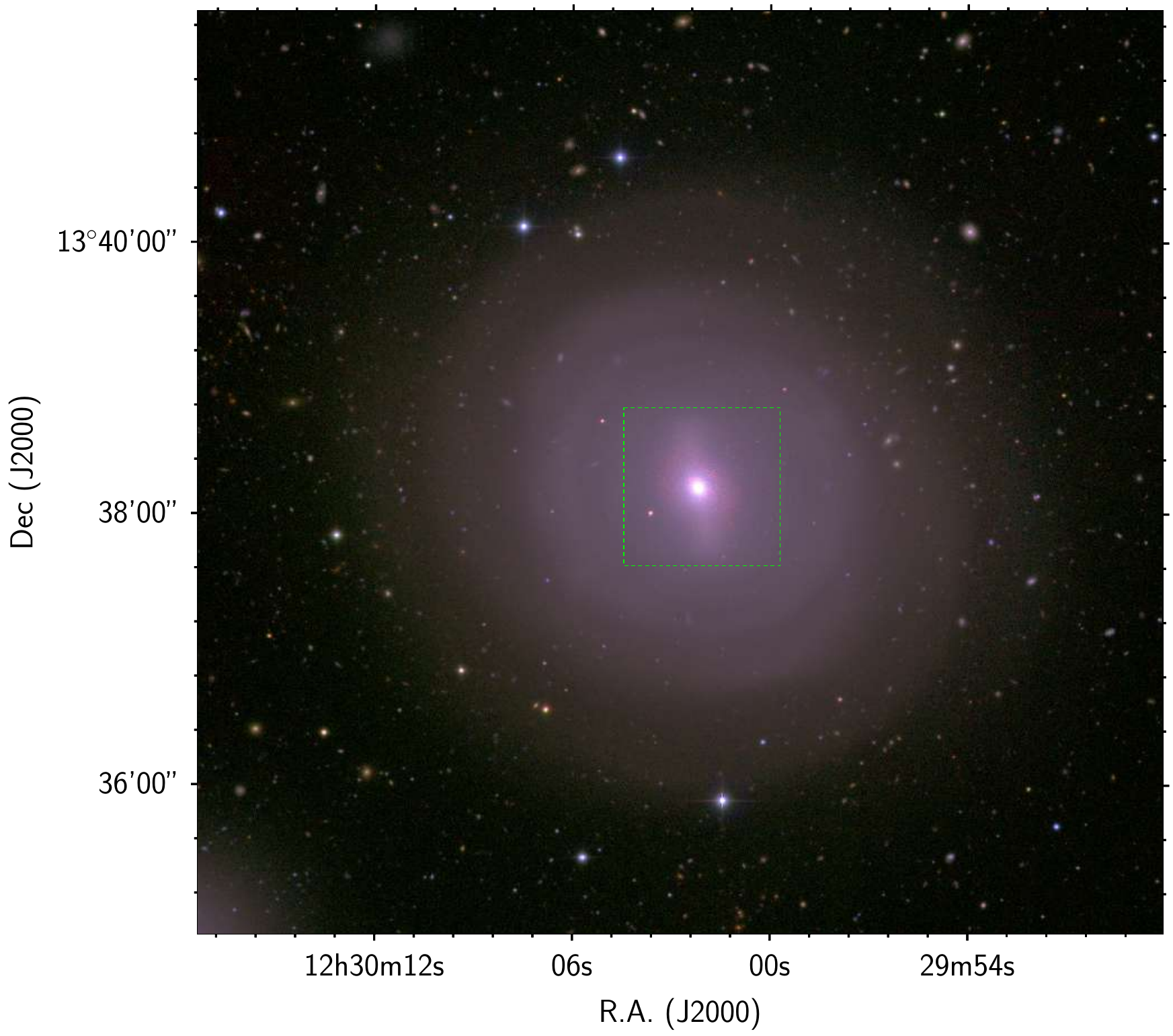}\\
   \includegraphics[width=0.39\textwidth]{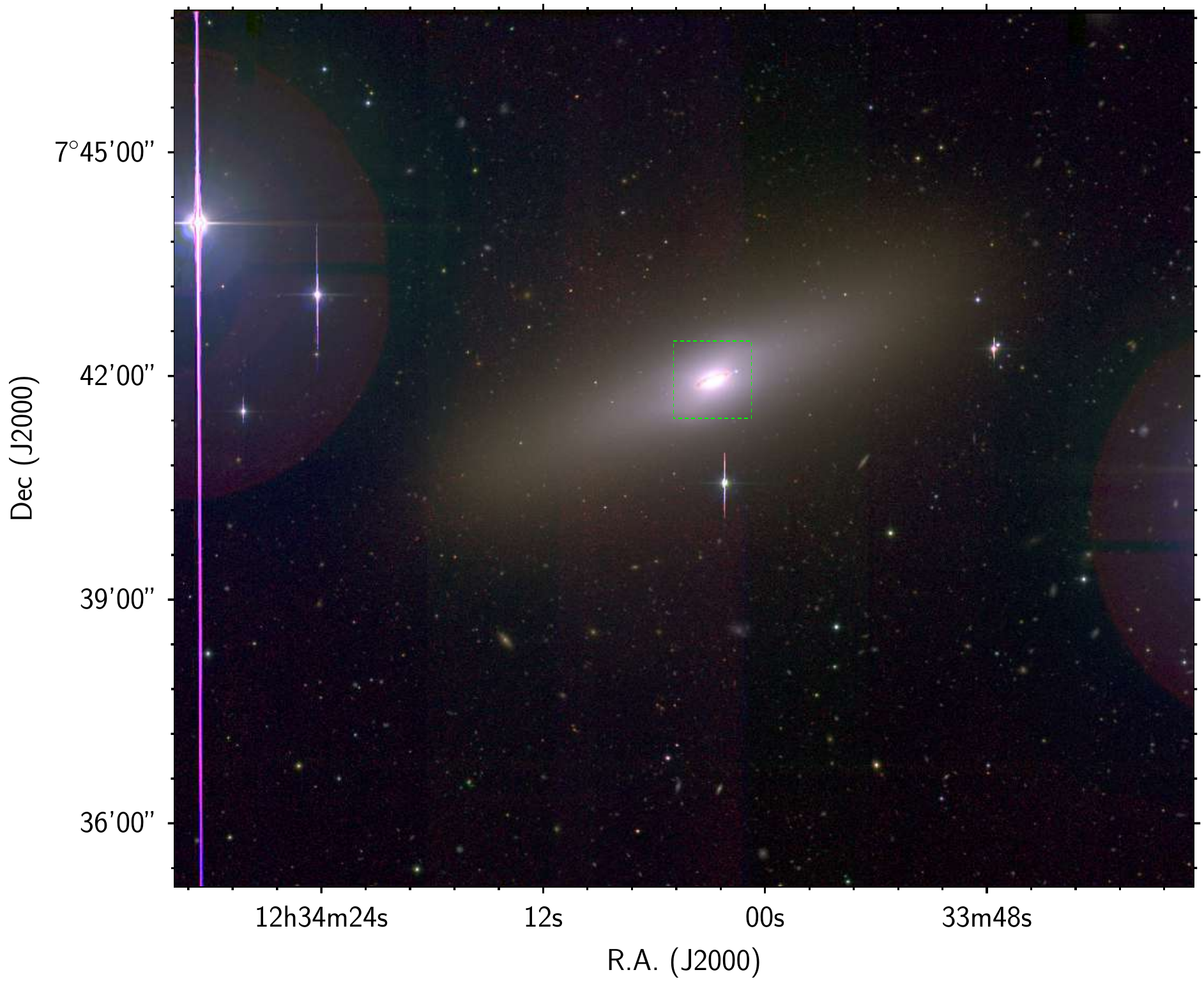}
   \includegraphics[width=0.44\textwidth]{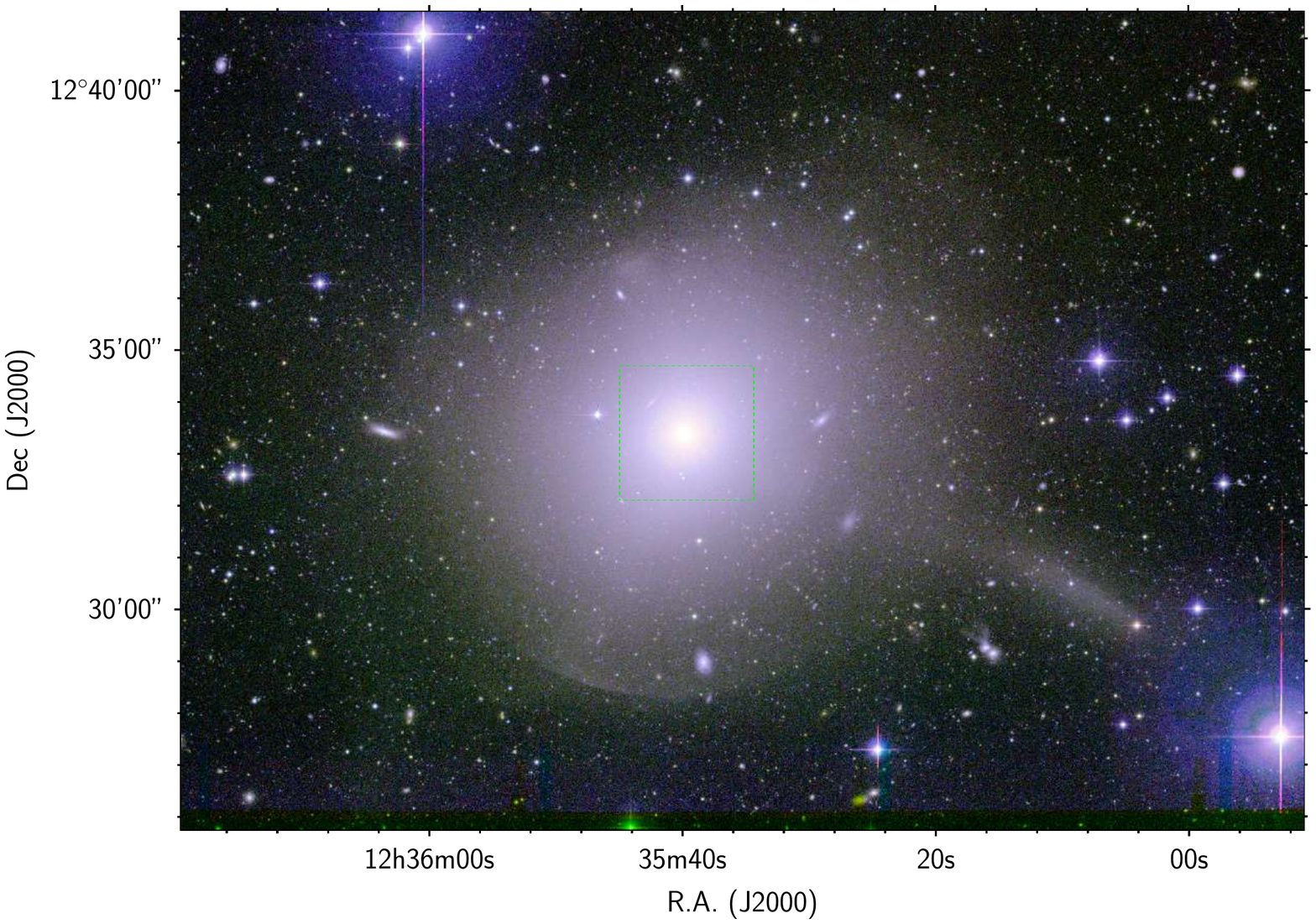}\\  
   \caption{Pseudo-colour images of the galaxies NGC 4262 (first row, left), 4429 (first row, right), 4459 (second row, left), 
   4469 (second row, right), 4476 (third row, left), 4477 (third row, right), 4526 (last row, left), and 4552 (last row, right) obtained combining the NGVS (Ferrarese et al. 2012) optical $u$ and $g$ in the blue channel, the $r$ and NB in the green, and the $i$ and the 
   continuum-subtracted H$\alpha$ in the red. The green dashed insetindicates the footprint of the continuum-subtracted
   H$\alpha$ images shown in Fig. \ref{Ha_image}. At the assumed distance of the galaxies (16.5 Mpc), 1 arcmin = 4.8 kpc.
 }
   \label{colour_image}%
   \end{figure*}

\begin{figure*}
   \centering
   \includegraphics[width=0.8\textwidth]{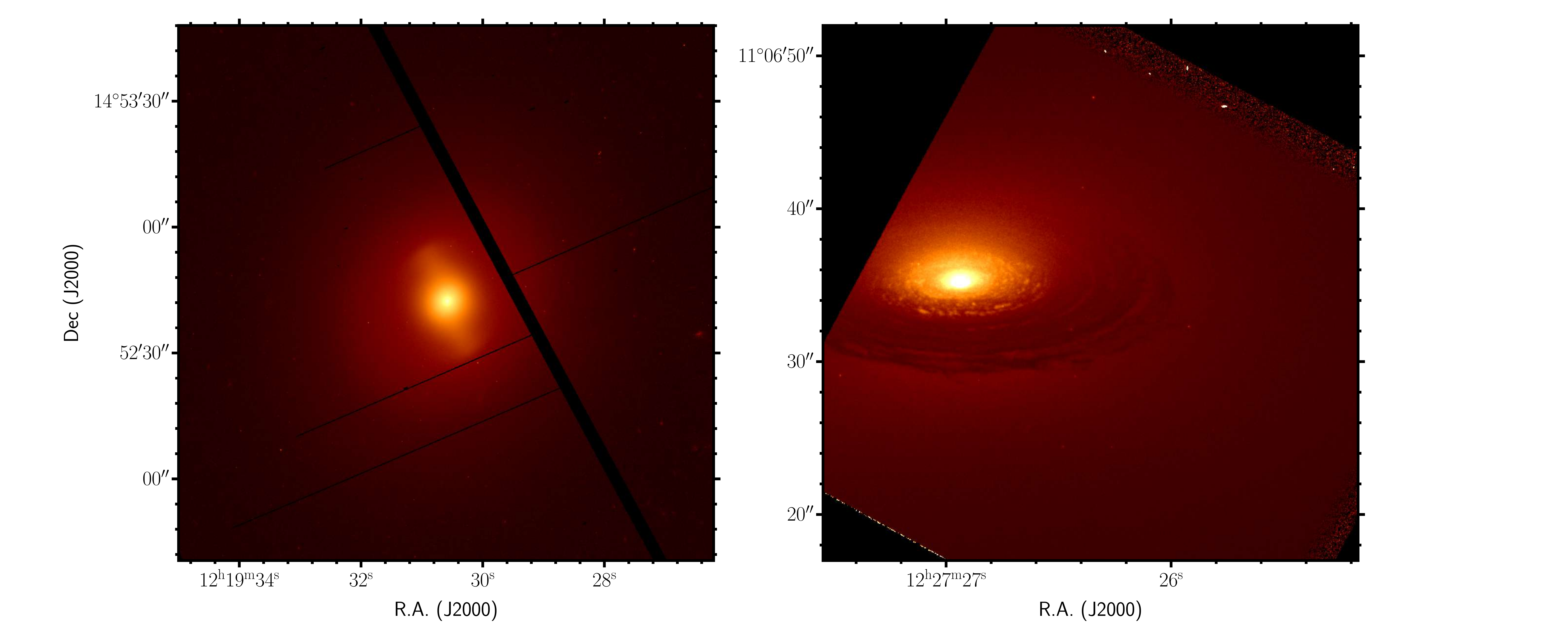}\\
   \includegraphics[width=0.8\textwidth]{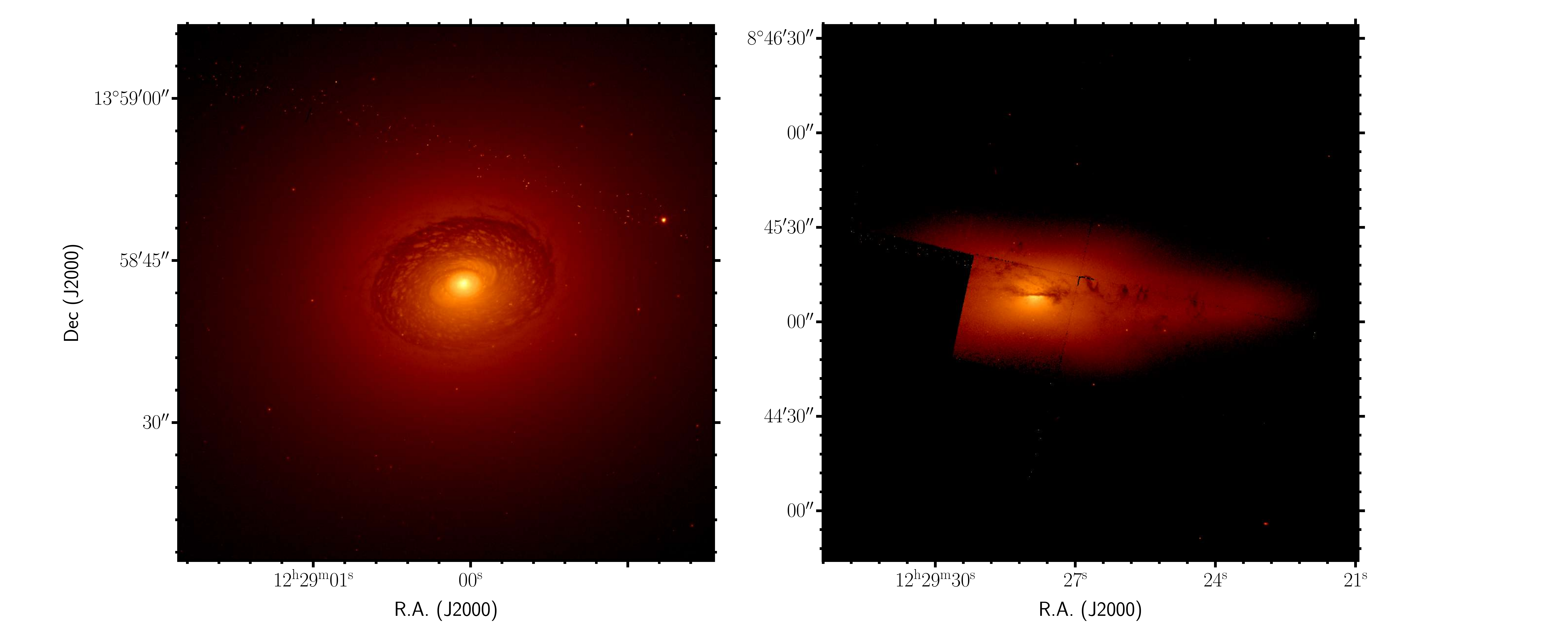}\\
   \includegraphics[width=0.8\textwidth]{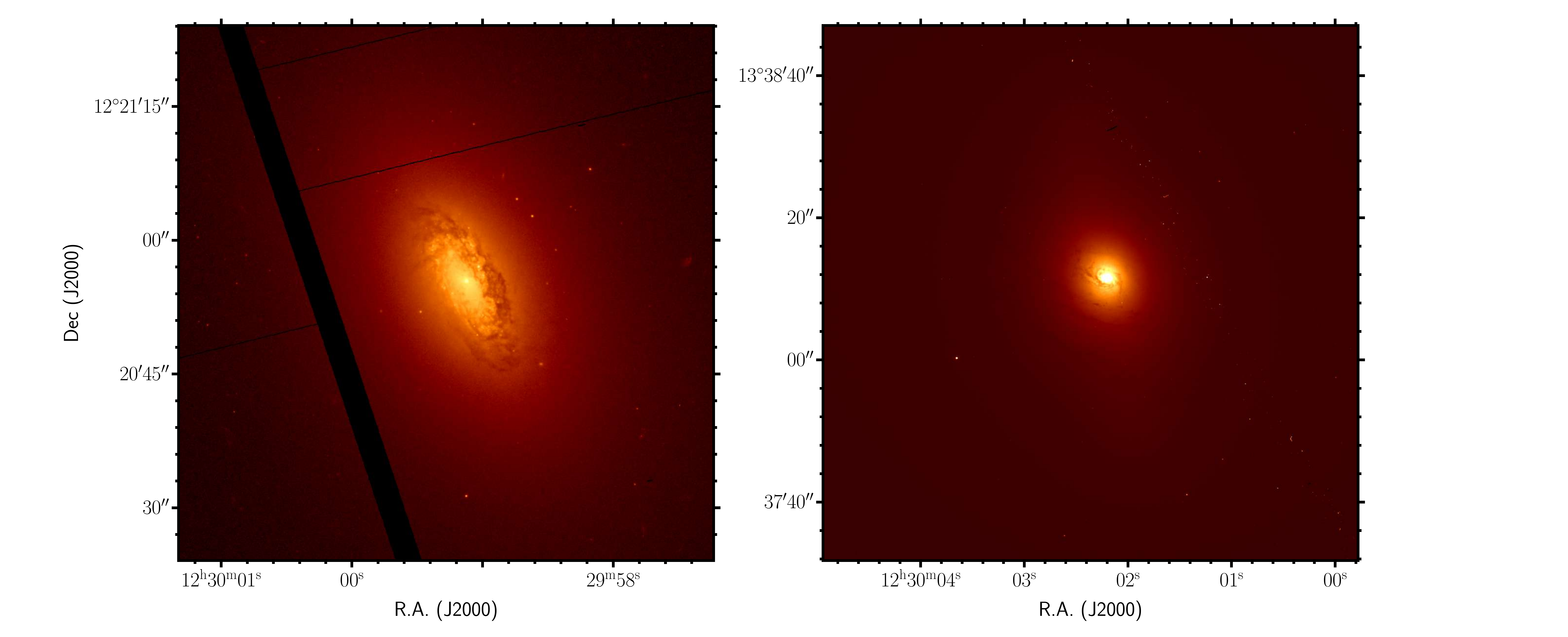}\\
   \includegraphics[width=0.8\textwidth]{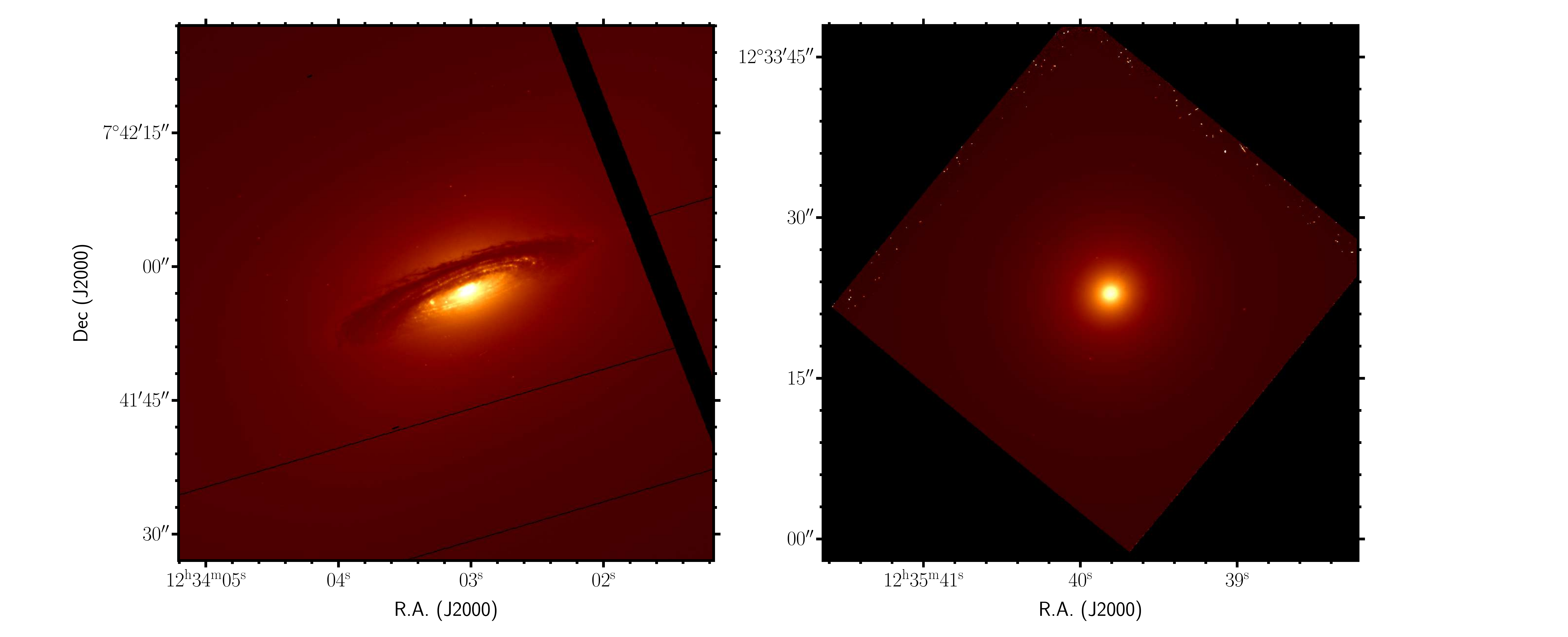}\\
   \caption{HST images of the galaxies NGC 4262 (F475W; first row, left), 4429 (F606W; first row, right), 4459 (F475W; second row, left), 
   4469 (F606W; second row, right), 4476 (F475W; third row, left), 4477 (F475W; third row, right), 4526 (F475W; last row, left), and 4552 (F555W;
   last row, right).
 }
   \label{HST}%
   \end{figure*}

The eight galaxies analysed in this work are all bright objects most of which were targets of previous dedicated studies. For these galaxies a large amount of multifrequency data
spanning the whole electromagnetic spectrum is available. 
They come from a number of untageted surveys of the cluster in the visible (NGVS, Ferrarese et al. 2012, see Fig. \ref{colour_image}), in the UV (GUViCS, Boselli et al. 2011), 
and in the far-infrared (HeViCS, Davies et al. 2010), 
or from dedicated observations, as summarised in Table \ref{data}. Data from targeted observations are also available: seven out of eight 
galaxies are included in the \textit{Herschel} Reference Survey (Boselli et al. 2010a), with data available on the HeDam database\footnote{https://hedam.lam.fr/HRS/}. 
Of particular interest for the present study are the 
HST images of the inner regions (see Fig. \ref{HST}) taken during the ACS Virgo cluster survey (C\^ot\'e et al. 2004) or in other targeted observations and available on the HST archive. 
Mid and far-IR data in the WISE (Wright et al. 2010) and MIPS/\textit{Spitzer} (Rieke et al. 2004) bands, necessary for an accurate 
dust attenuation correction and for the determination of the total dust content of the target galaxies, are available from Bendo et al. (2012a), Ciesla et al. (2014), and Boselli et al. (2014a).
Relevant for the present study are also the IFU spectroscopic data gathered by the ATLAS$^{3D}$ survey of Cappellari et al. (2011). Finally, $^{12}$CO(1-0) interferometric data obtained 
using the CARMA radiotelescope at an angular resolution significantly coarser than that reached by ALMA are available for five galaxies, as illustrated in Table \ref{ALMA}. These data
have been taken during the molecular gas survey of nearby early-type galaxies conducted by Alatalo et al. (2013). A few other objects have pointed observations 
taken at the IRAM radiotelescope (Combes et al. 2007). Six of these galaxies have been also observed at 21~cm with the WSRT during the HI follow-up of the ATLAS$^{3D}$ sample, 
but only one of them has been detected (Serra et al. 2012). Finally, X-ray data from \textit{Chandra} or XMM are available 
for all galaxies of the sample (see Table \ref{data}). Despite these X-ray data are available, they are not always relevant for the present work. We refer to dedicated works
when X-ray data are of interest for the following analysis.

\section{Physical parameters}

\subsection{Physical properties of the ionised gas}

\subsubsection{H$\alpha$ emission}

\begin{figure*}
   \centering
   \includegraphics[width=0.4\textwidth]{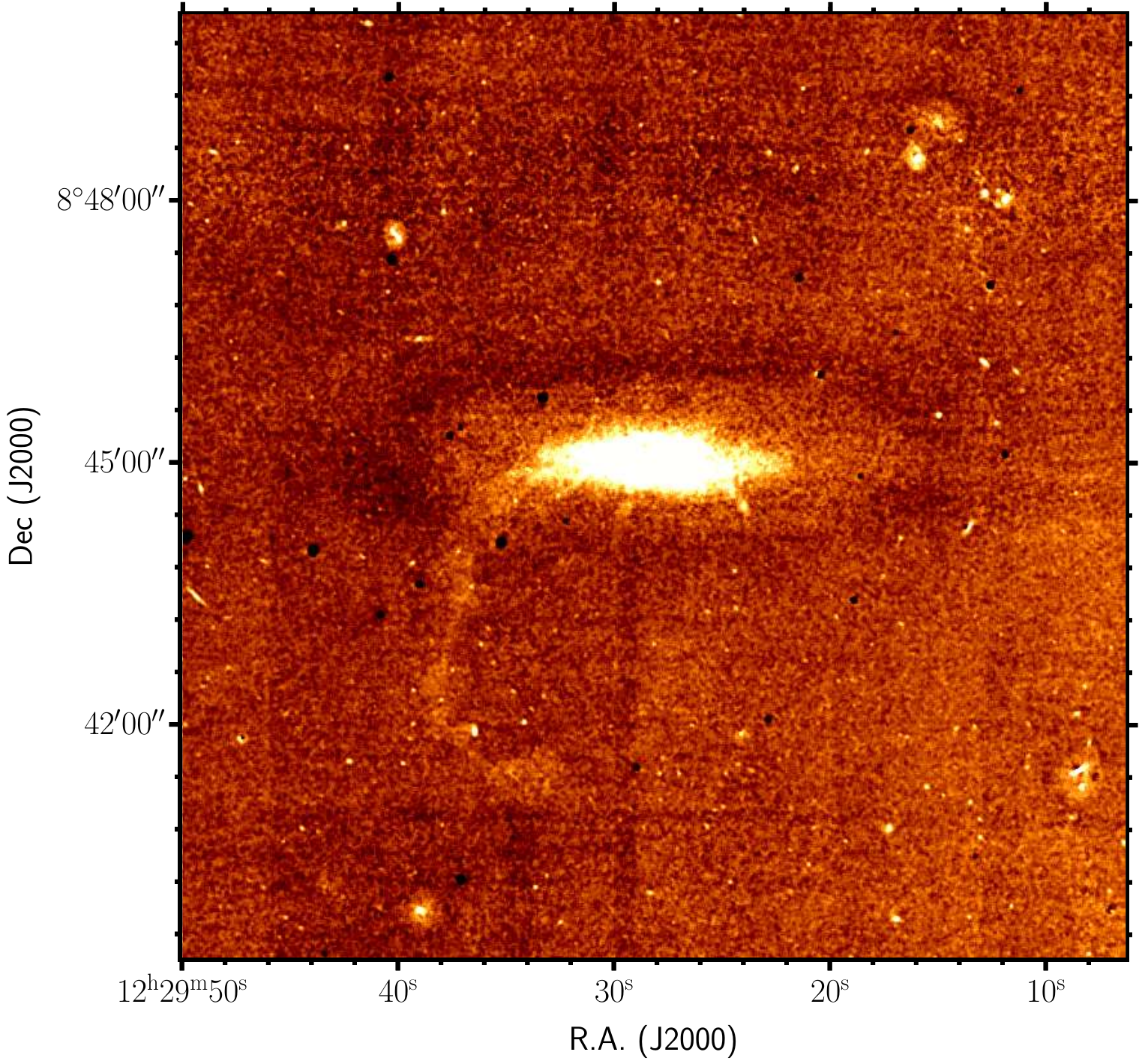}
   \includegraphics[width=0.52\textwidth]{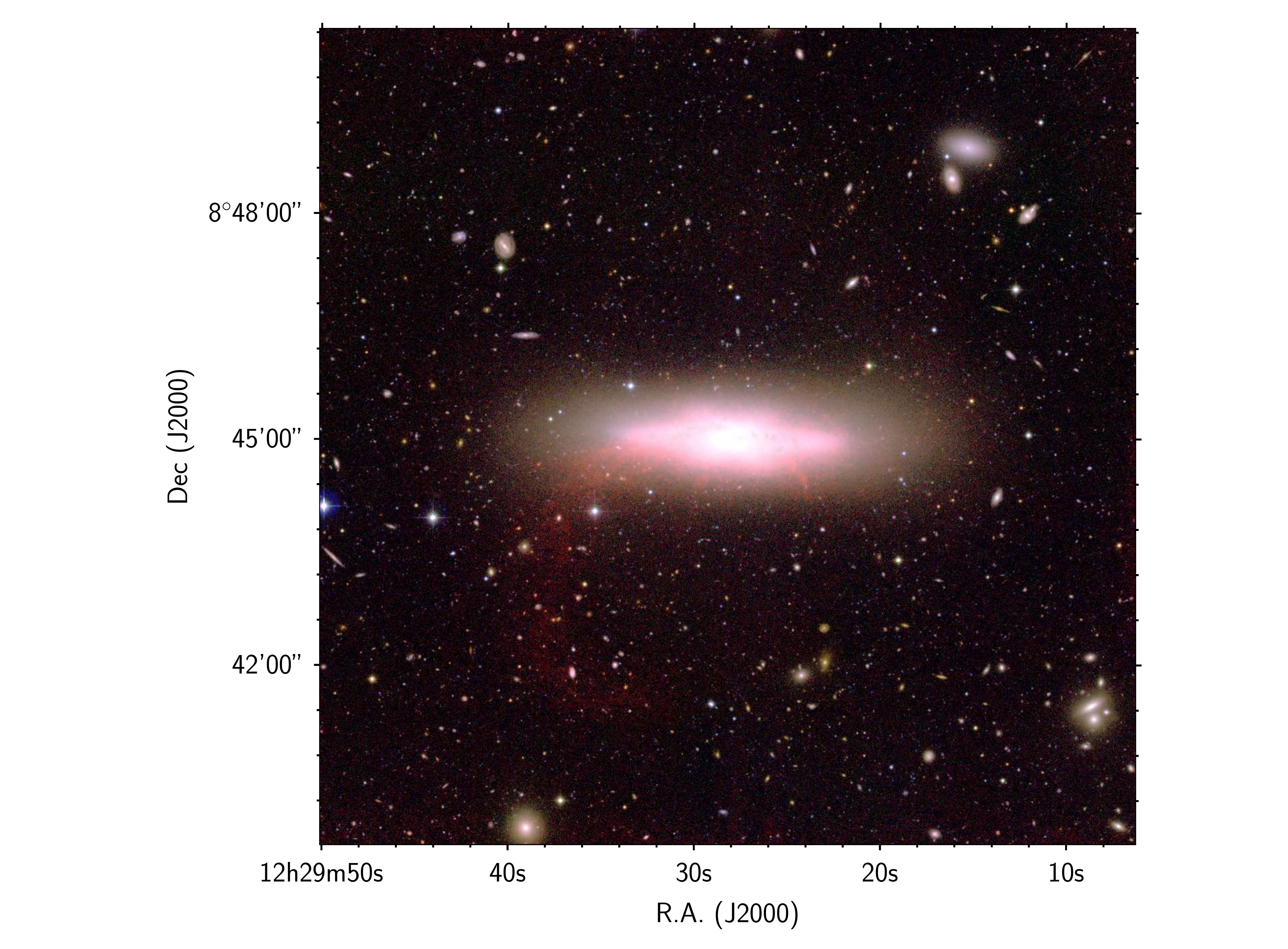}\\
  \caption{Left: continuum-subtracted H$\alpha$ image of the galaxy NGC 4469 smoothed to a resolution of 2.8\arcsec. The colour scale is arbitrarily chosen to 
  highlight the low surface brightness tail of ionised gas at the SE of the galaxy disc.
  Right: pseudo-colour image obtained combining the NGVS (Ferrarese et al. 2012) optical $u$ and $g$ in the blue 
  channel, the $r$ and NB in the green chanel, and the $i$ and the 
   continuum-subtracted H$\alpha$ image (this last smoothed to a resolution of 2.8\arcsec) in the red chanel. 
 }
   \label{N4469Hasm}%
   \end{figure*}

Thanks to the excellent quality of the VESTIGE data we can use 
the H$\alpha$ emission and distribution to derive physical quantities useful for the study of these intriguing objects. 
Figure \ref{Ha_image} shows the H$\alpha$ continuum-subtracted images of the eight lenticular galaxies analysed in this work.
Five of them (NGC`4429, 4459, 4476, 4477, 4526) have an ionised gas disc located in the very inner region, with a size 
(0.7 $\lesssim$ $R(H\alpha)$ $\lesssim$ 2.0 kpc) significantly smaller than that of the stellar component (Pogge \& Eskridge 1993). Their typical radial
extent is $\simeq$ 7-22\%\ that of the stellar disc (isophotal radius in the $i$-band as measured from the NGVS images)
and their typical surface brightness is 2.4 $\times$ 10$^{-16}$ $\lesssim$ $\Sigma(H\alpha)$ $\lesssim$ 6.8 $\times$ 10$^{-16}$ erg s$^{-1}$ cm$^{-2}$ arcsec$^{-2}$,
thus comparable to that of star-forming discs in late-type systems.
All images show the presence of resolved, compact regions from the centre to the periphery of the discs.
Tight, low surface brightness spiral arms are also visible in NGC 4477. The galaxies NGC~4262 and NGC~4552 
show rather a more diffuse and filamentary distribution of the ionised gas similar to that observed in cooling flows associated
to bright elliptical galaxies in the centre of rich clusters (Trinchieri \& Di Serego Alighieri 1991; Conselice et al. 2001; McDonald et al. 2010;
Gavazzi et al. 2000; Boselli et al. 2019).
These filaments have a relatively low surface brightness ($\Sigma(H\alpha)$ $\lesssim$ 10$^{-16}$ erg~s$^{-1}$~cm$^{-2}$~arcsec$^{-2}$), 
but do not extend outside the stellar component. Two unfolded, low surface brightness spiral arms and a few compact HII regions located along 
the ring structure seen in UV by Bettoni et al. (2010) on the western side of the galaxy are also present in NGC~4262.
The remaining boxy-shaped edge-on galaxy NGC~4469 has intermediate properties, with a very thin ($\simeq$ 100 pc) well defined circumnuclear disc of size $\simeq$ 400 pc in radius,
and with filamentary structures extending up to $\simeq$ 1 kpc from the plane of the galaxy which is seen edge-on. 
The most extended of them are coming out at the western and eastern edges of the stellar disc towards the southern 
direction (see Fig. \ref{N4469Hasm}). The eastern one, which has a typical surface brightness of 
$\Sigma(H\alpha)$ $\simeq$ 2 $\times$ 10$^{-18}$ erg s$^{-1}$ cm$^{-2}$ arcsec$^{-2}$ and is thus at the detection limit of the survey, extends out
to $\simeq$ 25 kpc in projected distance.

   \begin{figure*}
   \centering
   \includegraphics[width=0.48\textwidth]{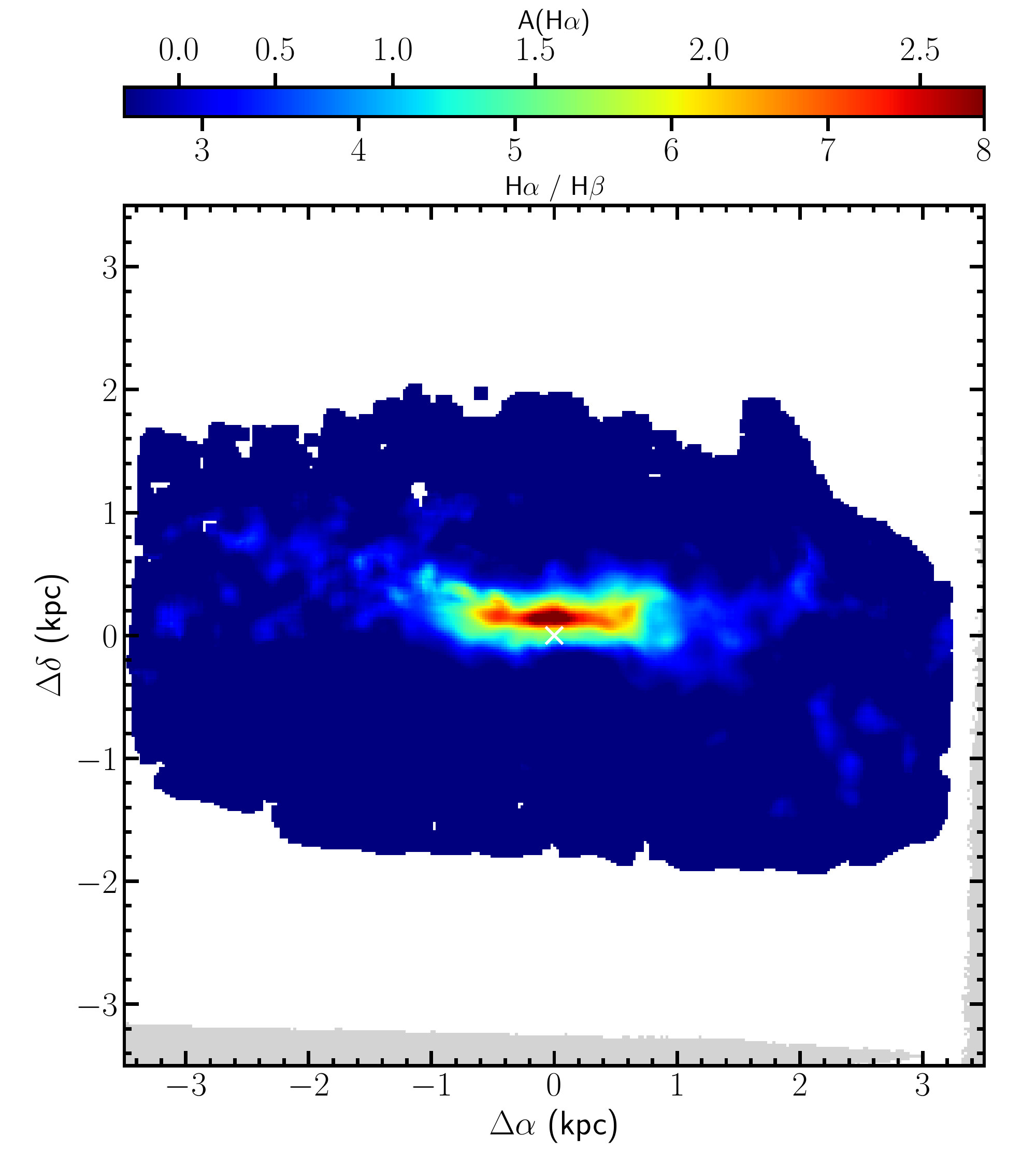}
   \includegraphics[width=0.48\textwidth]{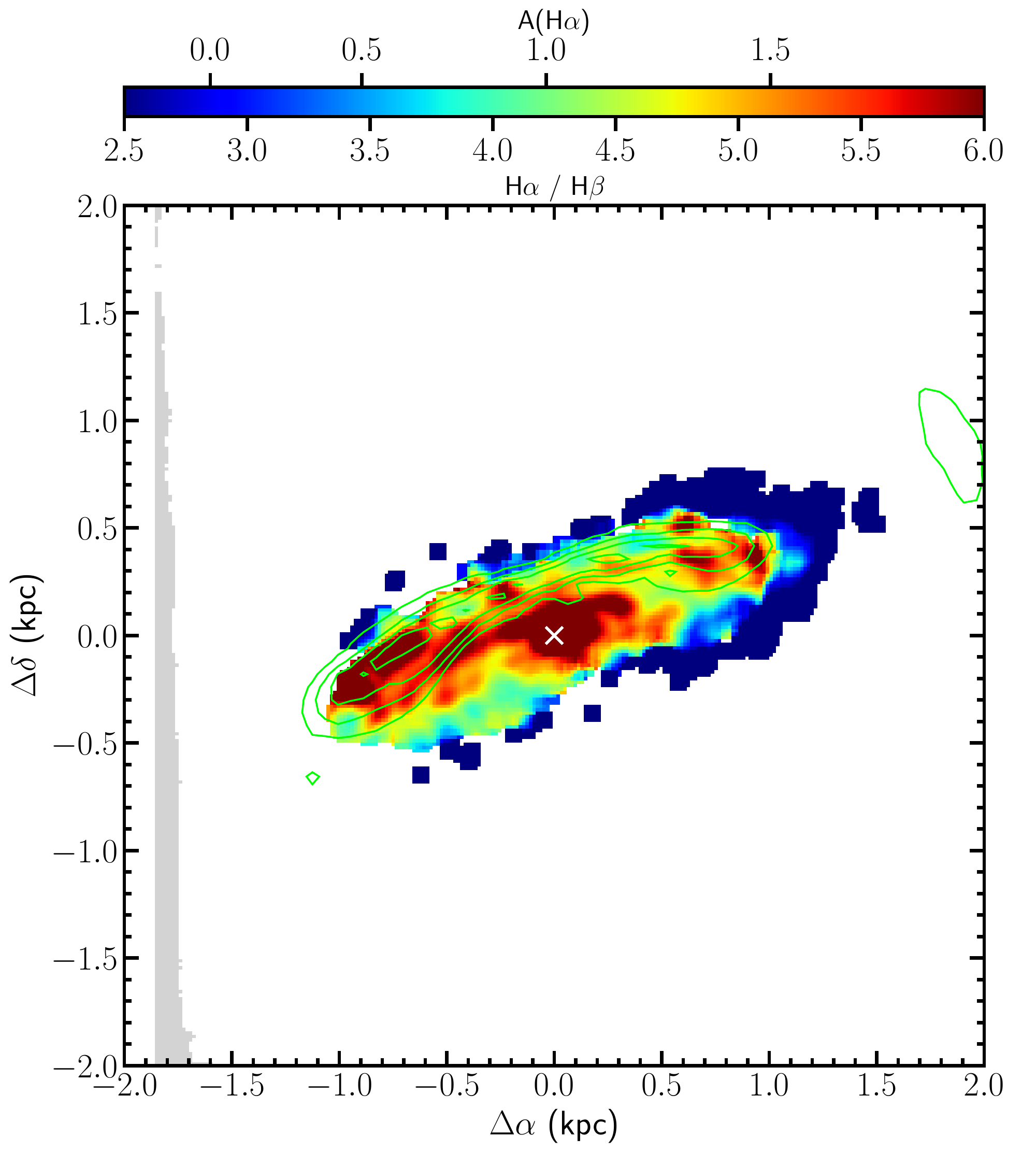}
   \caption{Distribution of the Balmer decrement derived using the H$\beta$ and H$\alpha$ lines with a $S/N$ $>$5 extracted from the MUSE data 
   in the galaxies NGC 4469 (left) and NGC 4526 (right). 
   The white and black crosses show the position of the photometric centre. The green contours in NGC 4526 indicate $A_g$ = 0.1, 0.3, 0.5, and 0.7 mag attenuation
   derived from the $g$-band NGVS image as described in Sect. 4.4. 
   }
   \label{BMD}%
   \end{figure*}

   \begin{figure*}
   \centering
   \includegraphics[width=0.48\textwidth]{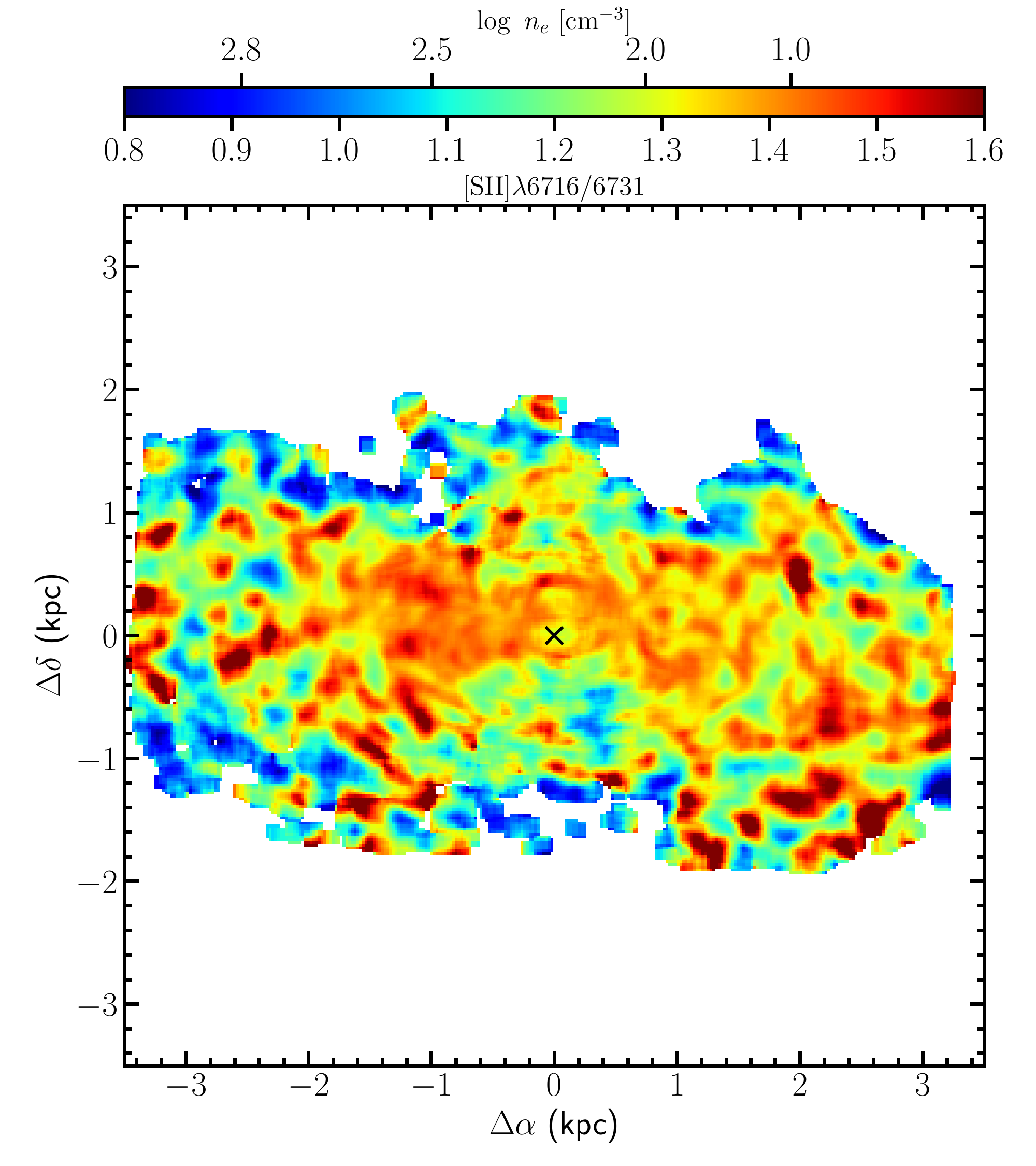}
   \includegraphics[width=0.48\textwidth]{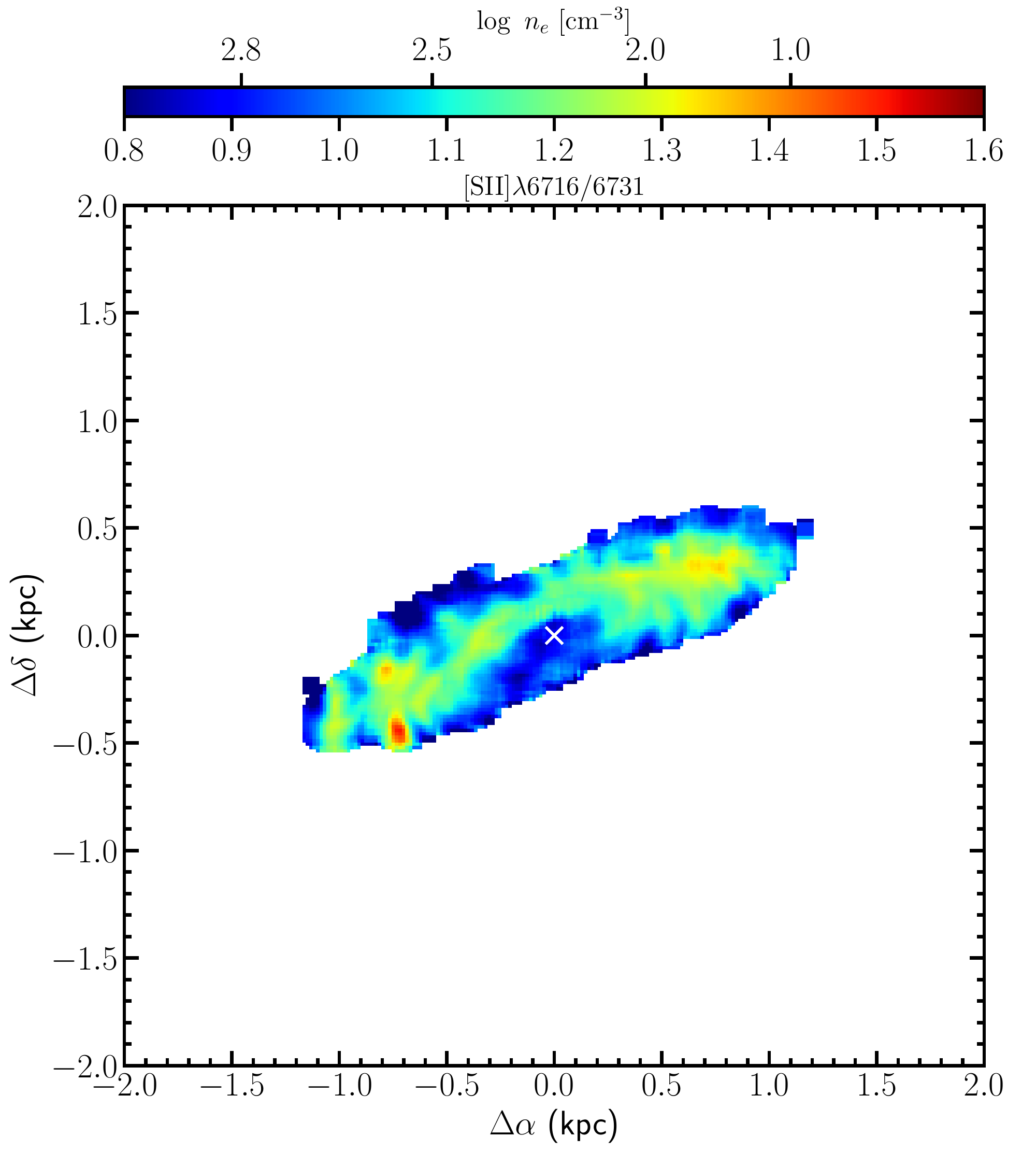}
     \caption{Distribution of the [SII]$\lambda$6716\AA/[SII]$\lambda$6731\AA\ ratio for emission lines with a $S/N$ $>$5
     extracted from the MUSE data of the galaxies NGC 4469 (left) and NGC 4526 (right). 
   The black and white crosses show the position of the photometric centre.
   }
   \label{SII}%
   \end{figure*}

\subsubsection{Balmer decrement and gas density}

Spectroscopic data can be used to derive other physical properties of the ionised gas. Figure \ref{BMD} shows the distribution of the Balmer decrement
in the two galaxies NGC 4469 and NGC 4526, which reaches values up to H$\alpha$/H$\beta$ $\simeq$ 6-8 in the inner discs, corresponding to $A(H\alpha)$ $\simeq$ 2-2.5 mag.
These dust features seen in the Balmer decrement correspond fairly well to those observed in absorption in the VESTIGE NB images.
Comparable values ($A(H\alpha)$ $\simeq$ 2.5 mag) are derived using the SOAR data in NGC 4429 and NGC 4476, although these estimates are very uncertain
because of the very low S/N in the H$\beta$ line.
Figure \ref{SII} shows
the distribution of the [SII]$\lambda$6716\AA/[SII]$\lambda$6731\AA\ in the ionised gas disc of NGC 4469 and NGC 4526 derived from the MUSE 
data. Figure \ref{SII} shows that in NGC 4469 the doublet ratio is [SII]$\lambda$6716\AA/[SII]$\lambda$6731\AA\ $\simeq$ 1.2-1.4 (corresponding to
$n_e$ $\simeq$ 50-200 cm$^{-3}$, Osterbrock \& Ferland 2006; Proxauf et al. 2014), as generally observed in late-type systems.
In NGC 4526, [SII]$\lambda$6716\AA/[SII]$\lambda$6731\AA\ $\simeq$ 1.2-1.3 (corresponding to
$n_e$ $\simeq$ 100-200 cm$^{-3}$) along the dusty 
structure crossing the ionised gas disc from the west to the east of the galaxy and located north from its nucleus, where 
most of the dust is seen in absorption (see Fig. \ref{HST}). On the rest of the disc the ratio drops to 
[SII]$\lambda$6716\AA/[SII]$\lambda$6731\AA\ $\simeq$ 0.9-1.0, corresponding to gas densities $n_e$ $\simeq$ 400-600 cm$^{-3}$.
These gas densities are significantly higher than those generally encountered in normal star-forming late-type galaxies
($n_e$ $\simeq$ 50 cm$^{-3}$, e.g. Rousseau-Nepton 2018), including perturbed objects within the Virgo cluster (Boselli et al. 2021b).
Similar values are derived for the discs of NGC 4429 ([SII]$\lambda$6716\AA/[SII]$\lambda$6731\AA\ $\simeq$ 1.1, corresponding to $n_e$ $\simeq$ 300 cm$^{-3}$)
and NGC 4476 ([SII]$\lambda$6716\AA/[SII]$\lambda$6731\AA\ $\simeq$ 1.4, corresponding to $n_e$ $\simeq$ 10-100 cm$^{-3}$) using the SOAR data.

\subsubsection{Oxygen abundance}

\begin{figure*}
   \centering
   \includegraphics[width=0.48\textwidth]{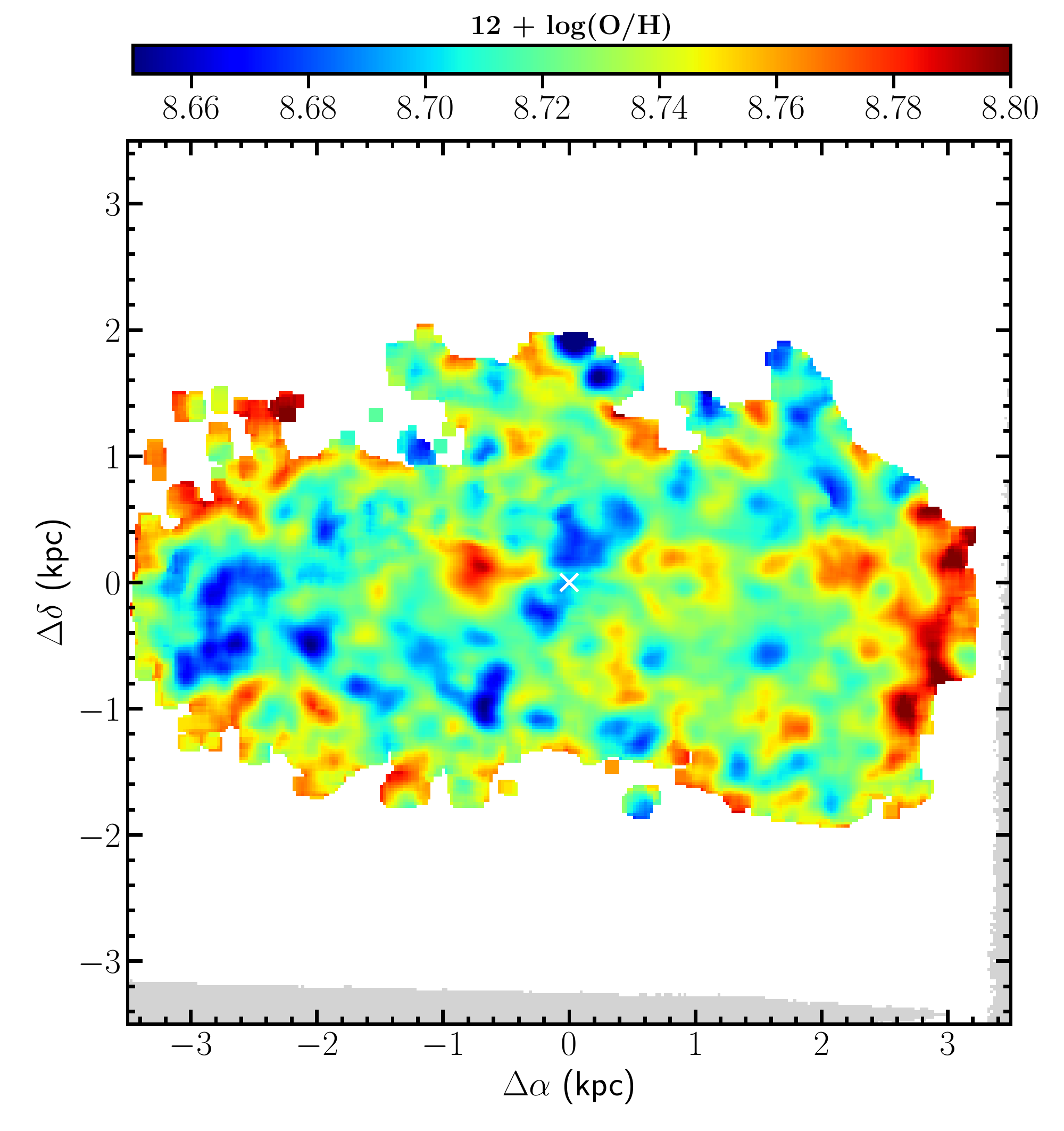}
   \includegraphics[width=0.48\textwidth]{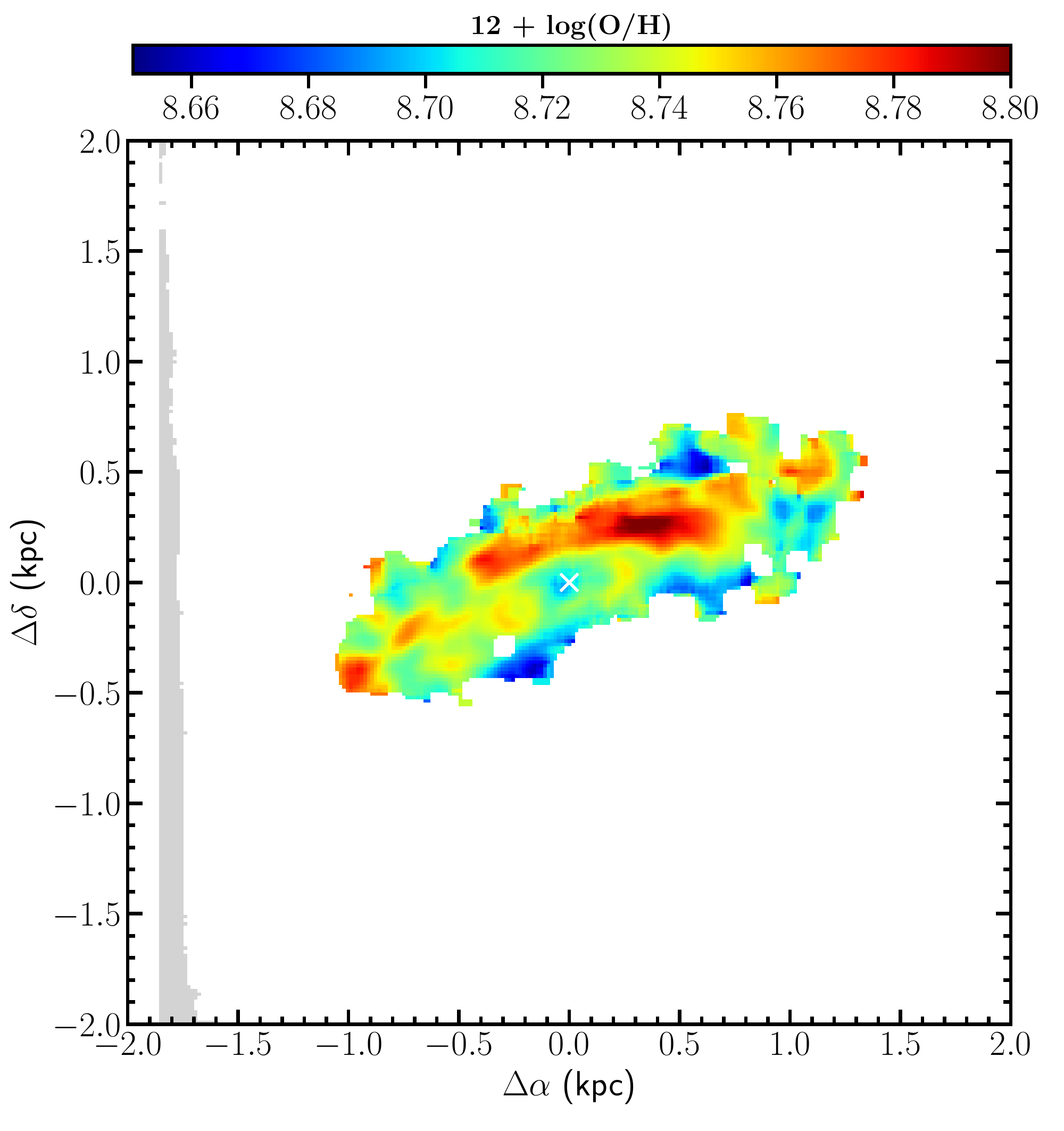}
   \caption{Distribution of the metallicity derived using the calibration of Curti et al. (2017) using the H$\beta$, [OIII], [NII], and H$\alpha$ lines with a $S/N$ $>$5
   from the MUSE data for the galaxies NGC 4469 (left) and NGC 4526 (right). 
   The white cross shows the position of the photometric centre.
   }
   \label{metal}%
   \end{figure*}

The MUSE data can also be used to trace the oxygen abundance distribution within the ionised gas disc of NGC 4469 and NGC 4526 (Fig. \ref{metal}).
For this purpose, we use the calibration of Curti et al. (2017) based on the H$\beta$, [OIII], [NII], and H$\alpha$ lines with a $S/N$ $>$ 5. 
Figure \ref{metal} indicates that the ionised gas disc of NGC 4469 is metal-rich in particular at its edges, where it reaches 
12+log O/H $\simeq$ 8.75-8.80, slightly less in its diffuse component (12+log O/H $\simeq$ 8.70).
NGC 4526 is also metal-rich, with oxygen abundances 8.70 $\lesssim$ 12+log O/H $\lesssim$ 8.80,
with the higher values present in the dusty structure crossing the galaxy from the west to the east at the north of the nucleus, visible also in 
absorption on the HST image (Fig. \ref{HST}).

\subsection{Source of ionisation}

\begin{figure*}
   \centering
   \includegraphics[width=0.5\textwidth]{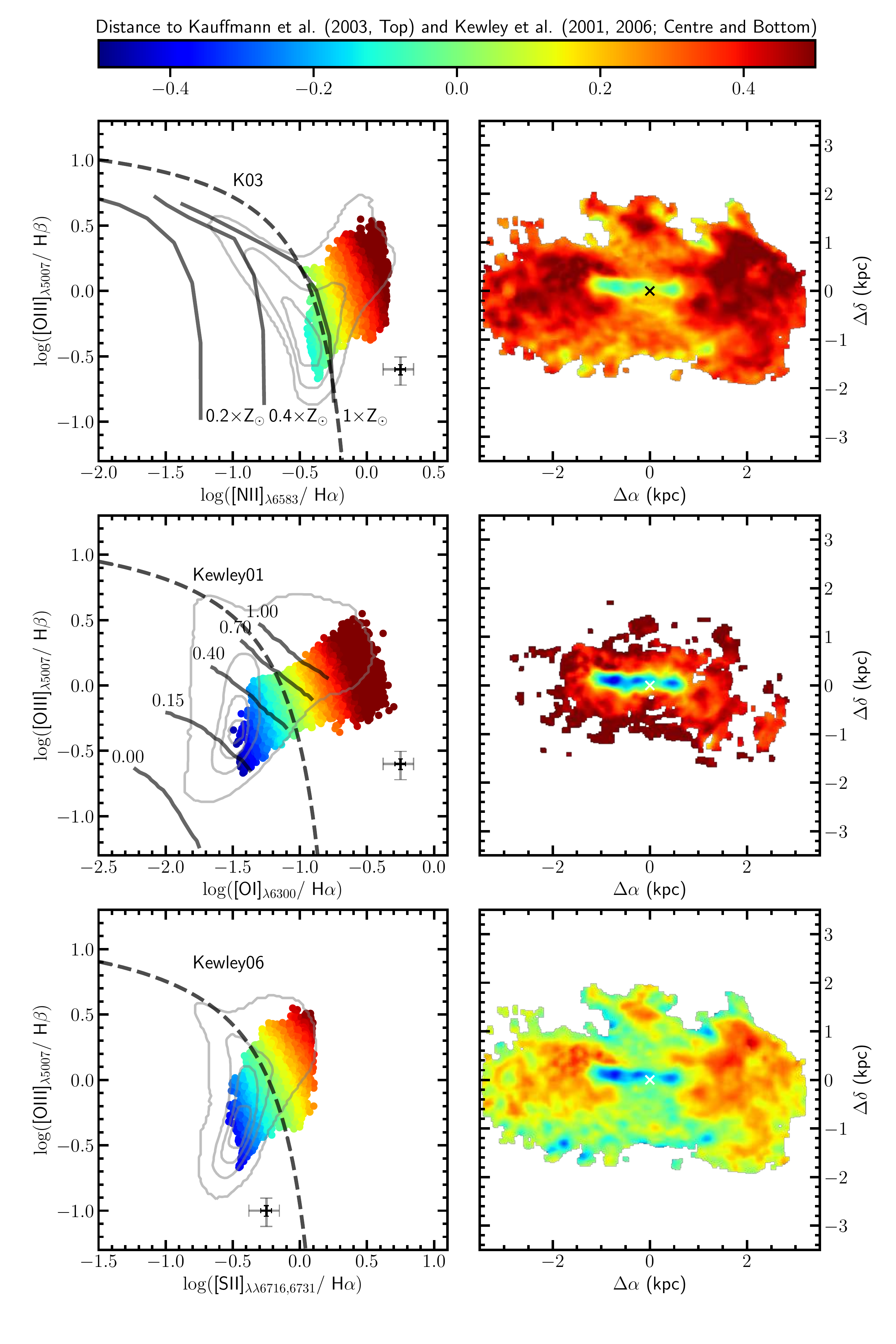}
   \caption{Left: [OIII]/H$\beta$ vs. [NII]/H$\alpha$ (top), [OIII]/H$\beta$ vs. [OI]/H$\alpha$ (centre), and [OIII]/H$\beta$ vs. [SII]/H$\alpha$ (bottom) BPT 
   diagrams for emission lines with a $S/N$ $>$5 in the galaxy NGC 4469. 
   The dashed curves separate AGN from HII regions (Kauffmann et al. 2003; Kewley et al. 2001, Kewley et al. 2006). 
   Data are colour coded according to their minimum distance 
   from these curves. The black and grey crosses indicate the typical error on the data for lines with S/N $\simeq$ 15 and S/N $\simeq$ 5, respectively.
   The grey contours show the distribution of a random sample of nuclear spectra of SDSS galaxies in the redshift range 0.01 - 0.1 and 
   stellar masses 10$^9$ $\leq$ $M_{star}$ $\leq$ 10$^{11}$ M$_{\odot}$.
   The thick solid lines in the upper left panel show three different photo-ionisation models at different metallicities (0.2, 0.4, 1 Z$_{\odot}$; Kewley et al. 2001),
   those in the middle left panel show the shock models of Rich et al. (2011) for increasing shock fractions (from left to right) in a twice-solar gas.
   Right: Map of the spaxel distribution colour-coded according to their position in the BPT diagram. The black and white crosses indicate the position of the photometric centre of the galaxy.
   }
   \label{BPT1}%
   \end{figure*}

\begin{figure*}
   \centering
   \includegraphics[width=0.5\textwidth]{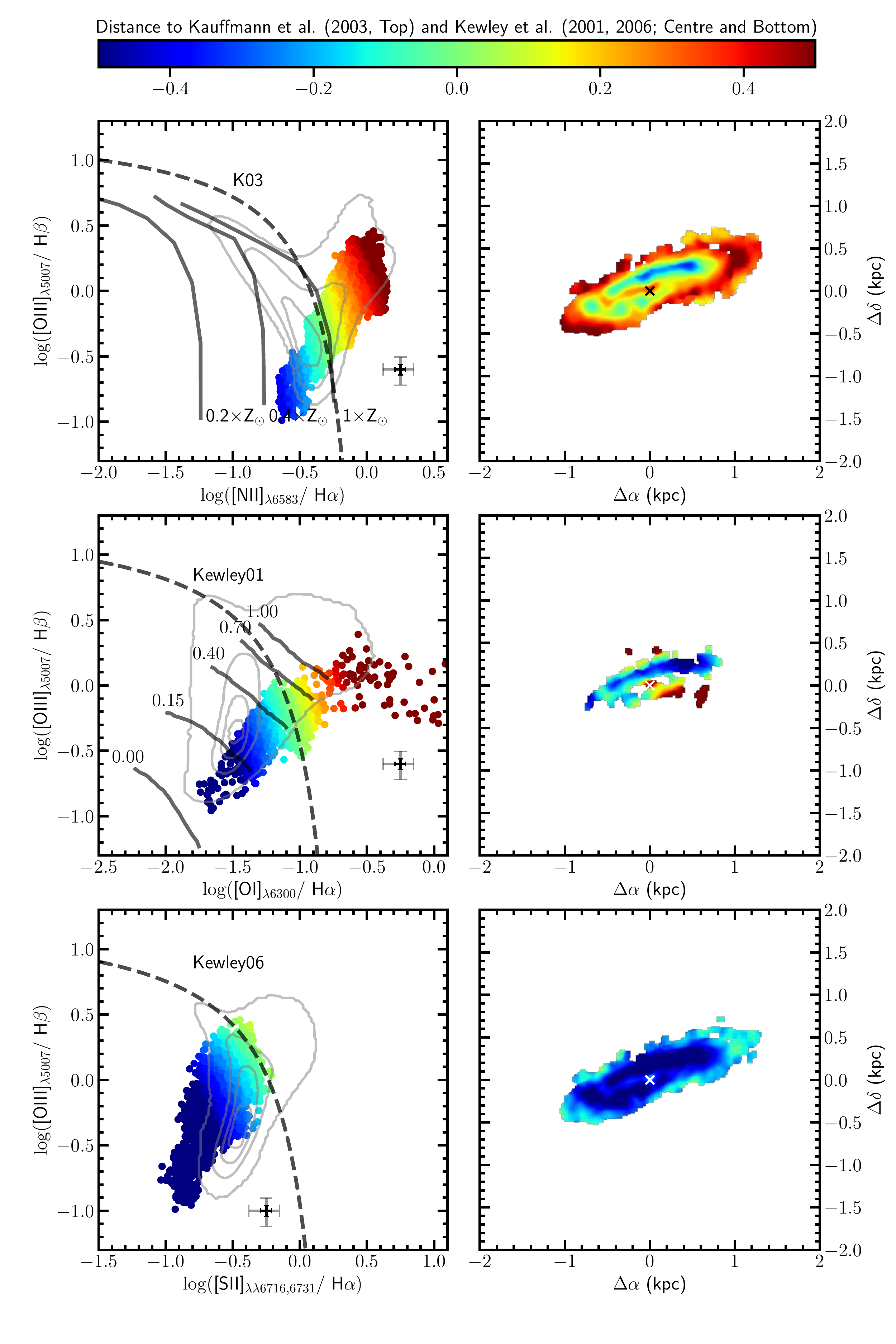}
   \caption{Same as Fig. \ref{BPT1} for the galaxy NGC 4526. 
   }
   \label{BPT2}%
   \end{figure*}

All sample galaxies are massive ($M_{star}$ $>$ 10$^{10}$ M$_{\odot}$) objects, most of which are characterised by a weak or strong nuclear 
activity (see Table \ref{gal}). Some of them have also an intense X-ray nuclear emission (NGC 4477, NGC 4526, NGC 4552; Kim et al. 2019).
It is thus possible that the observed gas, which is located in the innermost regions, 
is ionised by the nuclear activity. Being quiescent systems dominated by old stellar populations, the gas can also be ionsied by evolved stars (e.g. Sarzi et al. 2010). 
There is, however, convincing evidence that, at least in some of these objects, the 
H$\alpha$ emitting gas is photoionised by young stars and it is thus related to an ongoing activity of star formation.  
The first evidence comes from the spectroscopic data of NGC~4469 and NGC~4526 gathered with MUSE, which allow to construct 2D-diagnostic
diagrams (Fig. \ref{BPT1} and \ref{BPT2}).

Figure \ref{BPT1} clearly shows that in NGC~4469 the gas within the thin disc is photoionised by young stars, while the diffuse one 
surrounding the disc in the z-direction is excited by a high-energy source. Worth noticing is the fact that the line ratios consistently show
that all this diffuse gas lies below the Seyfert/LINER demarcation limit (Sharp \& Bland-Hawthorn 2010), indicating that
the gas is possibly shock-ionised. 
A similar, although less extreme, behaviour is observed in NGC 4526 (Fig. \ref{BPT2}),
with a possible exception in the nucleus.
AGN/shock-type features in the BPT diagrams have been observed in the periphery of the disc in a few  
star-forming galaxies undergoing an external perturbation induced by the hostile surrounding cluster environment (ram
pressure stripping, NGC~4424, Boselli et al. 2018c; CGCG 097-120, Fossati et al. 2019). 
We recall that, although the presence of an AGN is not indicated 
by spectroscopic data (Figs. \ref{BPT1} and \ref{BPT2}, see also Ho et al. 1997), NGC 4526 has a relatively strong X-rays point source in the nucleus which might
be associated to a hidden AGN (Kim et al. 2019). It is also worth mentioning that the BPT diagrams of NGC~4469 and NGC~4526 are different from those 
generally encountered in cooling flow galaxies with ionised gas filaments, which rather suggest ionisation by a combination 
of slow shocks and star formation (McDonald et al. 2012).

A complete and homogeneous set of IFU data is unfortunately unavailable for the other galaxies. The extremely uncertain measure of the highly attenuated [OIII]$\lambda$5007\AA\ line
prevents us to use the SOAR data to construct the same BPT diagrams for the galaxies NGC 4429 and NGC 4476. Their [NII]$\lambda$6583\AA/H$\alpha$ ($\simeq$ 0.95 and 
$\simeq$ 0.50 for NGC 4429 and NGC 4476, respectively) and [SII]$\lambda\lambda$6716,6731\AA/H$\alpha$ ($\simeq$ 0.85 and 0.61) and the [OIII]$\lambda$5007\AA/H$\beta$
derived from the ATLAS$^{3D}$ IFU spectroscopic data (see below) indicate that their ionised gas discs
have properties similar to those encountered in NGC 4469 and NGC 4526.
Integrated spectra in the optical domain (3700-7000 \AA, $R$ $\sim$ 1000) 
obtained by drifting the slit of the spectrograph over the disc of the galaxies are available for the remaining objects from Gavazzi et al. (2004) and Boselli et al. (2013), 
but the sensitivity of the instrument and the small size of the telescope prevent the detection of the faint ionised gas. IFU data in the blue spectral domain (4800-5380 \AA)
are available from the ATLAS$^{3D}$ survey (Cappellari et al. 2011)\footnote{For an accurate analysis of the 
ionised gas properties of the full ATLAS$^{3D}$ sample the reader is referred to Sarzi et al. (2010).}. The spectral coverage is not sufficient to construct BPT diagrams, but it allows us to measure the mean 
[OIII]$\lambda$5007\AA/H$\beta$ ratio in the inner 33$\times$41 arcsec$^2$ of 7/8 target galaxies. The mean values of the ratio derived after excluding all 
pixels included in the central 1 arcsec radius circle, which might be contaminated by a possible AGN activity, are given in Table \ref{atlas3d}.
High values of [OIII]$\lambda$5007\AA/H$\beta$, which might suggest the presence of a high ionisation source, are encountered in 
NGC 4262 ([OIII]$\lambda$5007\AA/H$\beta$=2.16$\pm$0.83), and with less extreme values in NGC~4477 ([OIII]$\lambda$5007\AA/H$\beta$=1.51$\pm$0.62)
and NGC 4552 ([OIII]$\lambda$5007\AA/H$\beta$=1.42$\pm$0.71). For comparison, the ratio measured in the diffuse gas of NGC~4469 is [OIII]$\lambda$5007\AA/H$\beta$=1.19$\pm$0.38.

\begin{table}
\caption{[OIII]$\lambda$5007\AA/H$\beta$ ratios}
\label{atlas3d}
{
\[
\begin{tabular}{ccc}
\hline
\noalign{\smallskip}
\hline
NGC	        & [OIII]$\lambda$5007\AA/H$\beta$	& Ref.	\\
\hline
4262		& 2.16$\pm$0.83				& 1	\\
4429		& 0.77$\pm$0.39				& 1	\\
4459		& 0.76$\pm$0.37				& 1	\\
4469$^a$	& 0.77$\pm$0.28 			& 2	\\
4476		& 0.59$\pm$0.25				& 1	\\
4477		& 1.51$\pm$0.62				& 1	\\  
4526		& 0.96$\pm$0.57, 0.78$\pm$0.42		& 1,2	\\
4552		& 1.42$\pm$0.71				& 1	\\    
\noalign{\smallskip}	
\hline
\end{tabular}
\]
References: 1) ATLAS$^{3D}$, 2) MUSE (both excluding the data within the inner 1 arcsec radius circle).\\
Notes: $\pm$ gives the dispersion in the distribution. a) the ratio is measured within the disc, excluding the diffuse extraplanar filaments,
where the ratio is significantly higher: [OIII]$\lambda$5007\AA/H$\beta$ = 1.19$\pm$0.38.
}
\end{table}

Further evidence of the presence of a young stellar population in the ionised gas discs of the galaxies NGC 4429 and NGC 4476 comes from the
imaging data in the FUV band gathered with the UVIT instrument on board of the ASTROSAT space telescope (see Fig. \ref{UVIT4429} and \ref{UVIT4476}). These
images show a bright structured and clumpy emission superimposed over a diffuse low surface brightness and extended emission, 
the former coming from recently formed HII regions associated to the ionised gas disc, the latter from the diffuse distribution of evolved stars 
in the thick and extended bulge of these lenticular galaxies (UV upturn, O'Connell 1999, Boselli et al. 2005). The same structured
and clumpy emission of resolved HII regions is also observable in the VESTIGE images of NGC~4429, NGC~4459, NGC~4476, NGC~4477, and NGC~4526 (Fig. \ref{Ha_image}),
or in absorption because of the presence of dust in the high resolution HST images (Fig. \ref{HST}). In the remaining galaxies NGC~4262 and NGC~4552
the ionised gas has a filamentary structure similar to that observed in cooling flows. We remark, however, that in NGC~4262 there are a couple of clumpy regions 
of ionised gas on the western side of the galaxy, the brightest of which is located at R.A.(J2000) = 12h19m26.6s, Dec(J2000) = 14$^o$52\arcmin30\arcsec with counterparts in
the FUV ASTROSAT/UVIT image. These regions are also visible in the ring-like structure discovered by Bettoni et al. (2010) in the GALEX FUV image, and are 
certainly associated to star-forming regions. NGC 4262 also has a two-arm spiral structure in the ionised gas close to the nucleus (see Fig. \ref{Ha_image} and \ref{UVIT4262}),
again suggesting a rotating star-forming disc of ionised gas. In NGC 4469, HII regions cannot be resolved because of the edge-on orientation of the disc. The disc, 
however, is very thin ($\simeq$ 100 pc) and clumpy, as expected if composed of star-forming regions.

\subsection{Star formation activity}

VESTIGE H$\alpha$+[NII] fluxes can be converted into star formation rates once corrected for dust attenuation and [NII] contamination. 
This can be easily done using the MUSE data in NGC 4469 and NGC 4526. 
Unfortunately, for the other VESTIGE sources accurate IFU spectroscopic data as those provided by MUSE are not available. 
Given their massive nature (see Table \ref{gal}), we expect that all these galaxies are metal-rich objects with characteristics similar to those encountered in NGC 4469 and NGC 4526, 
where [NII]/H$\alpha$ = 0.78 (derived excluding the nucleus of NGC 4526 because of a possible contamination 
of a nuclear activity, see Fig. \ref{BPT2}). Similar values are also observed in the discs of NGC 4429 ([NII]/H$\alpha$ $\simeq$ 1.27) and NGC 4476 ([NII]/H$\alpha$ $\simeq$ 0.67) using the SOAR data. 
All these numbers are consistent with what is observed in star-forming objects of similar mass (Boselli et al. 2009). 
For those galaxies without any available spectroscopic data we thus assume [NII]/H$\alpha$ = 0.78.

H$\alpha$ fluxes can be corrected for dust attenuation using two different methods. The first one uses the Balmer decrement,
while the second one uses the recipe of Calzetti et al. (2010) based on the 24$\mu$m $\textit{Spitzer}$/MIPS data (see Boselli et al. 2015 for a comparison of the two methods).

\subsubsection{$A(H\alpha)$ from the Balmer decrement}

An accurate estimate of the Balmer decrement is available only for NGC 4469 and NGC 4526 thanks to the MUSE data. In NGC 4429 and NGC 4476 the Balmer decrement is 
highly uncertain because of the weak H$\beta$ emission. However, given their very similar spectroscopic properties, we assume for these two objects the mean Balmer decrement 
measured within the discs of NGC 4469 and NGC 4526 (H$\alpha$/H$\beta$ $\simeq$ 5). For the remaining galaxies, we estimate the mean Balmer decrement from the  
2D dust distribution derived using the broad band NGVS images as described in Sec. 4.4. Indeed, the comparison between the 2D distribution of the dust attenuation 
derived using the Balmer decrement and that estimated after subtracting a model light distribution to the $g$-band image of NGC 4526\footnote{This exercise 
can be done only in this galaxy since the $g$-band image of NGC 4469 is saturated in the centre, and has a pronounced boxy-shape (Mosenkov et al. 2020) hardly reproducible
with a 2D fitting model, thus affecting the determination of $A(g)$ in the star-forming disc and in the outer, low surface brightness regions.} 
has shown a fairly good agreement, with:

\begin{equation}
{A(H\alpha) = 3.01 \times A(g)}
\end{equation}

\noindent
corresponding to:

\begin{equation}
{E(B-V)_{BD} = 4.87 \times E(B-V)_{star}}
\end{equation}

\noindent
where $E(B-V)_{BD}$ is the attenuation of the gaseous component as derived from the Balmer decrement and $E(B-V)_{star}$ that of the stellar component
derived from the $g$-band image. This relation is
steeper than the Calzetti's law ($E(B-V)_{BD}$ = 2.27 $\times$ $E(B-V)_{star}$; Calzetti et al. 2000) as expected for a screen dust distribution
for the gaseous component and a slab distribution for the stellar component. 
An accurate estimate of $A(g)$ is possible for the galaxies NGC 4459 and 4477. 
Assuming that the properties of their dust discs are comparable to that observed in NGC 4526, 
we use their mean values of $A(g)$ derived within the disc\footnote{This value is derived on the 
disc region where dust is seen in absorption, which for geometrical effects fairly corresponds to half of the disc, and extrapolated to the rest of the disc.} combined with eq. 1
to estimate $A(H\alpha)$ and to correct the observed H$\alpha$ fluxes for dust attenuation. 

\subsubsection{$A(H\alpha)$ from the 24$\mu$m $\textit{Spitzer}$/MIPS emission}
 
The dust attenuation correction based on the 24$\mu$m $\textit{Spitzer}$/MIPS emission assumes that the energy of the ionising radiation
absorbed by dust is re-radiated in the mid-infrared. This correction is given by the relation (Calzetti et al. 2010):

\begin{equation}
{L(H\alpha)_{Cor} = L(H\alpha)_{Obs} + 0.020 \times L(24\mu m)}
\end{equation}

\noindent
which is valid for $L(24\mu m)$ $\leq$ 4 $\times$ 10$^{42}$ erg s$^{-1}$, the range covered by the S0 galaxies analysed in this work.
For this purpose, we have derived the 24 $\mu$m flux densities of the eight target galaxies using the photometric extraction presented 
in Fossati et al. (2018) optimised to integrate within a given aperture and estimate a statistically robust uncertainty. This procedure has been
run on the homogeneously reduced dataset of Bendo et al. (2012a).

  \begin{figure*}
   \centering
   \includegraphics[width=0.99\textwidth]{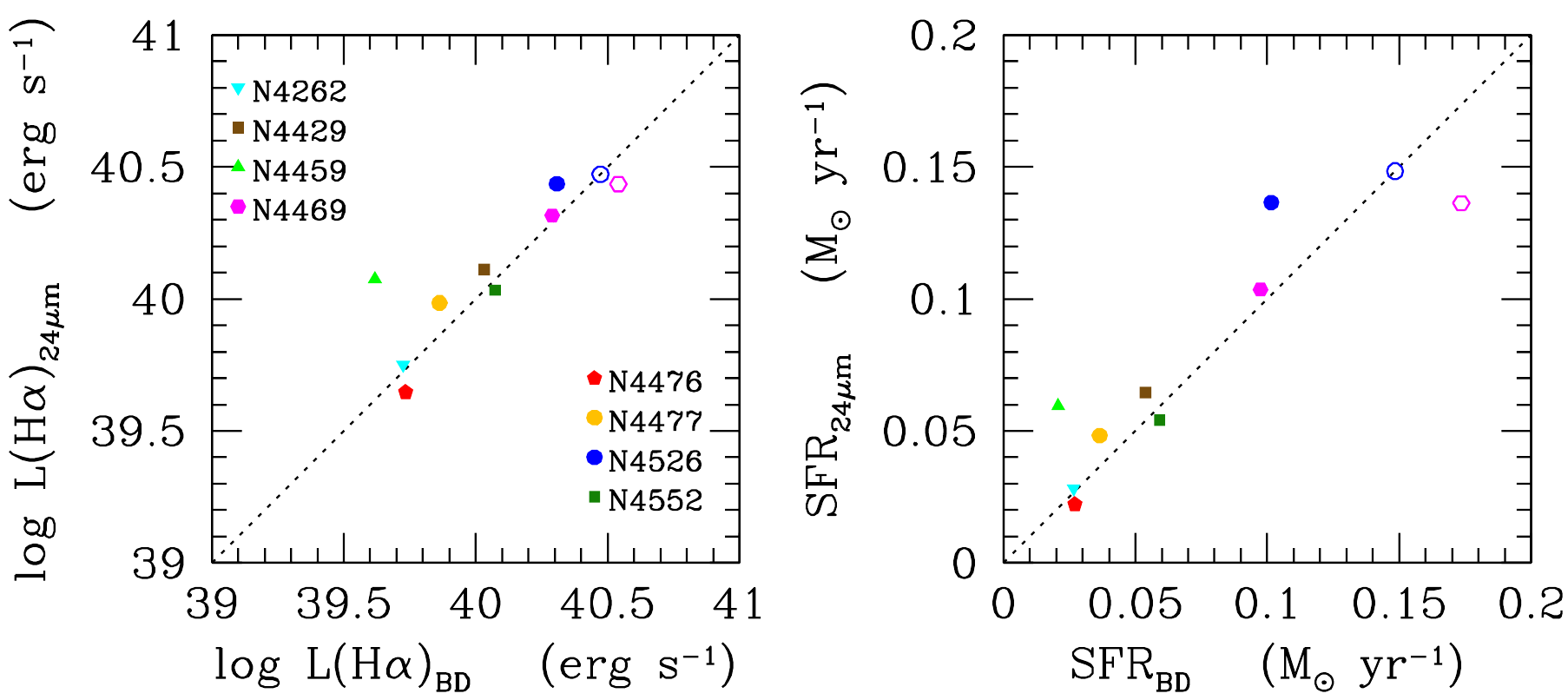}
   \caption{Left: comparison between the H$\alpha$ luminosity corrected for dust attenuation derived using the 24$\mu$m $\textit{Spitzer}$/MIPS
   emission ($L(H\alpha)_{24\mu m}$) and that derived using the Balmer decrement ($L(H\alpha)_{BD}$), as described in the text. Both luminosities
   are corrected for [NII] contamination. Different symbols indicate different galaxies. Filled symbols are for H$\alpha$ narrow-band data, open symbols for MUSE data.
   The dotted line shows the one-to-one relation. Right: same comparison for the 
   star formation rates derived using the Kennicutt (1998a) calibration adapted for a Chabrier (2003) IMF.
 }
   \label{sfr}%
   \end{figure*}

The H$\alpha$ luminosities derived using the two independent methods are fairly consistent, as indicated in Fig. \ref{sfr}. 
The dispersion in the relation is comparable to that observed in the calibration of Calzetti et al. (2007, 2010) ($\sim$ 0.3 dex)
or in the same relation observed in normal, late-type galaxies (Boselli et al. 2015).
Part of the scatter, however, can come from the large uncertainty on the derived Balmer decrement from the optical dust attenuation maps.

\subsubsection{Conversion of H$\alpha$ luminosities into SFR}

\begin{figure}
\centering
\includegraphics[width=9cm]{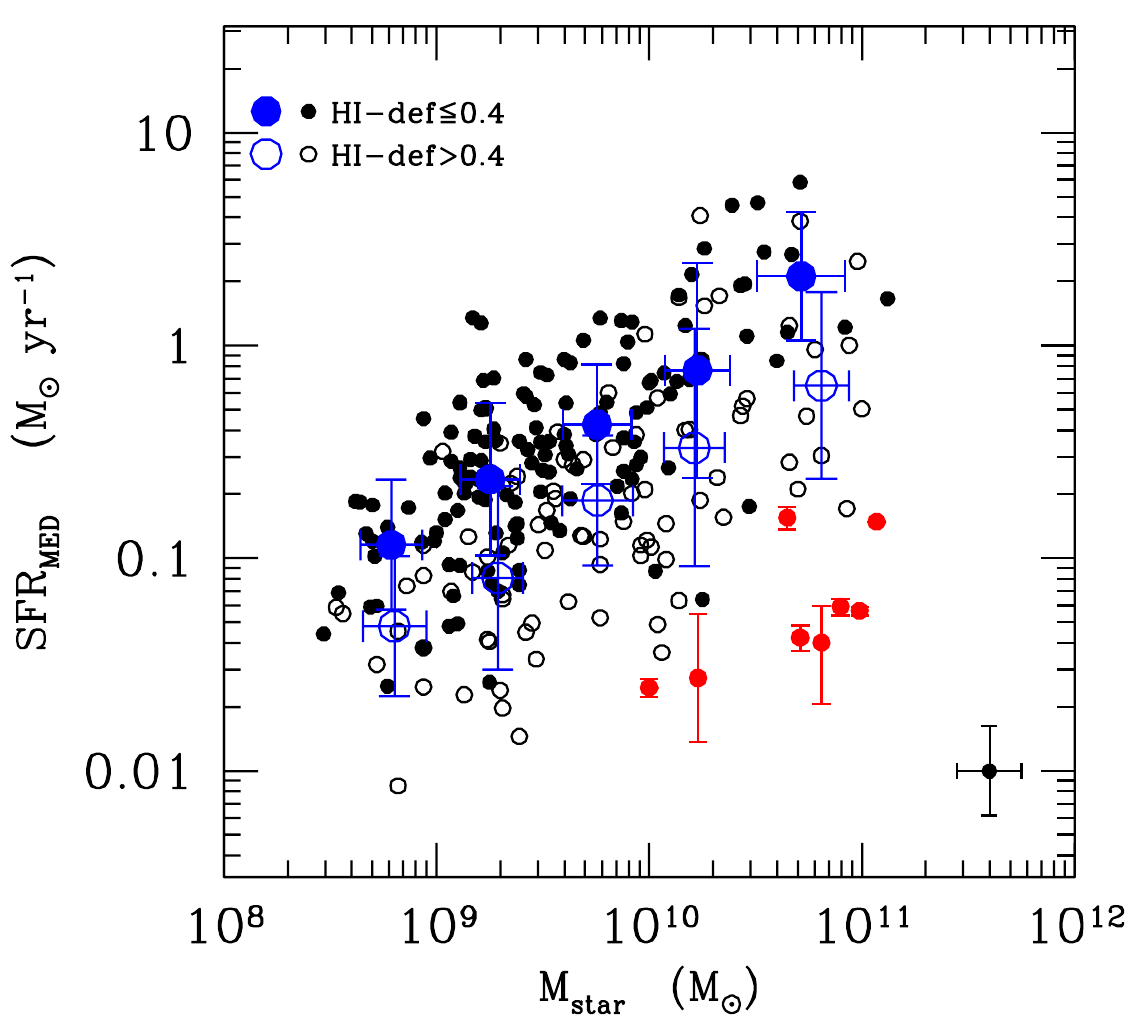}
\caption{Relationship between the star formation rate and the stellar mass for the \textit{Herschel} Reference Survey (HRS) late-type galaxies, adapted from Boselli et al. (2015). 
Black filled dots are for HI-normal ($HI-def$ $\leq$ 0.4)
field galaxies, black empty circles for HI-deficient ($HI-def$ $>$ 0.4) cluster objects. The blue large filled and empty circles give the mean values and the standard deviations
in different bins of stellar mass. The black error bar on the lower right corner shows the typical uncertainty on the data. The lenticular galaxies
analysed in this work are indicated by the red filled dots.}
\label{main}
\end{figure}

The H$\alpha$ luminosities of the target galaxies can be converted into star formation rates ($SFR$ in units of M$_{\odot}$ yr$^{-1}$) assuming that the ionising radiation comes
from young stars formed at a constant rate over $\gtrsim$ 10 Myr in the disc of ionised gas (e.g. Kennicutt 1998a, Boselli et al. 2009). 
This assumption is reasonable for those galaxies where the emission comes from clearly delimited discs 
(NGC 4429, 4459, 4469, 4476, 4477, 4526), while it 
might be unrealistic for the remaining two galaxies (NGC 4262, 4552), where the H$\alpha$ emission partly comes from diffuse filaments.
Assuming the Kennicutt (1998a) calibration, a Chabrier (2003) IMF, and a mean value for the attenuation as estimated above,
the total star formation rates of these galaxies are 0.02 $\lesssim$ $SFR$ $\lesssim$ 0.15 M$_{\odot}$ yr$^{-1}$
(see Table \ref{gal}). Their star formation activity is more than a factor of ten
lower than that of late-type galaxies of similar stellar mass, positioning them well below the main sequence (Boselli et al. 2015; see Fig. \ref{main}). 
Given the small size of the discs (0.7 $\lesssim$ $R(H\alpha)$ $\lesssim$ 2.0 kpc), however, their typical star formation surface density is 
0.004 $\lesssim$ $\Sigma(SFR)$ $\lesssim$ 0.02 M$_{\odot}$ yr$^{-1}$ kpc$^{-2}$, thus comparable to that of larger
discs characterising gas-rich spiral galaxies (Kennicutt 1998b; Bigiel et al. 2008).

\subsubsection{Physical properties of individual HII regions}

Using the same procedures adopted in Boselli et al. (2020, 2021b) based on the \textsc{HIIphot} data reduction pipeline (Thilker et al. 2000),
we identify the HII regions and derive their properties in the discs of the galaxies NGC 4429, 4459, 4476, 4477, and 4526. 
We refer to these papers for an accurate description of the code and of its use on the VESTIGE NB imaging data. 
We exclude the galaxies NGC 4262
and NGC 4552 because of their filamentary ionised gas emission, not unambiguously associated to any well identified star-forming complex, 
and NGC 4469 since it is an edge-on system, where individual HII regions cannot be resolved for projection effects.
Given that the stellar continuum in the inner regions of these lenticular galaxies is bright,
we limit this analysis to the HII regions with $L(H\alpha)$ $\geq$ 10$^{37}$ erg s$^{-1}$ and equivalent radii
$r_{eq}(H\alpha)$ $\gtrsim$ 50 pc, defined as the radii of the circles of surface equivalent to the area of the detected HII region 
down to a surface brightness limit of $\Sigma(H\alpha)$ = 3 $\times$ 10$^{-17}$ erg s$^{-1}$ cm$^{-2}$ (Helmboldt et al. 2005).
As in Boselli et al. (2021b), equivalent radii and diameters are corrected for the effects of the point-spread function.
This analysis is intended to trace the statistical properties of the HII regions in the discs of these lenticular galaxies.
\textsc{HIIphot} detects 42, 61, 14, 28, and 48 HII regions satisfying these criteria in the galaxies NGC 4429, 4459, 4476, 4477, and 4526, respectively.


\begin{figure*}
   \centering
   \includegraphics[width=1.00\textwidth]{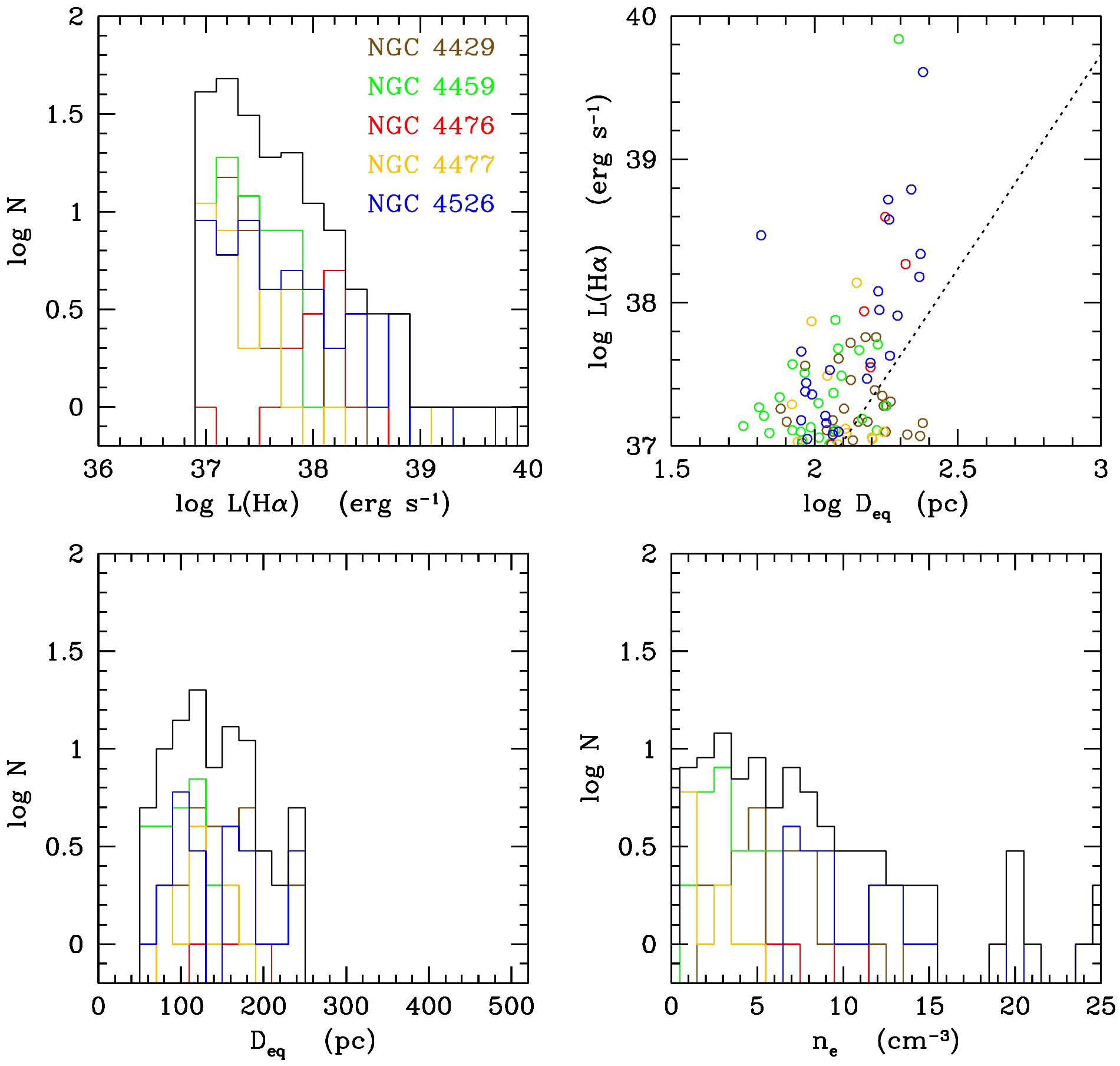}\\
   \caption{Statistical properties of the HII regions of the S0 galaxies with luminosity $L(H\alpha)$ $\geq$ 10$^{37}$ erg s$^{-1}$ and $S/N$ $>$ 5. 
   Upper left: H$\alpha$ luminosity function. Upper right: relationship between the H$\alpha$ luminosity and the equivalent diameter. 
   Lower left: distribution of the equivalent diameter. Lower right: distribution of the electron density. Galaxies are colour coded as follows: brown for NGC 4429,
   green for NGC 4459, red for NGC 4476, orange for NGC 4477, and blue for NGC 4526. The black histogram gives the cumulative distribution for all objects. 
   All the H$\alpha$ luminosities are corrected for 
   [NII] contamination assuming [NII]/H$\alpha$ = 0.78. Electron densities are also corrected for dust attenuation for comparison with other studies.
 }
   \label{HII}%
   \end{figure*}

Figure \ref{HII} shows the principal statistical properties of the
HII regions in the discs of the five lenticular galaxies. All the H$\alpha$ fluxes have been corrected for [NII] contamination assuming 
[NII]/H$\alpha$ = 0.78 as measured on the disc of NGC 4469 and 4526 using the MUSE data. For comparison with other works, H$\alpha$ 
luminosities are not corrected for dust attenuation, except for the electron density $n_e$,
where H$\alpha$ luminosities are estimated using the same $A(H\alpha)$ derived in Sect. 4.3.1 ($A(H\alpha)$ = 1.433, 0.418, 1.433, 0.186, and 1.543 mag
for NGC 4429, 4459, 4476, 4477, and 4526, respectively). Compared to those observed in the ram pressure 
perturbed dwarf galaxy IC 3476 (Boselli et al. 2021b), 
the HII regions in the discs of these lenticular galaxies have typically a lower luminosity and a smaller size. 
The only galaxy with HII regions of luminosity $L(H\alpha)$ $>$ 10$^{39}$ erg s$^{-1}$ is NGC 4526 (two regions).
Bright HII regions can be formed by the compression of the gas induced by a ram pressure stripping event during the first interaction 
of the galaxy ISM with the surrounding ICM (e.g. Boselli et al. 2021b). The phase-space diagram shown in Fig. \ref{phase}
suggests that NGC 4526 is a first infaller, consistent with this interpretation.
A complete and detailed comparative analysis with the physical properties of HII regions in a large sample of 
spiral galaxies spanning a wide range in morphological type and stellar mass
will be done once the VESTIGE survey is completed.

\subsection{Dust content and distribution}

The high quality HST images (Fig. \ref{HST}) and the NGVS/VESTIGE optical images (Fig. \ref{colour_image}), show dust structures in absorption. 
Prominent dusty discs with a clear spiral structure are present in the HST images of NGC 4429, 4459, 4476, and NGC 4526. A thin, edge-on dusty disc with 
filaments extending in the z-direction are visible in NGC 4469. Thin and extended filaments spiraling around the nucleus of the galaxy are visible
in NGC 4477 (F475W). A very smooth and limited filament of dust in absorption is also seen in the northern direction in NGC 4552 (F555W), while NGC 
4262 does not have any clear dust structure.

  \begin{figure*}
   \centering
   \includegraphics[width=0.4\textwidth]{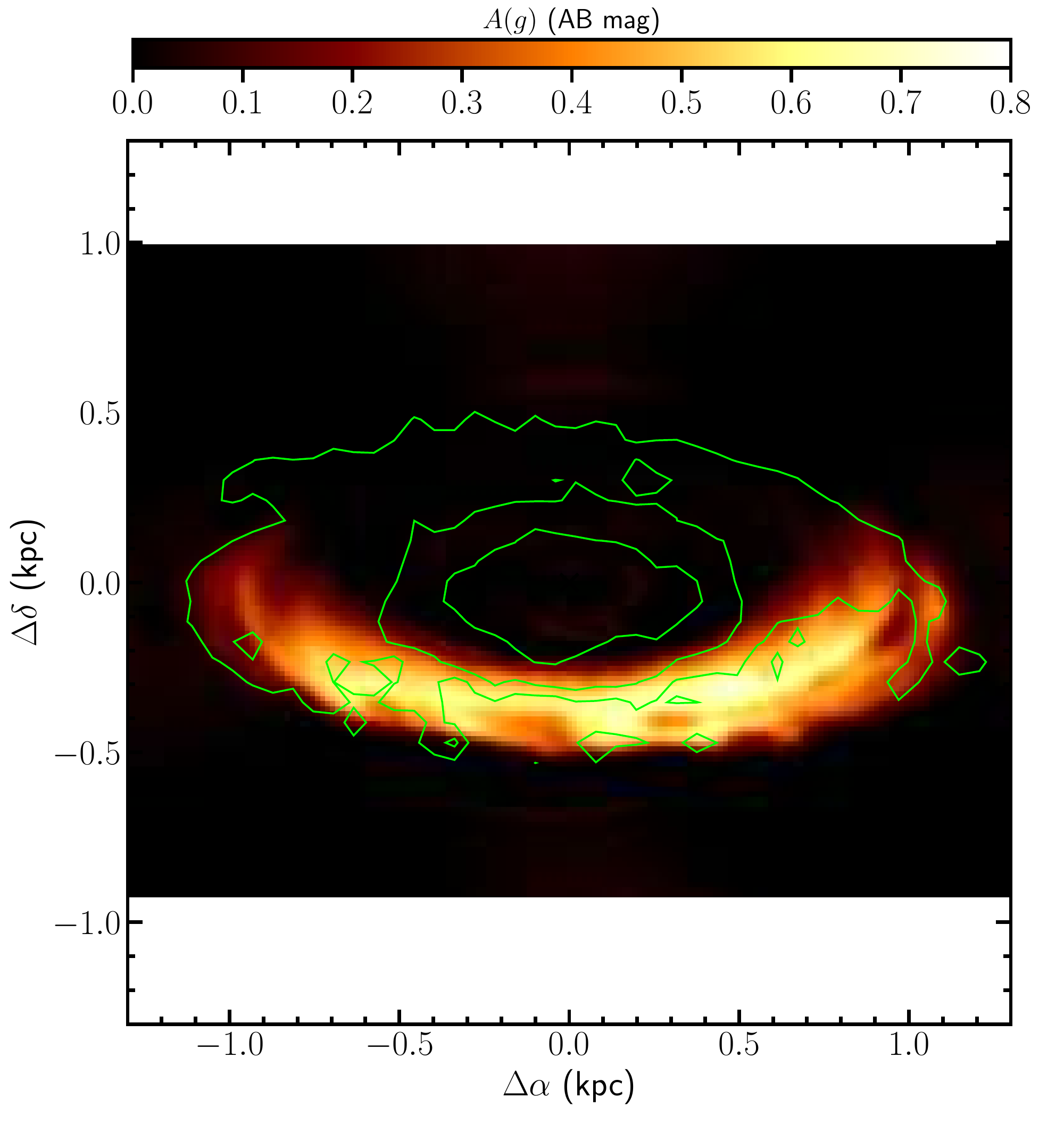}
   \includegraphics[width=0.4\textwidth]{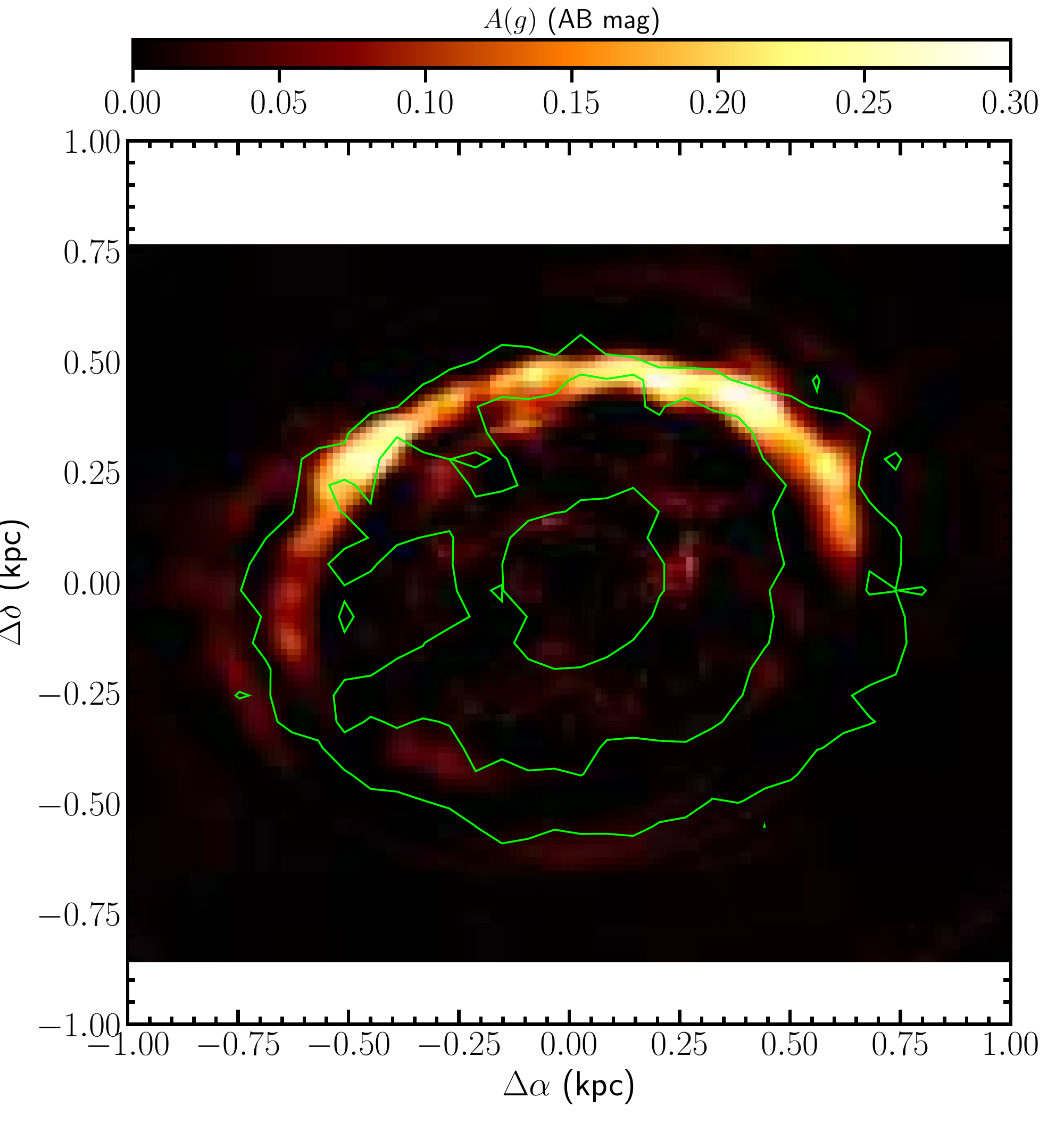}\\
   \includegraphics[width=0.4\textwidth]{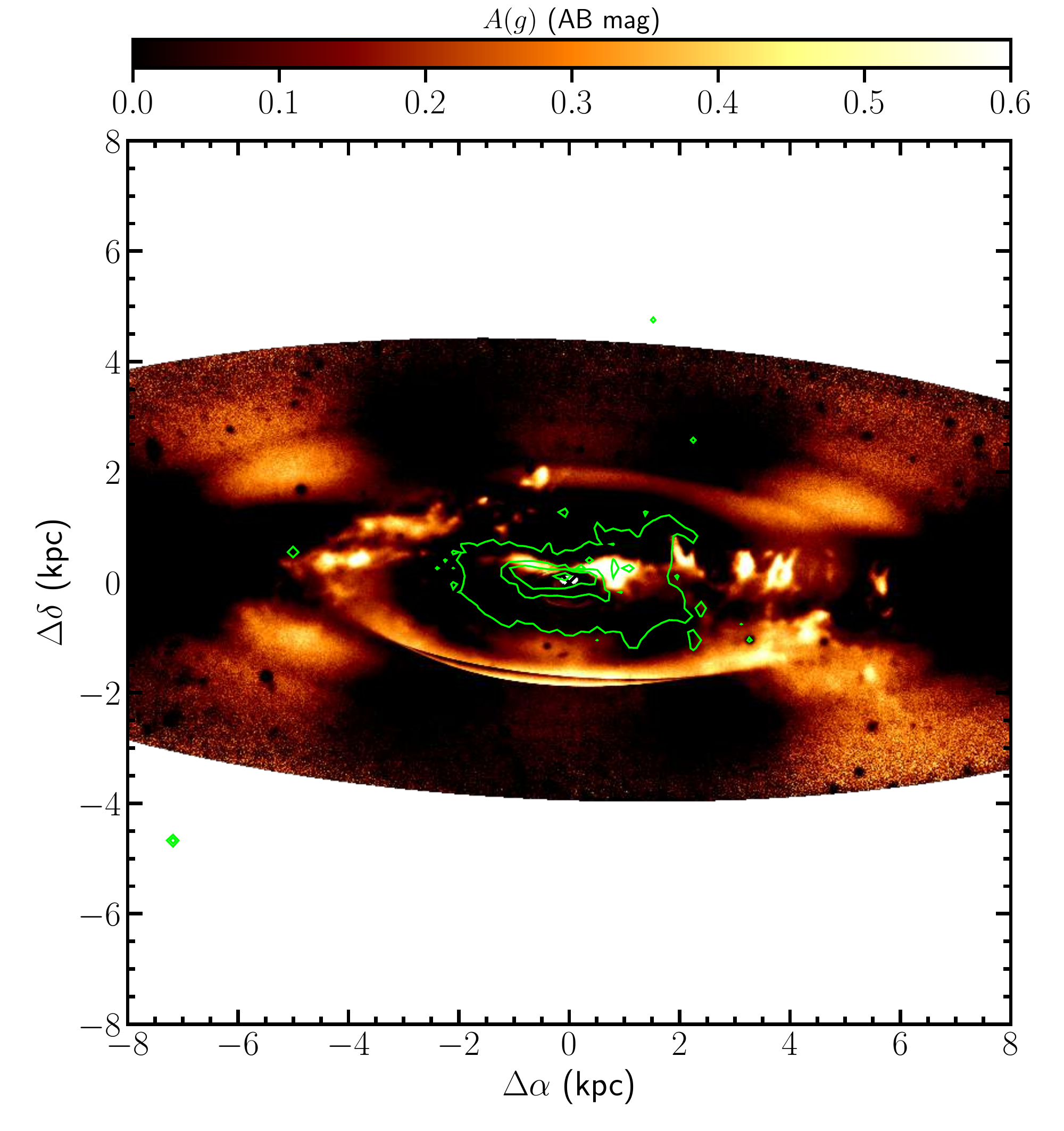}
   \includegraphics[width=0.4\textwidth]{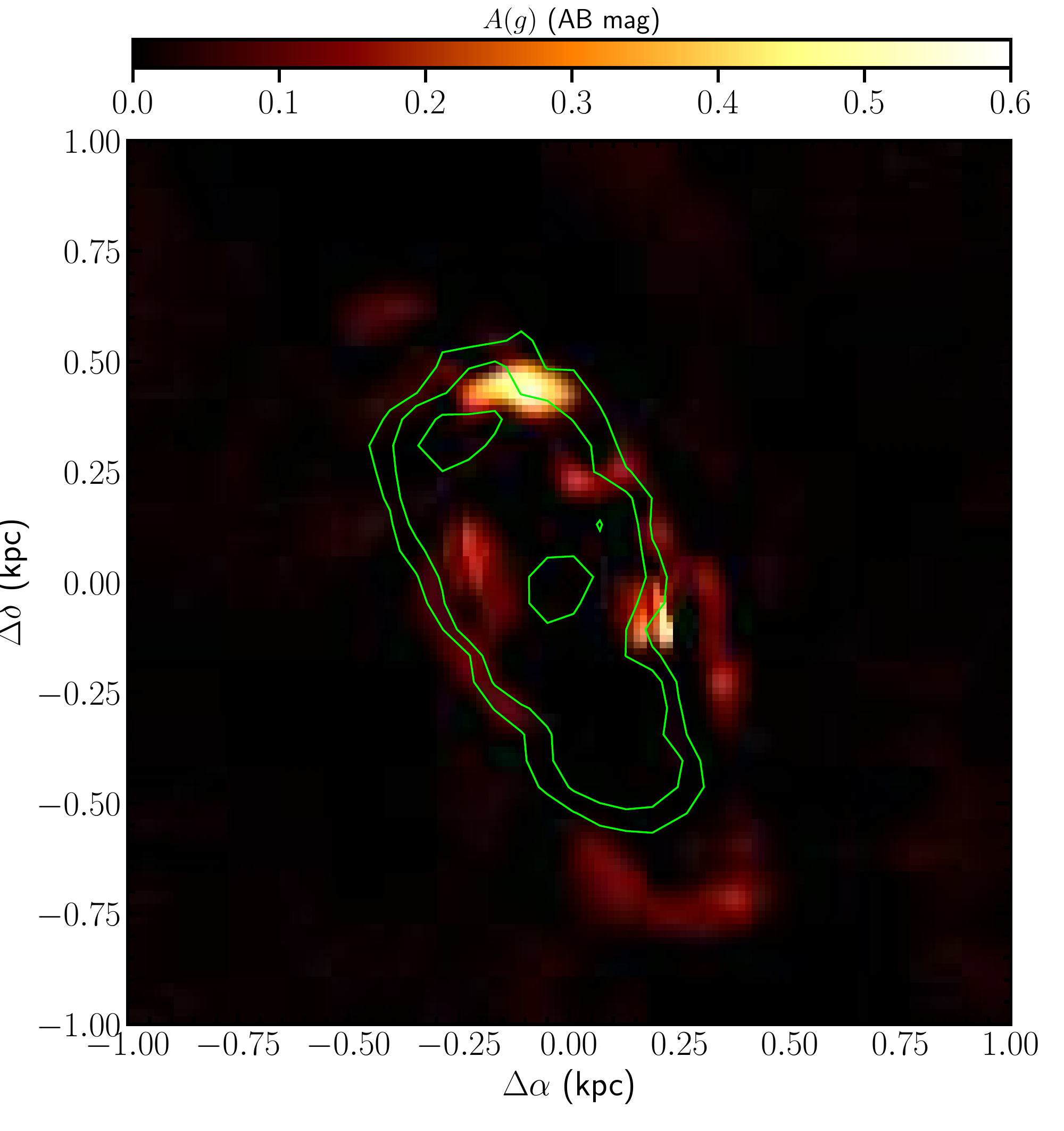}\\
   \includegraphics[width=0.4\textwidth]{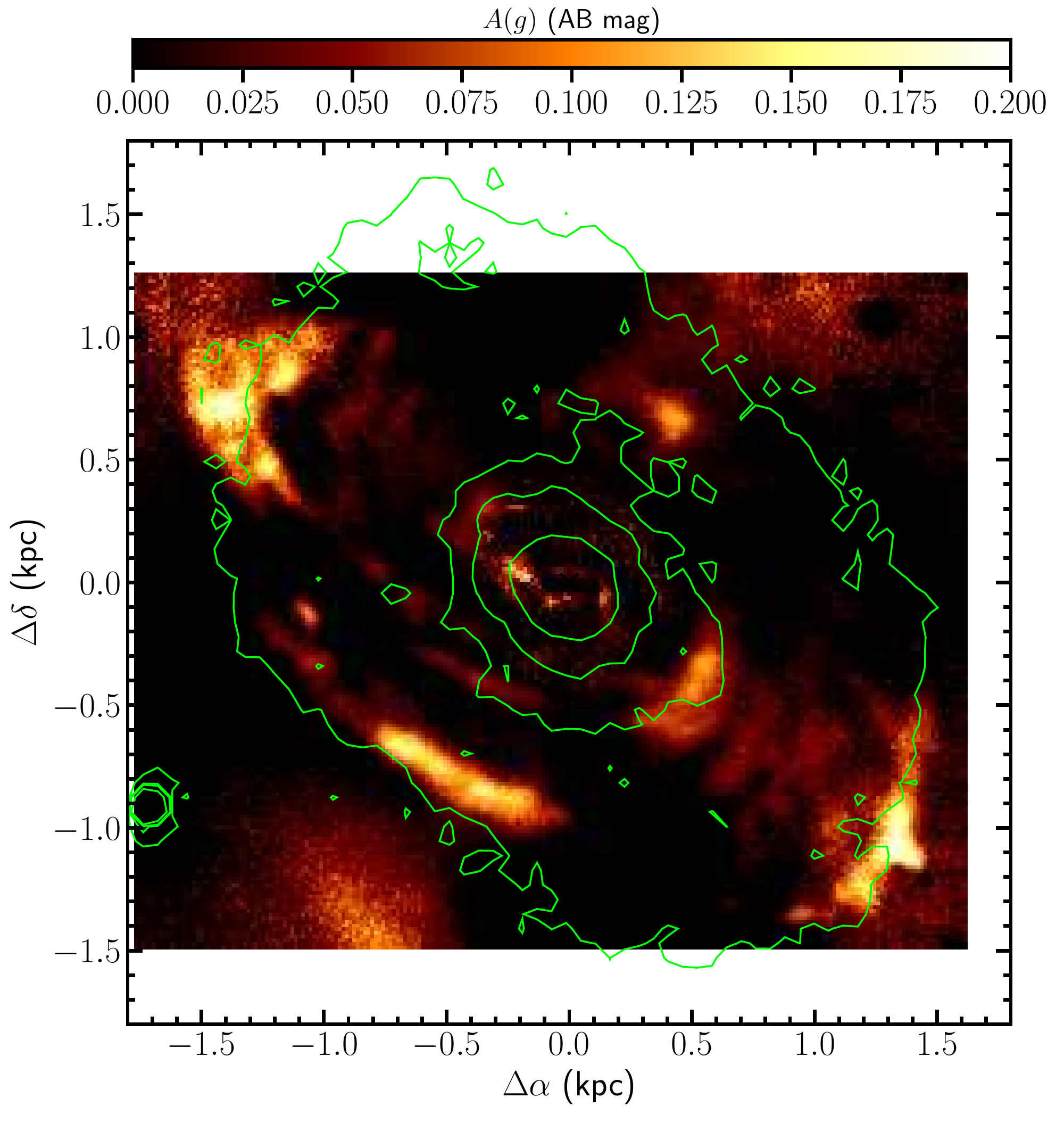}
   \includegraphics[width=0.4\textwidth]{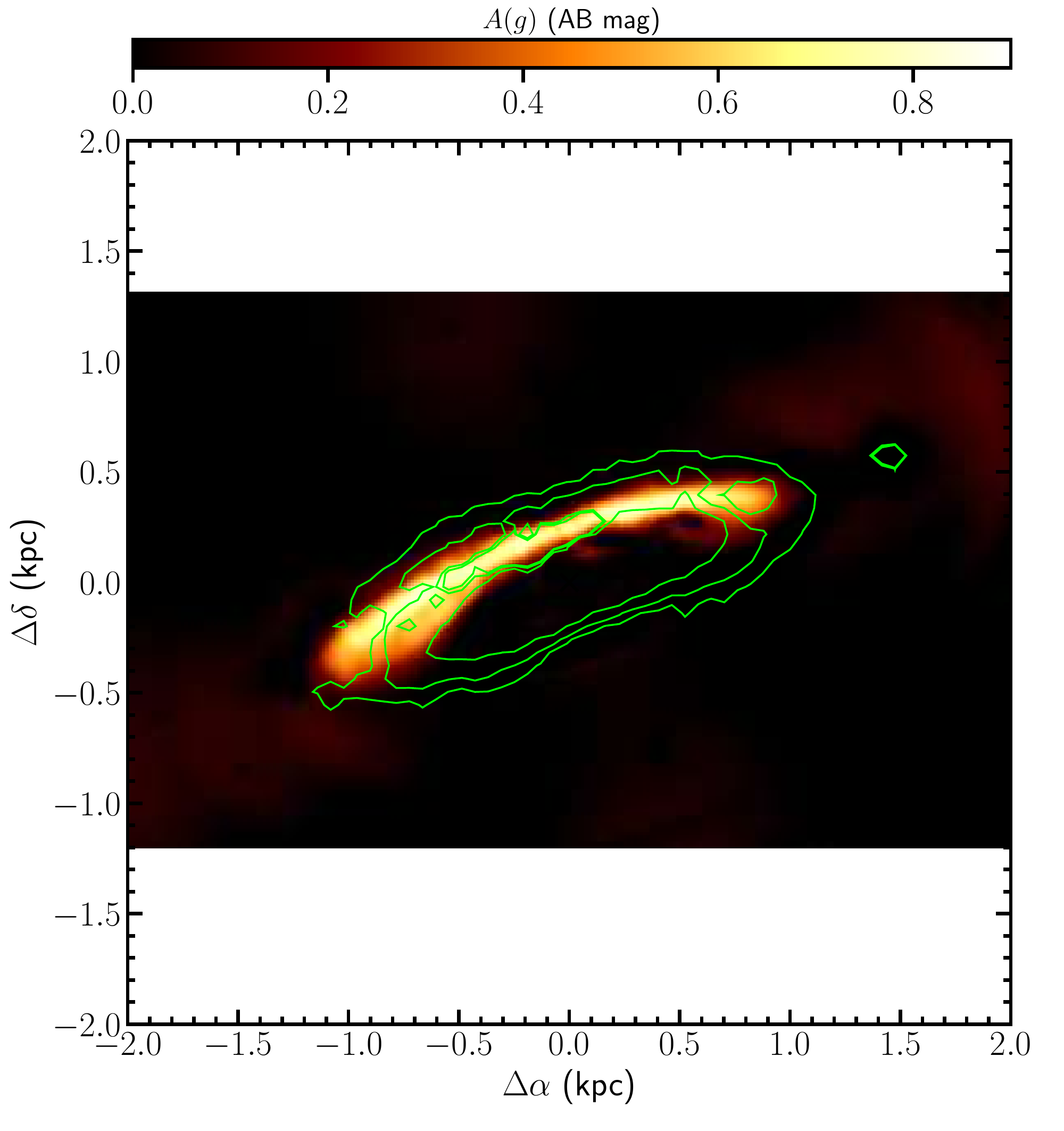}\\
   \caption{$A(g)$ extinction maps of the galaxies NGC 4429 (first row, left), 4459 (first row, right), 4469 (second row, left), 4476
   (second row, right), 4477 (last row, left), and 4526 (last row, right), with contour levels of the VESTIGE H$\alpha$ emission. 
Contours are at 3$\times$10$^{-16}$, 6$\times$10$^{-16}$, 1.5$\times$10$^{-15}$ erg s$^{-1}$ cm$^{-2}$ for the galaxies NGC 4429, 4459, 4476, and 4526, and at 
10$^{-16}$, 3$\times$10$^{-16}$, 6$\times$10$^{-16}$, 1.5$\times$10$^{-15}$ erg s$^{-1}$ cm$^{-2}$ for the galaxies NGC 4469 and 4477. 
The colour scale of the $A(g)$ extinction maps of the galaxy 
NGC 4469 goes to negative values to take into account the low frequency gradients unremoved by the fitted model and relative to the 
prominent boxy shape of the bulge.
 }
   \label{Ag}
   \end{figure*}

For each object we construct a 2D model of the stellar distribution and subtract it from the NGVS $g$-band optical images (Ferrarese et al. 2012).
These images are covering the whole extent of the galaxies and not only a fraction of them as in the HST images, thus  they can be used 
to derive the dust distribution seen in absorption, exactly as done for M87 in Boselli et al. (2019), 
or for other lenticular galaxies with ionised gas emission and prominent dust features (e.g. Patil et al. 2007, Finkelman et al. 2010, Kulkarni et al. 2014). 
For this exercise, we integrated the model subtracted
image within a region positioned over the disc of dust as seen in absorption (see Fig. \ref{Ag}).
We then transform the attenuation map $A(g)$ first in $A(V)$ with a Milky Way extinction curve then in $\Sigma_{dust}$ using the relation (Spitzer 1978; Kitayama et al. 2009):

\begin{equation}
{A(V) = 0.011 \left(\frac{Q_{ext}}{2}\right)\left(\frac{\Sigma_{dust}}{10^3 \rm{M_{\odot}~kpc^{-2}}}\right)\left(\frac{\rho_{dust}}{3 \rm{g~cm^{-3}}}\right)^{-1}\left(\frac{a}{0.1 \rm{\mu m}}\right)^{-1}}
\end{equation}

\noindent
where $Q_{ext}$ is the extinction coefficient factor, $\rho_{dust}$ is the dust mass density, and $a$ is the grain radius, adopting the standard values given in the equation.
The total dust mass is then derived assuming that the dust is uniformly distributed over the disc associated to the ionised gas. The extinction maps give
dust masses of the order of 6 $\times$ 10$^3$ $\lesssim$ $M_{dust}$ $\lesssim$ 9 $\times$ 10$^4$ M$_{\odot}$. 

To check the reliability of this method, we compare the total dust masses derived in this way with those calculated by fitting the observed spectral energy distribution of each 
galaxy with the CIGALE SED fitting code (Noll et al. 2009, Boquien et al. 2019; see Table \ref{dust}). 
Given that the dusty disc is located only within the very inner regions compared to the much more extended stellar emission, we determine the total dust mass by fitting only the spectral bands where the emission is
dominated by dust, including WISE at 22 $\mu$m, \textit{Spitzer}/MIPS at 24 and 70 $\mu$m, \textit{Herschel}/PACS and SPIRE at 100, 160, 250, 350, and 500 $\mu$m (8 bands in total). 
The emission of early-type galaxies in the mid-IR might be contaminated by that of stellar envelopes (Boselli et al. 1998, Temi et al. 2009, Ciesla et al. 2014). We thus first remove the contribution from the stellar emission
to the WISE 22$\mu$m and \textit{Spitzer}/MIPS 24$\mu$m bands as in Ciesla et al. (2014). We estimate this contamination by fitting the stellar SED (from the $FUV$ to the $K$-band, in ten photometric bands) of each galaxy again with CIGALE.
For this purpose, we use Bruzual \& Charlot (2003) stellar population models, a Chabrier (2003) IMF, and a delayed star formation history as the one described in Boquien et al. (2019).
We then fit the infrared SED using the Draine \& Li (2007) dust models, consistently with Ciesla et al. (2014) and Boselli et al. (2016b). 
The dust masses of these galaxies are 6 $\times$ 10$^5$ $\lesssim$ $M_{dust}$ $\lesssim$ 10$^7$ M$_{\odot}$ (see Table \ref{dust}), a factor of 
$\simeq$ 50 lower than those of star-forming systems of similar stellar mass (Cortese et al. 2012).

These dust masses are a factor of 14-380 higher than those derived from dust absorption. These differences are too large to be ascribed to the uncertainties in 
the determination of the dust attenuation map, or to the determination of the dust distribution within the disc. 
As already widely discussed in the literature, this discrepancy might have different origins:\\
1) As first remarked by Goudfrooij \& de Jong (1995), dust masses derived from optical dust absorption maps should be taken as 
lower limits because eq. 4 is valid in a screen model, i.e. when the emitting stars are all located behind the dust. This is obviously
not the case in lenticular galaxies, where the disc of dust is within the plane of the galaxy (sandwich model), thus around half of the stars of the prominent bulge
are located in between the dust and the observer.\\
2) Dust masses derived using SED fitting techniques might include a diffuse and unstructured dust component. If present, this component
would be subtracted along with the stellar component from the optical images since included in the fitted model.
The excess of dust mass derived from SED fitting might thus indicate the presence of diffuse dust produced, injected into the ISM (Goudfrooij \& de Joung 1995, 
Dwek 1998, Zhukovska et al. 2008, Valiante et al. 2009, Triani et al. 2020), and heated by the old stellar population, dominant in early-type systems (Boselli et al.
2010b, 2012; Bendo et al. 2012b, 2015).

\begin{figure*}
\centering
\includegraphics[width=9.0cm]{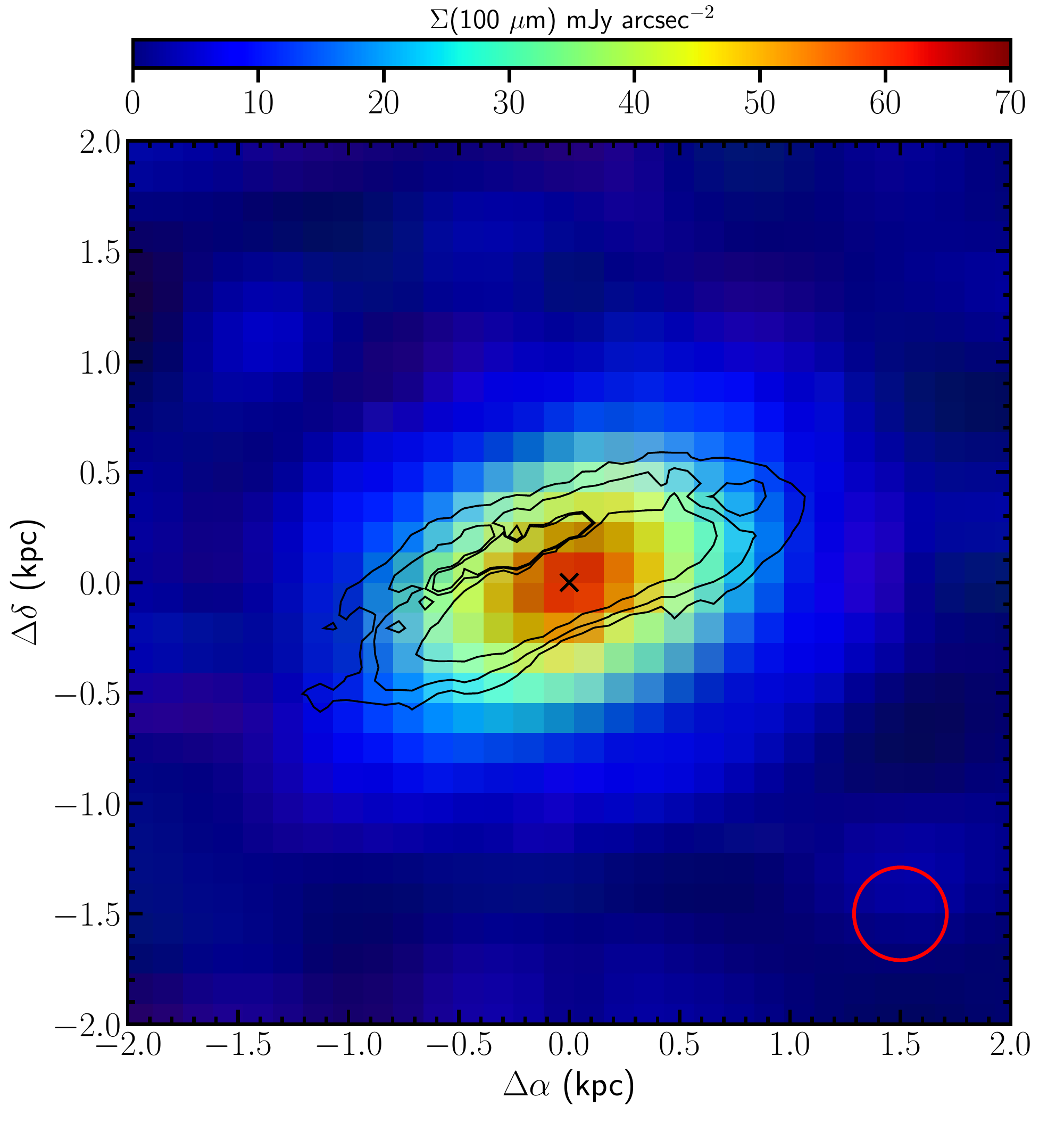}
\includegraphics[width=9.0cm]{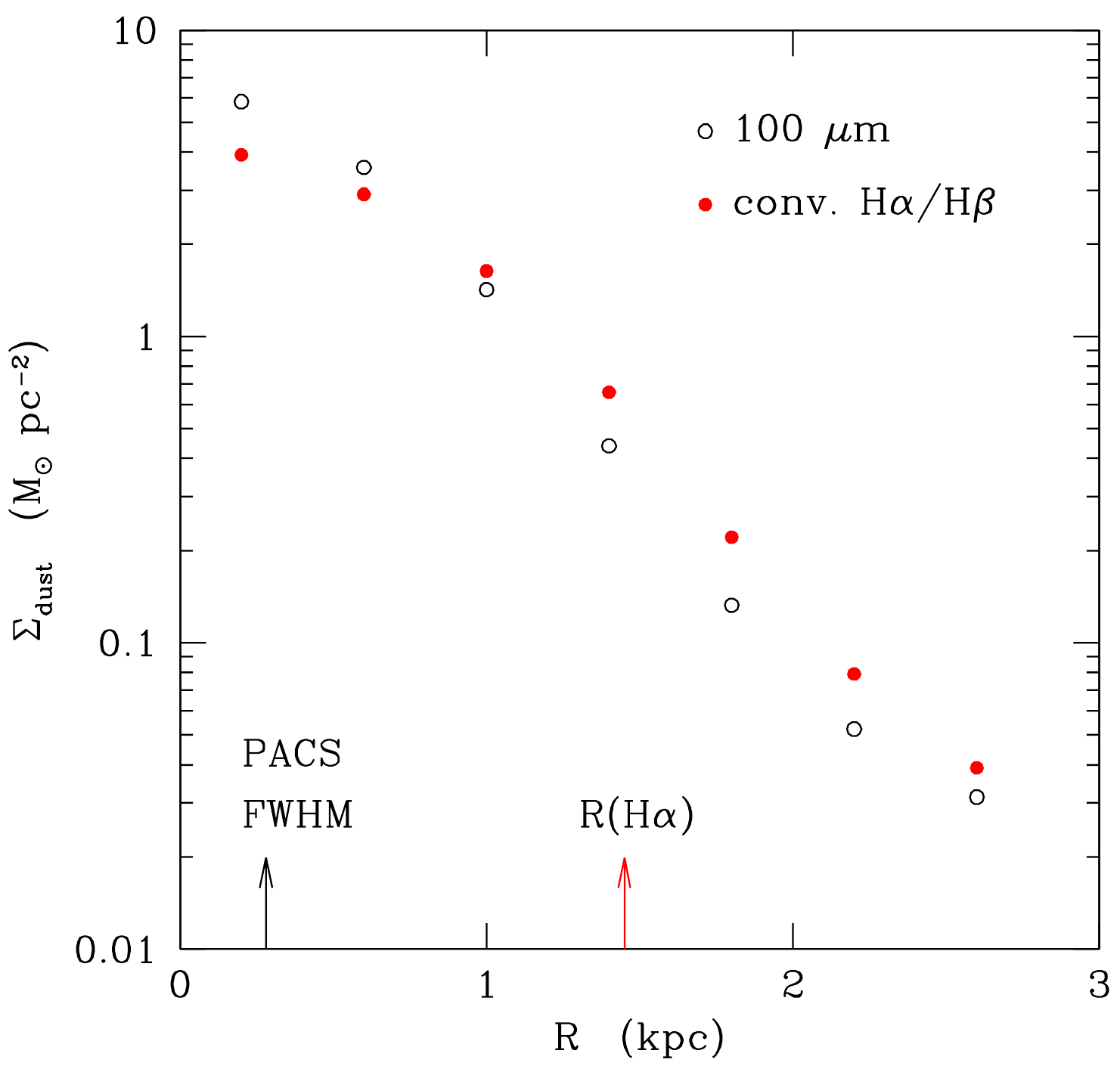}\\
\caption{Left: \textit{Herschel}/PACS image of the galaxy NGC 4526 at 100 $\mu$m (from Cortese et al. 2014), in units of mJy arcsec$^{-2}$, with contour levels of the VESTIGE H$\alpha$ emission 
(3$\times$10$^{-16}$, 6$\times$10$^{-16}$, 1.5$\times$10$^{-15}$ erg s$^{-1}$ cm$^{-2}$). The black cross indicates the position of the centre of the galaxy. 
The red circle on the lower right corner gives the FWHM of the PACS data (7\arcsec). Right: radial variation of the dust surface density profile of NGC 4526
derived from the PACS 100 $\mu$m image (black empty circles)
and from the $E(B-V)$ 2D-map derived using the Balmer decrement extracted from the MUSE datacube an brought to the same resolution than PACS as described in the text (red filled circles).
The black and red vertical arrows indicate the PACS FWHM and the extension of the H$\alpha$ emitting disc.}
\label{N4526_PACS}
\end{figure*}

To see whether the extended emission observed in the PACS bands is due to a diffuse dust component or to the one associated to the ionised gas disc 
we convolve the $E(B-V)_{BD}$ 2D map of NGC 4526 derived using the Balmer decrement extracted from the  
MUSE data with the 100 $\mu$m PACS PSF given in Bocchio et al. (2016) for fast scan observations in parallel mode (Cortese et al. 2014). 
We then extract the radial profile of both the 100 $\mu$m PACS
emission and of the convolved $E(B-V)_{BD}$ 2D map in concentric elliptical annuli centred on the nucleus of the galaxy and of size increasing by 0.4 kpc. 
We then convert the measured 100 $\mu$m surface brightnesses into dust surface densities using the relation (e.g. Spitzer 1978, Boselli 2011):

\begin{equation}
{M_{dust} = \frac{S(\nu)D^2}{K_{\nu}B(\nu,T)}}
\end{equation}

\noindent
where $S(\nu)$ is the PACS flux density at 100 $\mu$m, $D$ the distance of the galaxy. We then assume a dust grain opacity at 100 $\mu$m 
$K_{100}$=27.15 cm$^2$ g$^{-1}$ (Darine 2003) and a dust temperature $T$ = 22.6 K, which is the temperature necessary to make the total dust mass
estimated from the CIGALE SED fitting analysis ($M_{dust}$ = 9.95$\times$ 10$^6$ M$_{\odot}$) equal to the one given in eq. 5 once integrated over all the 
increasing elliptical annuli. We then compare this dust radial profile derived from the 100 $\mu$m PACS data to the one derived from the convolved MUSE
Balmer decrement in Fig. \ref{N4526_PACS}, where $E(B-V)_{BD}$ is converted into $\Sigma_{dust}$ using eq. 4 and normalised to the total dust content of the galaxy
as for the PACS profile. The radial variation of the two profiles are very similar, clearly indicating that the extended 100 $\mu$m emission of NGC 4526
is due to the dust frozen into the circumnuclear ionised gas disc. Since NGC 4526 is the galaxy of the sample with the most extended
100 $\mu$m emission with respect to that of the diffuse ionised gas, we argue that this conclusion is probably valid also for the other objects.

In agreement with the results obtained in other studies based on the analysis done using 3D radiative transfer models of early-type galaxies
with similar dust features (Viaene et al. 2015, 2019), we conclude that in these lenticular galaxies 
the observed difference between the dust content estimated from optical absorption features and 
SED fitting analysis is mainly due to geometrical effects related to the relative distribution of the absorbing dust and the emitting stars. 
We notice that the importance of geometry is corroborated by two other results mentioned above: a) the dust
features seen in absorption in the HST images (Fig. \ref{HST}) and in the $A(g)$ attenuation maps (Fig. \ref{Ag}) are asymmetric and strongly depend on the 
orientation of the galaxies's discs with respect to the line of sight, and b) in NGC 4526, where the data allowed a direct comparison, 
$E(B-V)_{BD}$ = 4.87 $\times$ $E(B-V)_{star}$, thus significantly flatter than the canonical Calzetti's law observed in starburst 
galaxies (Calzetti et al. 2000, see Sect. 4.3.1), where the dust layer has a thickness comparable to that of the stellar disc (slab model).
Furthermore, $E(B-V)_{BD}$ is fairly symmetric because the young and massive stars responsible for the ionising radiation and the 
absorbing dust are well mixed within the disc as in a slab model, while this is not the case for the dominant old stellar population 
which is distributed in a much thicker and extended lenticular bulge (sandwich model).

\begin{table}
\caption{Dust masses}
\label{dust}
{
\[
\begin{tabular}{cccc}
\hline
\noalign{\smallskip}
\hline
NGC	        & $M_{dust}(SED)$				& $M_{dust}(A(g))$	& G/D$^b$	\\
\hline
Units		& M$_{\odot}$					& M$_{\odot}$		&		\\
\hline
4262		& -						& -			& -		\\
4429		& 1.83$\times$10$^6$ $\pm$2.9$\times$10$^5$	& 5.0$\times$10$^{4}$	& 86.1 ~(226.4)	\\
4459		& 1.50$\times$10$^6$ $\pm$1.1$\times$10$^5$	& 2.4$\times$10$^{4}$	& 90.5 ~(318.4)	\\
4469$^a$	& 7.67$\times$10$^6$ $\pm$3.8$\times$10$^5$	& 2.0$\times$10$^{4}$   & -		\\
4476		& 1.06$\times$10$^6$ $\pm$1.4$\times$10$^5$	& 5.7$\times$10$^{3}$	& 67.1 ~(138.1)	\\
4477		& 5.81$\times$10$^5$ $\pm$1.3$\times$10$^5$	& 4.0$\times$10$^{4}$	& 30.9		\\  
4526		& 9.95$\times$10$^6$ $\pm$1.6$\times$10$^6$	& 9.4$\times$10$^{4}$	& 40.8 ~(31.6)	\\
4552		& 6.02$\times$10$^6$ $\pm$3.8$\times$10$^6$	& -			& -		\\    
\noalign{\smallskip}	
\hline
\end{tabular}
\]
Notes: a) $A(g)$ measured only within the star-forming disc, thus the dust mass coming from $A(g)$ is a lower limit given that filaments of dust are seen outside 
the selected region. b) Gas-to-dust ratio measured using dust masses derived from the SED fitting analysis and molecular gas masses with CARMA (and ALMA) data.
}
\end{table}

\subsection{Cold gas content}

Interferometric data in the CO lines can be used to derive the molecular gas content and distribution within the target galaxies. 
Five galaxies (NGC 4429, 4459, 4476, 4477, and 4526) have been observed with CARMA in the $^{12}$CO(1-0) line with a typical angular resolution of 
3\arcsec$\lesssim$FWHM$\lesssim$ 9\arcsec\ at sufficient sensitivity to detect molecular gas in diffuse and extended structures. Assuming a constant CO-to-H$_2$ conversion factor 
$X_{CO}$ = 2.3 $\times$ 10$^{20}$ cm$^{-2}$/(K km s$^{-1}$) (Strong et al. 1988), appropriate for massive galaxies such as these (e.g. Boselli et al. 2002; Bolatto et al. 2013), 
we derive the molecular gas masses given in Table \ref{ALMA}.  
These molecular gas masses can be compared to those derived using different sets of ALMA 
data, which have the advantage of being at a much higher angular resolution than the CARMA data. ALMA data for two galaxies are in the $^{12}$CO(2-1) line
and for one in the $^{12}$CO(3-2) line.
Using the same procedure applied to the CARMA data, and assuming the ratios $^{12}$CO(2-1)/$^{12}$CO(1-0) = $0.65^{0.83}_{0.50}$ 
and $^{12}$CO(3-2)/$^{12}$CO(1-0) = $0.31^{0.42}_{0.20}$ from Leroy et al. (2021), quite consistent with previous values (e.g. Leroy et al. 2009, Wilson et al. 2012), we estimate
total molecular hydrogen masses from the ALMA data. The two sets of data are fairly consistent (see Table \ref{ALMA}). The values given in Table \ref{gal}
are, in order of priority, the 12~m ALMA or CARMA data, whenever two sets of data are available, with a typical uncertainty of $\sim$ 0.1 and 0.2 dex, respectively,
estimated from the comparison of the 12~m and 7~m ALMA data, or the comparison of the ALMA and CARMA data. We recall that the ALMA data might suffer from an extra systematic 
uncertainty due to the poorly constrained $^{12}$CO(2-1)/$^{12}$CO(1-0) and $^{12}$CO(3-2)/$^{12}$CO(1-0) ratios which, given the peculiar physical properties
of the discs of gas analysed in this work, might be different than those assumed here and derived from the observation of normal, star-forming galaxies (e.g. Hafok \& 
Stutzki 2003). For the galaxies NGC 4469 and NGC 4552, molecular gas masses are derived as indicated in the next session.

\subsection{Gas-to dust ratio}

The same set of data can be used to estimate the typical gas-to-dust ratio of these lenticular galaxies and compare it to that of late-type galaxies
with similar metallicity.
Indeed, it has been shown that in the Milky Way, in the nearby Magellanic clouds, or in other nearby galaxies, where the angular distribution of the 
data allows us to trace the ratio of the gas-to-dust column density in different environments, $\Sigma_{gas}$/$\Sigma_{dust}$ varies only as a function of the 
metallicity (e.g. Dwek 1998, Issa et al. 1990, Sandstrom et al. 2013, Roman-Duval et al. 2014, Mattsson \& Andersen 2012, Mattsson et al. 2014, Giannetti et al. 2017).
Since all the analysed objects are massive and thus probably metal rich, if the gas and dust have an internal origin, we expect a constant gas-to-dust ratio close to the one measured 
in the solar neighbourhood ($\Sigma_{gas}$/$\Sigma_{dust}$(Z$_{\odot}$) = 100-160, Sodroski et al. 1994, Draine et al. 2007), where the oxygen abundance is 12+log O/H(Z$_{\odot}$) = 8.69 (Asplund et al. 2009). 
Since we measure dust and gas masses rather than column densities, we have to assume that the different phases of the ISM have a similar distribution, as indeed the observations 
indicate (see Fig. \ref{HST}). The total gas mass can be derived combining the molecular and the atomic components.
Among the eight galaxies, only three have been detected in HI (NGC 4262, Serra et al. 2012; NGC 4469 and NGC 4526, Haynes et al. 2018). For all the others, the measured upper limits 
to their HI mass are significantly lower than the measured molecular gas masses. We can thus confidently assume that the total gas is dominated 
by the molecular gas phase. This is consistent with the observed physical association between the dust distribution and the molecular gas distribution which is stronger than that with 
the atomic gas phase whenever the molecular gas is the dominant phase (Dame et al. 2001, Sandstrom et al. 2013)\footnote{We recall that the galaxies NGC 4459, NGC 4477, NGC 4526,
and NGC 4552 have hot gas masses of 4$\times$10$^7$ $\lesssim$ M$_{hot}$ $\lesssim$ 2.3$\times$10$^8$ M$_{\odot}$ (Su et al. 2015), thus comparable to the molecular gas masses. 
Since diffuse and generally not associated to the dust component, this hot gas phase is not accounted for in the estimate of the gas-to-dust ratio.}. 

  \begin{figure}
   \centering
   \includegraphics[width=0.49\textwidth]{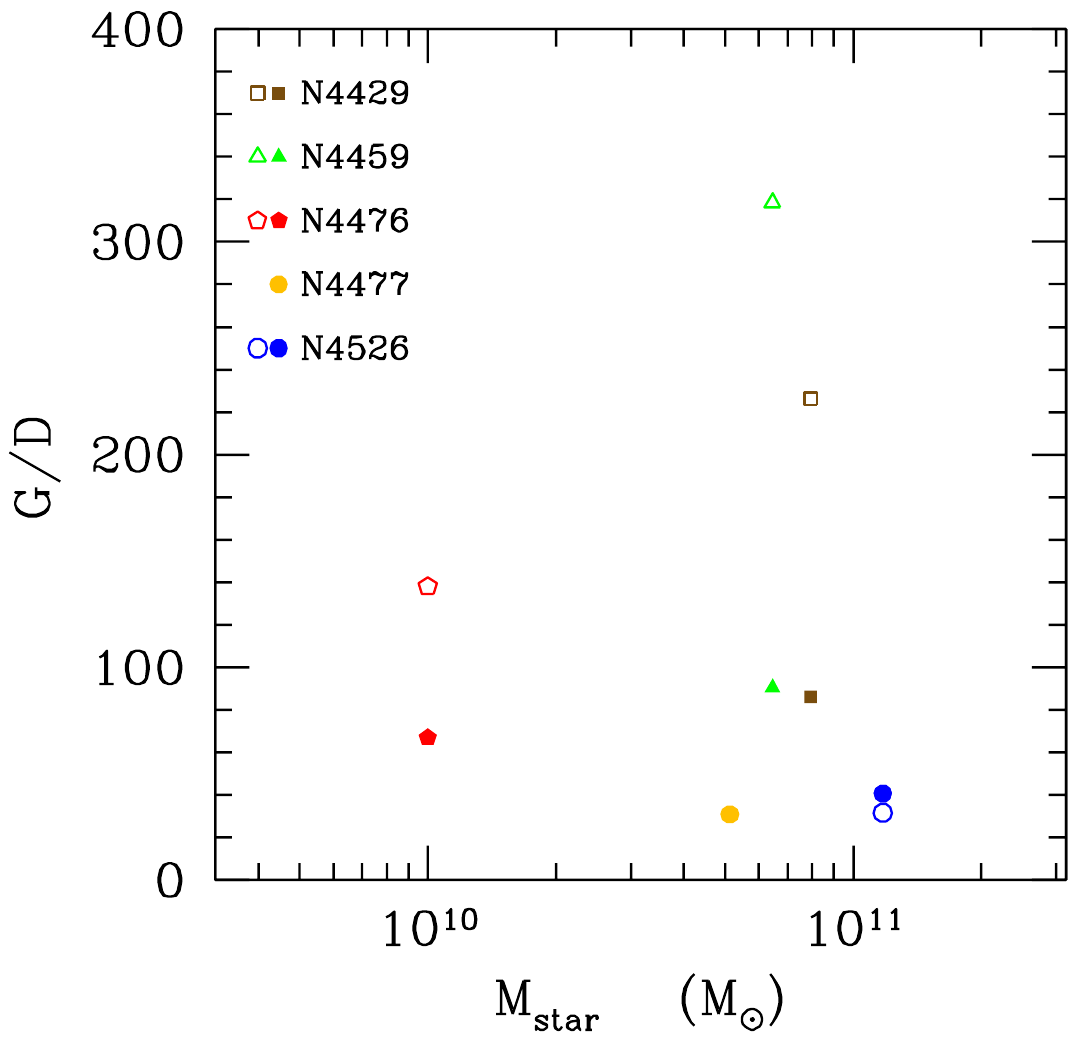}
   \caption{Relationship between the volumetric gas-to-dust (G/D) ratio of the target galaxies measured using the CARMA (filled symbols)
   and ALMA (empty symbols) data and their stellar mass. Different symbols indicate different galaxies.
 }
   \label{GDratio}%
   \end{figure}

\begin{figure}
   \centering
   \includegraphics[width=0.49\textwidth]{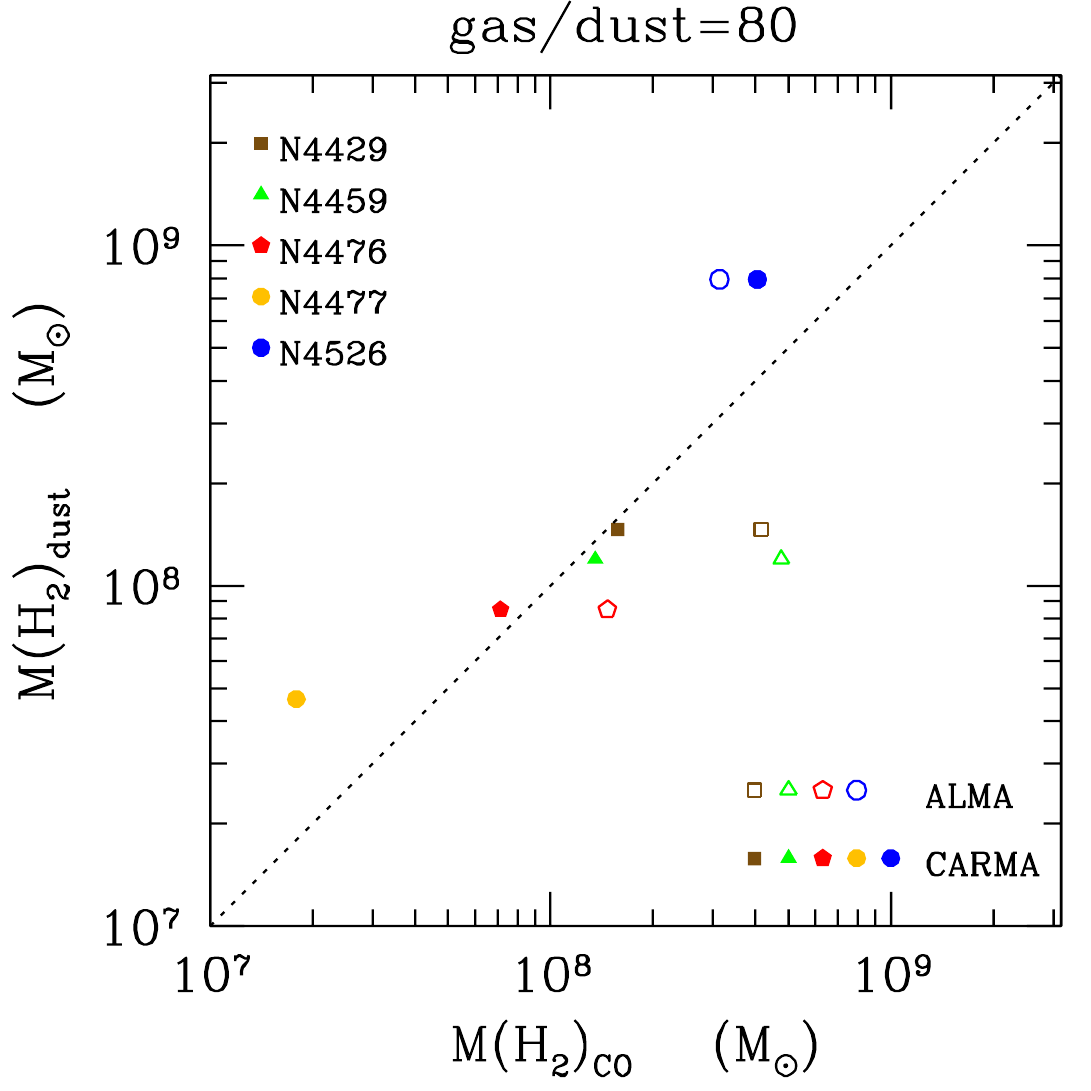}
   \caption{Relationship between the molecular gas mass derived using the total dust mass and assuming a gas-to-dust ratio of $G/D$=80 and the molecular gas
   mass derived using CO observations for galaxies with ALMA (empty symbols) and CARMA (filled symbols) data. The different galaxies are indicated by different symbols.
   The dotted line gives the 
   expected relation for a gas-to-dust ratio $G/D$=80
 }
   \label{MH2}%
   \end{figure}

Figure \ref{GDratio} shows the relationship between the volumetric gas-to-dust (G/D) ratio of the target galaxies (see Table \ref{dust}) and their stellar mass, 
with dust masses derived from the SED fitting analysis. Figure \ref{GDratio} indicates that the gas-to-dust ratio of 
these lenticular galaxies ranges between 30 $\lesssim$ $G/D$ $\lesssim$ 320, and thus corresponds to the low range of the envelope sampled by similar 
massive late-type systems (20 $\lesssim$ $G/D$ $\lesssim$ 1000) by Remy-Ruyer et al. (2014). A value of $G/D$ $\simeq$ 80 is very consistent with the oxygen abundance
of the gas in NGC 4469 and NGC~4526 (12+log O/H $\simeq$ 8.75-8.80) as derived with the MUSE data, which is slightly higher then solar. The only exception is NGC 4262
which has a large amount of HI gas (5.6 $\times$ 10$^8$ M$_{\odot}$) and is undetected in CO (Boselli et al. 2014b) and in the far-IR (Ciesla et al. 2012, Cortese et al. 2014).
The large amount of HI gas suggests that in this galaxy the different phases of the ISM might not be distributed within the
same region. Indeed, the HI map of NGC~4262 reveals that the atomic gas is mainly located on a disc much more extended than the stars ($\simeq$ 30 kpc in diameter, Serra et al. 2012), 
it is thus conceivable that part of this HI, the one located far from the core of the galaxy, is metal poor and thus dust and molecular gas poor. Furthermore, this
gas has a very low column density not sufficient to collapse into giant molecular clouds. 
This analysis indicates that we can get a rough estimate of the molecular gas mass of the galaxies NGC 4469 and NGC 4552 from their total dust masses
assuming a typical gas-to-dust ratio of $G/D$ = 80, as depicted in Fig. \ref{MH2}, where the molecular gas masses
derived using a constant gas-to-dust ratio and those measured using CO data are confronted. These values are given in Table \ref{gal}, with a typical uncertainty of 0.3 dex,
roughly corresponding to the scatter in the relation shown in Fig. \ref{MH2}. The value derived for NGC 4552, log $M(H_2)$ = 8.68$\pm$0.3 M$_{\odot}$, is slightly larger than the upper limit 
derived from CO observations in Boselli et al. (2014b), log $M(H_2)$ $<$ 8.37 M$_{\odot}$, but still within the assumed uncertainty. 

\subsection{Schmidt relation}

\begin{figure}
   \centering
   \includegraphics[width=0.49\textwidth]{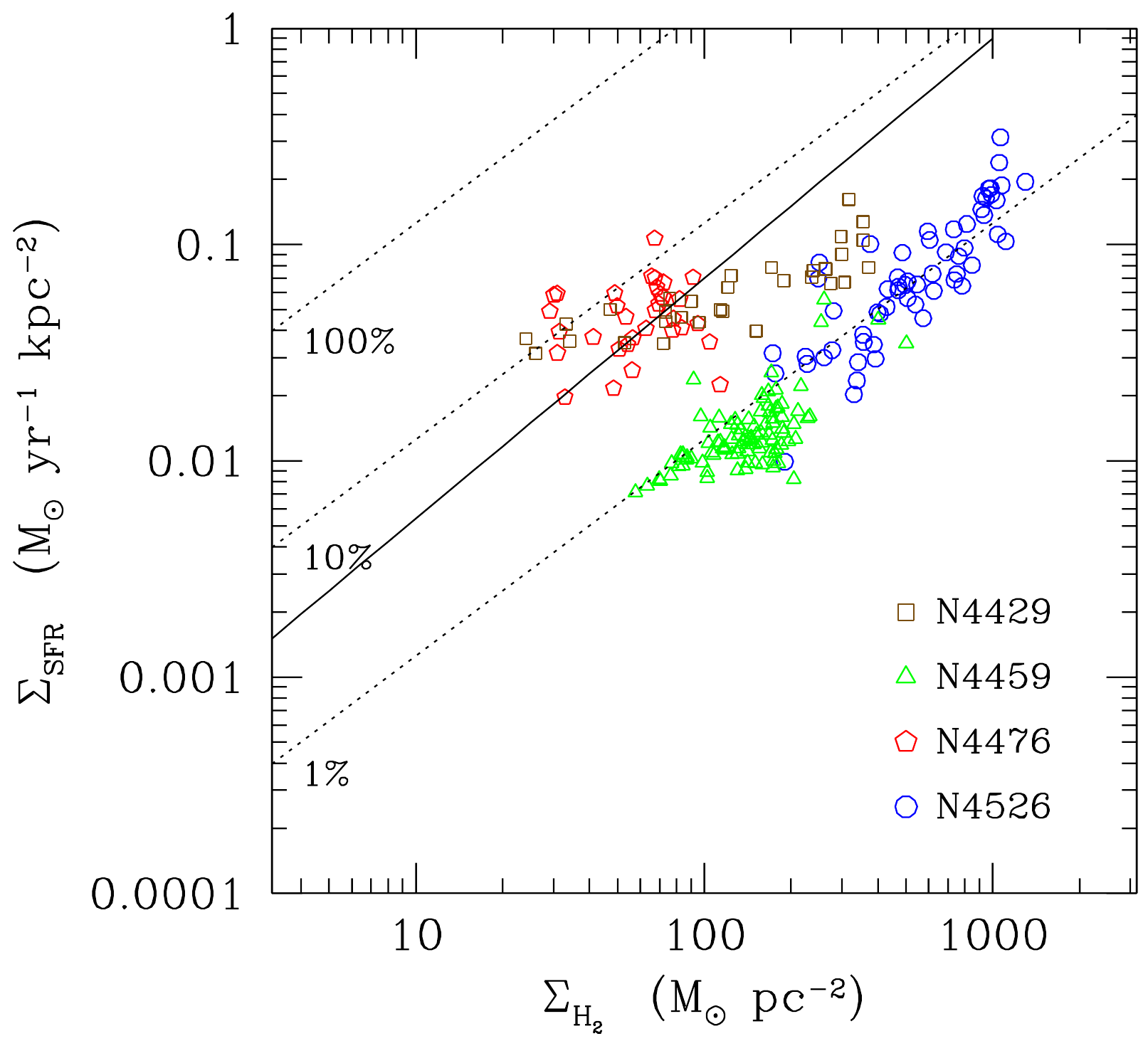}
   \caption{Relationship between the star formation rate surface density and the molecular gas surface density at scales of $\simeq$ 100 pc. Different symbols indicate different galaxies.
   The black solid line shows the best fit obtained by Bigiel et al. (2008) (scaled to the same $X_{CO}$ conversion factor) for the Heracles sample of 
   nearby galaxies using star formation rates derived from H$\alpha$+24$\mu$m flux surface densities. The dotted lines show lines of constant star formation efficiency,
   indicating the level of $\Sigma_{SFR}$ needed to consume 1\%\ , 10\%\ , and 100\%\ of the gas reservoir (including helium) in 10$^8$ years, corresponding to
   depletion times of 10$^8$, 10$^9$, and 10$^{10}$ years.
 }
   \label{SLallalma}%
   \end{figure}

The previously discussed data set allow us to trace the relationship between the star formation rate surface density, $\Sigma_{SFR}$, and the molecular gas surface density, 
$\Sigma_{gas}$, generally called the Schmidt law (Schmidt 1959, Kennicutt 1998b), in five galaxies (NGC 4429, 4459, 4476, 4477, and 4526) at kpc scales using the CARMA data, 
and in NGC 4429, NGC 4459, NGC 4476, and NGC 4526 at $\sim$ 100 pc scales using the ALMA data (Fig. \ref{SLallalma}). For this purpose, we convolved the ALMA
moment 0 images of NGC 4429, NGC 4459, and NGC 4476 to the resolution of the H$\alpha$ NB images (0.70\arcsec, 0.86\arcsec, and 0.86\arcsec, respectively), and the VESTIGE
image of NGC 4526 to the resolution of the ALMA moment 0 image (1.14\arcsec$\times$0.76\arcsec). We then rebinned the images to 1.2\arcsec pixels to have an independent 
signal in each corresponding position, and removed the low signal-to-noise pixels ($S/N$ $<$ 5). We also removed the central ($\lesssim$ 2\arcsec) pixels
which might be contaminated by a nuclear activity or by artefacts due to the continuum subtraction procedure which is highly uncertain in these regions where 
the emission is very peaked and where saturated pixels might be present.

Figure \ref{SLallalma} indicates that the star-forming discs of these four galaxies follow the Schmidt relation drawn by late-type galaxies, albeit with higher values
of star formation (0.01 $\lesssim$ $\Sigma_{SFR}$ $\lesssim$ 0.3 M$_{\odot}$ yr$^{-1}$ kpc$^{-2}$) and molecular gas surface densities
(20 $\lesssim$ $\Sigma_{H_2}$ $\lesssim$ 1000 M$_{\odot}$ pc$^{-2}$). These values, which at a first glance might seem very high for these quiescent systems, 
can be explained considering that the inner discs are very dust-rich as indicated by their HST images (Fig. \ref{HST}) and by the other multifrequency data presented above.
Given the tight relation between dust and gas column density, these discs have also very high molcular gas column densities. Figure \ref{SLallalma}, however, also shows that  
the molecular gas is transformed into stars with an efficiency $\sim$ 2.5 times lower than that measured in the discs of star-forming galaxies but on 
a spatial scales significantly smaller (100 pc vs. 0.75 kpc, Bigiel et al. 2008), in line with the results of Davis et al. (2014) end Ellison et al. (2021). A similar result
is obtained using the CARMA data at a much coarser angular resolution. The star formation efficiency observed in these S0 galaxies is, however, significantly
higher than that observed in early-type objects formed after a minor merging event (e.g. van de Voort et al. 2018). This result has been interpreted by Davis et al. (2014) as due to dynamical effects
related to the location of the molecular gas in the inner region, on the rising part of the rotation curve, where shear is high. Although here the star formation 
per unit gas mass is reduced with respect to that observed in gas-rich, rotating spirals, the amount of star formed per free-fall time is constant (Davis et al. 2014).
This result might be tightly connected to the more general ``morphological quenching" scenario proposed by Martig et al. (2009), 
where gas discs can be stabilised against star formation when they are embedded in prominent bulges such as those observed in 
lenticular galaxies (see however Koyama et al. 2019). Although our results are in line with these interpretations, we caution that both the star formation rates
and the molecular gas column densities used in Fig. \ref{SLallalma} might suffer from systematic effects in the dust attenuation correction,
which is highly uncertain whenever 2D-Balmer decrement maps are not available (NGC 4429, NGC 4459, and NGC 4476), and in the assumed CO(3-2)/CO(1-0) (NGC 4429) 
and CO(2-1)/CO(1-0) (NGC 4459 and NGC 4476) flux ratios, which are still poorly constrained in these extreme environments. We can also add that the claimed low
star formation efficiency strongly depends on the assumed CO-to-H$_2$ ($X_{CO}$) conversion factor which is known to depend on the physical properties of the 
interstellar medium (e.g. Boselli et al. 1997, Kaufman et al. 1999, Bolatto et al. 2013). These discs are metal-rich, dense in molecular gas, and are very dusty, and are thus typical environments where self-shielding 
against the surrounding ISRF, mainly dominated by the emission of the evolved stellar population of the prominent bulges, 
is very efficient. In these environments $X_{CO}$ is expected to decrease, as suggested by models (Wolfire et al. 1995, Kaufman et al. 1999) 
and observations (Boselli et al. 2002, Bolatto et al. 2013).
Furthermore, hot X-rays emitting gas is also present in some of these quiescent massive objects (NGC 4459, 4477, 4526, 4552). It is thus conceivable that the $X_{CO}$
conversion factor is systematically different than the one measured in the discs of massive, star-forming spirals, and thus explain the observed decrease of the 
star formation efficiency by a factor of $\sim$ 2.5 present in these inner discs of massive lenticular galaxies.

\section{Discussion and conclusion}

The VESTIGE H$\alpha$ NB imaging survey has revealed the existence of eight massive (10$^{10}$ $\lesssim$ $M_{star}$ $\lesssim$ 10$^{11}$ M$_{\odot}$)
lenticular galaxies in the Virgo cluster with ionised gas emission in their inner (a few kpc) regions. The ionised gas is generally located on a disc with a
morphology very similar to that of the dust seen in absorption in the high-resolution HST images, as observed in other lenticular galaxies 
(e.g. Goudfrooij et al. 1994, Sarzi et al. 2006, Finkelman et al. 2010, 2012, Kulkarni et al. 2014). In two galaxies the ionised gas has a filamentary structure
similar to that observed in cooling flows at the centre of massive clusters. These eight objects are a 
significant fraction ($\sim$ 25\%) of the lenticular galaxies of the Virgo cluster in the same stellar mass range. Since
the VESTIGE survey is still incomplete with a partial coverage of the VCC footprints ($\sim$ 60\%\, with full depth only on $\sim$ 25\%\ of the VCC), this  work suggests 
that these ionised gas features are quite
common in this cluster environment. The presence of ionised gas in early-type galaxies, and in particular in lenticular galaxies, was already known since the works of 
Pogge \& Eskridge (1987, 1993), Trinchieri \& di Serego Alighieri (1991), and Macchetto (1996), but it has never been studied in such a detail due to the lack
of high resolution multifrequency data as those presented here. It is also worth noting the increase in image quality reached by the VESTIGE  
data with respect to those gathered by previous works using 2~m class telescopes, for which data for a few galaxies in common are available 
(e.g. Gavazzi et al. 2018b). This is thanks to the extraordinary seeing conditions at the CFHT 
combined with the accurate stellar continuum subtraction based on broad-band colour images.

The analysis presented in this work has clearly shown that in most cases the gas is photoionised by young stars newly formed in the disc.
These results suggest that this Virgo cluster sample is significantly different from the ATLAS$^{3D}$ sample, where the gas is mostly ionised by evolved stars 
and only in 10\%\ of them by young stars (Sarzi et al. 2010). Systematic differences between cluster and isolated systems are however expected, as indeed observed 
in their kinematical (e.g. Davis et al. 2011, Krajnovic et al. 2011) and dust vs. gas properties (Kokusho et al. 2019). In these Virgo galaxies 
the H$\alpha$ luminosity is a direct tracer of star formation and indicates that these objects lie well below the main sequence relation driven by
normal, star-forming discs, with star formation rates of the order of 0.02 $\lesssim$ $SFR$ $\lesssim$ 0.15 M$_{\odot}$ yr$^{-1}$. 
We also derived the statistical properties of individual HII
regions detected in these star-forming discs, showing that, on average, they correspond to those encountered in normal, star-forming galaxies. 
In NGC 4262 and NGC 4552, where the emission is principally due to filamentary structures, the gas is 
probably shock-ionised and might be hot gas cooling and infalling into the nucleus after a merging event. The lack of strong AGNs makes us exclude 
nuclear activity as the major source of ionisation in all the targets. 

CO millimetric CARMA and ALMA data allowed us to show that the total molecular gas content of these objects (2$\times$10$^7$ $\lesssim$ $M(H_2)$ $\lesssim$ 5$\times$10$^8$ M$_{\odot}$)
is lower than that found in spiral galaxies of similar mass 
(e.g. Boselli et al. 2014c). The molecular gas is distributed within discs with properties similar to those observed in H$\alpha$, and forms stars 
following the Schmidt relation but with an efficiency $\simeq$ 2.5 lower than that observed in spiral galaxies, as previously found in similar objects by Davis et al. (2014).
The low star formation efficiency observed in these systems might be due to a variation in the shear or in the global stability of 
the gas which is here located in the inner regions, where the rotation curve of the galaxies rises rapidly (Davis et al. 2014) and the gas becomes pressure supported (Bertola et al. 1995). 

Both optical (HST and NGVS) and far-IR (\textit{Herschel} and \textit{Spitzer}) images indicate the presence of large amounts of dust in these lenticular galaxies
as often observed in early-type systems (e.g. Goudfrooij \& de Jong 1995).
The multifrequency data allowed us to estimate their total dust content using two independent techniques, the first
one based on the SED fitting analysis including UV-to-far-IR data, the second one based on the measure of the optical attenuation derived from the NGVS
broad-band images after subtracting a model to trace the intrinsic distribution of stars. The two techniques give very different results, with the SED fitting analysis 
providing dust masses $\sim$ 100 times higher than those obtained from the 2D dust attenuation maps. The comparison of the \textit{Herschel}/PACS 100 $\mu$m
image of NGC 4526 with the Balmer decrement map obtained using the MUSE data convolved to a similar resolution has indicated a similar distribution of
the emitting and absorbing dust in this galaxy. This result thus excludes any major contribution of diffuse dust locked up in the atmosphere of evolved stars which
might be missed in the visible 2D dust attenuation maps as the origin of this major discrepancy. This analysis suggests that the 2D dust attenuation maps
derived from the optical images strongly underestimate the total dust content of these lenticular galaxies just because only a fraction of the stellar emission, 
which is geometrically located along the line-of-sight behind the dust screen, is efficiently absorbed by dust. This result agrees with the one found by the 
radiation transfer analysis of two early-type galaxy in the Virgo and Fornax clusters with similar properties (Viaene et al. 2015, 2019). Again, the total dust mass 
of these galaxies, which ranges between 6$\times$10$^5$ $\lesssim$ M$_{dust}$ $\lesssim$ 10$^7$ M$_{\odot}$, is fairly low compared to that of spiral galaxies 
of similar stellar mass (e.g. Cortese et al. 2012). It matches, however, the typical dust mass of lenticulars estimated using \textit{Herschel} data 
from the \textit{Herschel} Reference Survey (Smith et al. 2012). 

While in a few objects the origin of the ionised gas seems clear, in others the multifrequency data give contradictory results.
In NGC 4262 and NGC 4552 the filamentary structure of the ionised gas, the presence of a cold gas and young stellar ring misaligned with 
the stellar main body of the galaxy (NGC 4262), and the presence of prominent shells and tidal tails in the deep NGVS broad-band optical images (NGC 4552)
clearly suggest an external origin of the gas, accreted after a merging event. NGC 4262 is a S0 galaxy with a gaseous ring, which hydrodynamic simulations indicate 
might have been formed after a minor merging event with a gas rich spiral galaxy (Mapelli et al. 2015). NGC 4552 is a slow rotator and might have been 
formed after a major merging episode able to heat the stellar disc and produce typical gravitation perturbations as those seen in the optical image (Penoyre et al. 2017, Pop et al. 2018).
In these objects, gas and dust might have been accreted after a strong gravitational interaction with a gas-rich system. Once in contact with the hot coronal gas emitting in X-rays,
the cold accreted gas is heated and becomes ionised, while the dust remains shielded within the stripped cold gas which gradually evaporates (evaporation flow scenario, 
Sparks et al. 1989, de Joung et al. 1990, Finkelman et al. 2012). This scenario predicts that the ionised gas and the dust are located in similar filamentary structures 
as those observed in these two galaxies.

MUSE spectroscopic data of NGC 4469 and NGC 4526 have shown that the typical metallicity of the gas over the star-forming discs
is $\simeq$ solar, as expected for these massive galaxies. The analysis presented in Sec. 4 has also shown that most of the galaxies 
have a gas-to-dust ratio of $G/D$ $\simeq$ 80, typical of star-forming systems of similar stellar mass (e.g. Remy-Ruyer et al. 2014).
As extensively discussed in Davis et al. (2011), this evidence suggests that the gas has an internal origin and might be either recycled gas 
produced by the old stellar population dominating in these early-type systems, or the remnant of gas locked-up in the inner regions 
in star-forming galaxies stripped of their ISM by a ram pressure stripping event (e.g. Boselli \& Gavazzi 2006, Boselli et al. 2021a).
The first hypothesis (gas produced by stellar mass loss), however, is questioned because the amount of associated 
dust is too low compared to that observed in these early-type systems (Patil et al. 2007).
Ram pressure stripping removes the gas outside-in, producing radially truncated discs since the gas is hardly perturbed in the inner regions whecdre 
the gravitational potential well of the galaxy is sufficiently deep to keep it anchored to the disc (e.g. Boselli et al. 2006). 
Evidence of an ongoing ram pressure stripping event is present in NGC 4469, which has a prominent tail of ionised gas coming out from the
disc without any stellar counterpart, typical of galaxies undergoing this perturbing mechanism (e.g. Boselli et al. 2021a). Similar features,
although observed with a much lower angular resolution in the X-rays images, are also present in NGC 4477 and NGC 4552, suggesting that ram pressure is
perturbing the gas in the outer halo of these objects (Li et al. 2018, Machacek et al. 2006b).
Recently accreted gas via infall or minor merging events would be unpolluted or low metallicity gas, thus with different properties than those
observed in most of these objects. 

Indications on the possible origin of the ionised gas can be taken from the analysis of the 2D stellar distribution and of the relative kinematics of the gas and of the stars.
Clear, as mentioned above, is the merging origin of NGC 4552 suggested by the deep NGVS broad-band image which shows prominent shells and tidal remnants. 
In this object the gas has been accreted during a mergin event, but its observed distribution is also affected by an ongoing
ram pressure stripping episode. Another interesting 
example is the X-shaped, boxy stellar morphology with a tilted outer disc of NGC 4469. This peculiar morphology has been interpreted as a possible result of harassment (Mosenkov et al. 2020),
or of a recent merging event (Mihos et al. 1995), 
although the presence of an ionised gas tail at the SW of the galaxy due to a ram pressure stripping event is a further evidence that different 
mechanisms can be at place simultaneously. Three out of eight galaxies (NGC 4262, 4469,
and 4477) have a prominent bar, often indicating a recent instability of the stellar disc resulting after a strong gravitational perturbation (e.g. Moore et al. 1998).
An external origin of the ionised gas (gas accretion, minor merger) is also suggested by the misalignement of the stellar and ionised gas kinematics 
observed in NGC 4477 (Sarzi et al. 2006, Crocker et al. 2011) or by the peculiar stellar kinematics of NGC 4429 (Cortes et al. 2015) and NGC 4526 
(Sarzi et al. 2006, Krajnovic et al. 2011, Foster et al. 2016). On the contrary, Bertola et al. (1995) interpreted the high velocity dispersion of the ionised gas
observed in some lenticular galaxies as an indication that this gas has been produced from the mass lost of evolved stars in the bulge. 

\begin{figure}
   \centering
   \includegraphics[width=0.49\textwidth]{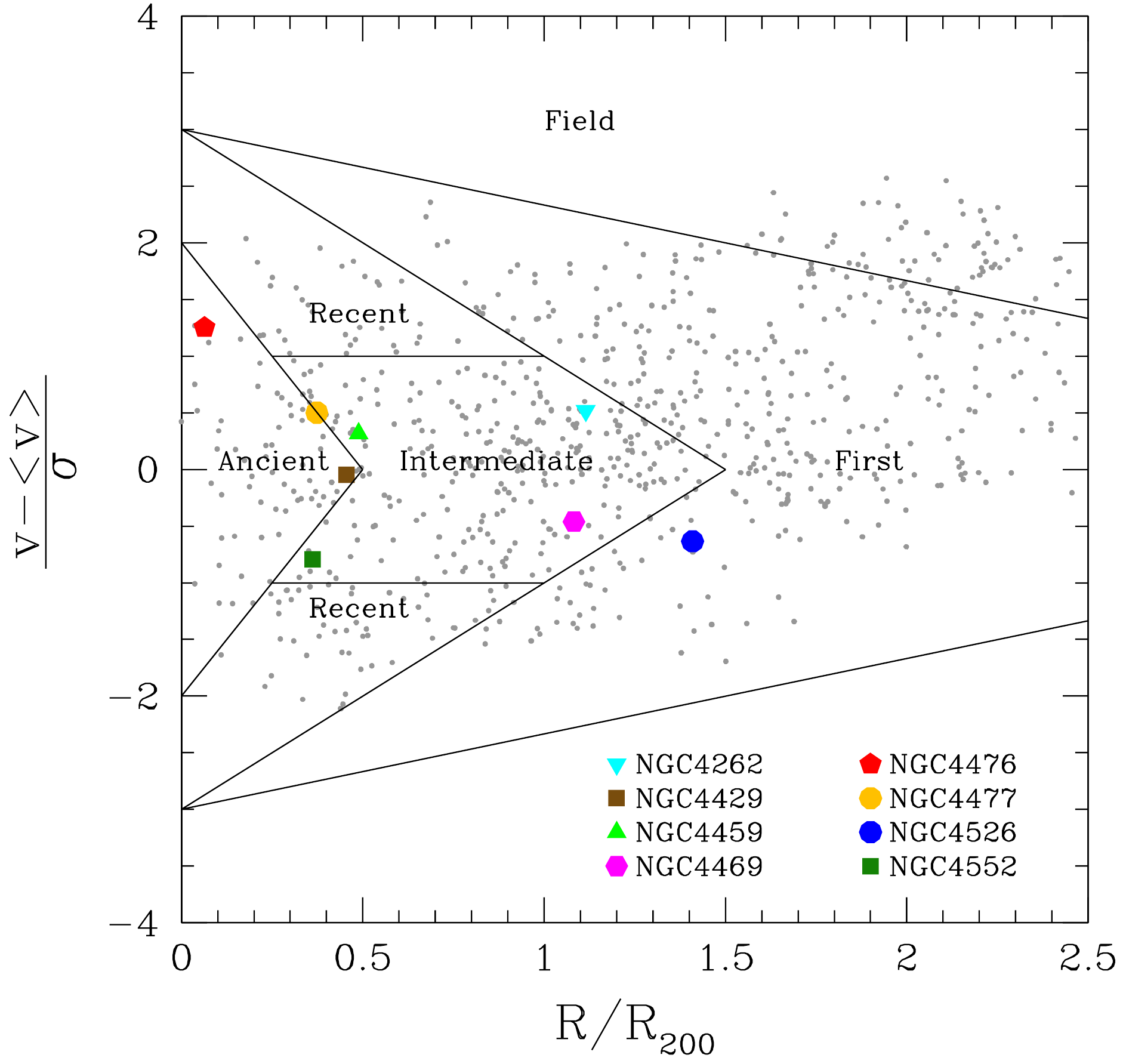}
   \caption{Distribution of the eight lenticular galaxies within the phase-space diagram traced by galaxies within the Virgo cluster. Grey dots indicate
   the Virgo cluster galaxies with a recessional velocity $<$ 3500 km~s$^{-1}$ extracted from the sample of Boselli et al. (2014a). The solid lines 
   delimit the different regions extracted from the simulations of Rhee et al. (2017) to identify galaxies in different phases of their infall into the cluster: 
   First (not fallen yet), recent (0 $<$ $\tau_{inf}$ $<$ 3.6 Gyr), intermediate (3.6 $<$ $\tau_{inf}$ $<$ 6.5 Gyr), and ancient (6.5 $<$ $\tau_{inf}$ $<$ 13.7 Gyr) infallers. 
 }
   \label{phase}%
   \end{figure}

All these physical properties derived from the analysis of each single object should be considered along with more general arguments based on the 
expected evolution of these lenticular galaxies within the Virgo cluster environment. This can be done by tracing their position within the phase-space diagram
(see Fig. \ref{phase}) as already done for these and other objects in Boselli et al. (2014a, 2016b). Figure \ref{phase} clearly shows that most, if not all, of
these lenticulars are located within the ``intermediate" region composed by objects accreted by the cluster 3.6 $<$ $\tau_{inf}$ $<$ 6.5 Gyr ago, as indicated by the 
simulations of Rhee et al. (2017). These timescales are slightly longer than those derived from the SED fitting analysis of Boselli et al. (2016b) for the
quenching of the star formation activity of early-type galaxies within the Virgo cluster (1 $\lesssim$ $\tau_{quench}$ $\lesssim$ 3 Gyr), consistent with a picture where these lenticular 
galaxies are accreted as gas-rich systems and later quench their star formation activity after their interaction with the hostile surrounding environment.

Inspite of the exquisite quality of the multifrequency data in our hand and of the careful analysis carried oou in this work, we have not been able to 
identify a unique origin for the ionised gas in these eight lenticular galaxies located within the Virgo cluster. This result is due to the fact
that lenticular galaxies in rich environments have been formed following different formation paths since they might have suffered several perturbing mechanisms
concurring at different degrees and affecting their star formation history in a complex and unpredictable way. This is also the case for the galaxies analysed in this work.
For this reason we plan to study in detail the structural
and kinematical properties of the stars and of the gas using the unique set of mutlifrequency data available or gathered with targeted IFU spectroscopic observations
recently undertaken at the Observatoire de Haute-Provence. We also plan to use the complete VESTIGE blind survey to statistically quantify the frequency of these objects
in a cluster such as Virgo, and compare it to the predictions of semi-analytic models and simulations (e.g. Lagos et al. 2014).
The results of these analyses will be presented in forthcoming communications.

\begin{acknowledgements}

We are grateful to the whole CFHT team who assisted us in the preparation and in the execution of the observations and in the calibration and data reduction: 
Todd Burdullis, Daniel Devost, Bill Mahoney, Nadine Manset, Andreea Petric, Simon Prunet, Kanoa Withington.
We also thank S. Torres and the SOAR team for their support during the observations and data reduction, and the referee for
constructive comments and suggestions.
We acknowledge financial support from ``Programme National de Cosmologie and Galaxies" (PNCG) funded by CNRS/INSU-IN2P3-INP, CEA and CNES, France,
and from ``Projet International de Coop\'eration Scientifique" (PICS) with Canada funded by the CNRS, France.
This research has made use of the NASA/IPAC Extragalactic Database (NED) 
which is operated by the Jet Propulsion Laboratory, California Institute of 
Technology, under contract with the National Aeronautics and Space Administration
and of the GOLDMine database (http://goldmine.mib.infn.it/) (Gavazzi et al. 2003).
The UVIT project is collaboration between the following institutes from India: Indian Institute of Astrophysics (IIA), 
Bengaluru,  Inter  University  Centre  for  Astronomy  and  Astrophysics  (IUCAA),  Pune,  and  National  Centre  for 
Radioastrophysics (NCRA) (TIFR), Pune, and the Canadian Space Agency (CSA). 
This paper makes use of the following ALMA data: ADS/JAO.ALMA 2013.1.00493.S. ALMA is a partnership of ESO (representing its member states), 
NSF (USA) and NINS (Japan), together with NRC (Canada), MOST and ASIAA (Taiwan), and KASI (Republic of Korea), in cooperation with the Republic 
of Chile. The Joint ALMA Observatory is operated by ESO, AUI/NRAO and NAOJ. The National Radio Astronomy Observatory is a facility of the National Science Foundation operated under cooperative 
agreement by Associated Universities, Inc.
MF has received funding from the European Research Council (ERC) (grant agreement No 757535).
N.Z.D. acknowledges partial support from FONDECYT through project 3190769.
L.G. acknowledges financial support from the Spanish Ministry of Science and Innovation (MCIN) under the 2019 Ram\'on y Cajal program RYC2019-027683 and from the Spanish MCIN project HOSTFLOWS PID2020-115253GA-I00.
MB gratefully acknowledges support by the ANID BASAL project FB210003 and the FONDECYT regular grant 1211000.

\end{acknowledgements}

\begin{appendix}

\section{Comments on individual objects}

\subsection{NGC4262 (VCC 355)}

The H$\alpha$ image of NGC 4262 (Fig. \ref{Ha_image}) shows an extended ring-like structure characterised by a few HII regions oriented along a NE-SW major axis.
This ring-like structure is visible also in the ASTROSAT/UVIT FUV image (Fig. \ref{UVIT4262}) and in the GALEX images (Bettoni et al. 2010), 
signifying the presence of young stars. The ring has been also detected in HI by Krumm et al. (1985) and more recently by Oosterloo et al. (2010) with the WSRT.
The thin HI ring structure, its orientation with respect to the main stellar disc (polar ring, Serra et al. 2014), and its extension (a factor of $\sim$ 2 larger than the stellar disc)
led Krumm et al. (1985) conclude that the HI gas ring is either the remnant of the primordial proto-galaxy cloud, or HI gas recently accreted from the intergalactic medium or
from a captured gas-rich dwarf. By studying the relative kinematics of the HI gas within the ring and that of the ionised gas in the inner regions, 
strongly kinematically decoupled from that of the stars within the disc (Sarzi et al. 2006), Bettoni et al. (2010) proposed that the atomic gas
has been removed from the main body of the galaxy after a gravitational interaction with another Virgo cluster member, possibly NGC 4254. The NGVS (Fig. \ref{colour_image}) and HST
(Fig. \ref{HST}) images indicate the presence of a prominent bar, whose origin is often associated with gravitational perturbations (e.g. Moore et al. 1996, 1998). 
Hydrodynamic simulations indicate that galaxies with gas-rich polar rings are generally formed after a minor merging event (Mapelli et al. 2015). All this
observational evidence, combined with that presented in this work, consistently suggests that NGC 4262 has recently suffered a gravitational perturbation which is probably 
at the origin of the ionised gas filaments observed in the inner regions and of the HI gas in the polar ring structure.

\subsection{NGC4429 (VCC1003)}

NGC 4429 is characterised by a high surface brightness ionised gas disc located in the inner 0.47 kpc region (radius), and a more extended (1.1 kpc)
low surface brightness disc (Fig. \ref{Ha_image}). A very similar morphology is visible in absorption in the HST F606W (Fig. \ref{HST}) and NGVS (Fig. \ref{Ag}) images, 
which reveal the presence of a dust disc with a folded spiral structure associated with the ionised gas disc. The dust disc also contains cold rotating molecular gas 
(Alatalo et al. 2013, Davis et al. 2018).
IFU spectroscopy of NGC 4429 shows a regular stellar velocity field suggesting a pure circular motion with possibly 
a cold circumnuclear stellar disc (Cortes et al. 2015). The velocity dispersion pattern, however, is fairly asymmetric, again suggesting the presence of a
circumnuclear disc (Cortes et al. 2015). The steep rising rotation curve also suggests a high central mass concentration. NGC 4429 is classified as a fast rotator (Emsellem et al. 2011)
despite its high central velocity dispersion ($\sigma$ $\simeq$ 200 km s$^{-1}$, Cortes et al. 2015). The galaxy also shows a kinematically distinct component, i.e.
with the stellar kinematic PA aligned with the photometric PA in the inner ($r$ $\leq$ 2 kpc) and outer galaxy, but misaligned by $\sim$ 10$^o$ at intermediate radii (Cortes et al. 2015).
These kinematical properties led Cortes et al. (2015) conclude that NGC 4429 might result from a minor (3:1-6:1)
merger, where the gas has been accreted in the inner regions after this event.

\subsection{NGC4459 (VCC1154)}

NGC 4459 has an ionised gas disc in its inner region, which corresponds to the dust disc observed in absorption in the HST F475W (Fig. \ref{HST}) and NGVS (Fig. \ref{Ag})
images and to the molecular gas disc seen with ALMA. The high resolution HST image shows that the dust is distributed on two concentric almost face-on discs
with folded spiral structures. The same disc has been tentatively detected in HI by Lucero \& Young (2013). Crocker et al. (2011) 
derives a star formation of $\simeq$ 0.05-0.12 M$_{\odot}$ yr$^{-1}$, slightly higher than the one measured in this work (see Table \ref{gal}). They have also shown a surprisingly low 
radio continuum activity with respect to the galaxy far-infrared emission.
The molecular gas kinematics (Young et al. 2008) does not give any strong indication to the possible origin of the gas since the gaseous disc is kinematically coupled with the rotating stellar disc.
The stellar kinematics derived using SAURON indicates that the galaxy is a fast rotator (Sarzi et al. 2006) with a regular stellar kinematics (Krajnovic et al. 2006).
The kinematics of the galaxy derived using globular clusters indicates that the high velocity rotation is conserved also well outside the effective radius (Bellstedt et al. 2017).
After a comparison with hydrodynamic simulations, the same authors conclude that the kinematical properties of NGC 4459 are consistent with a gas-rich minor merger
origin, although not excluding a major merging event. Our results, however, combined with the agreement between the stellar and gas kinematics, rather suggest an internal origin,
with gas stripped in the outer disc by ram pressure. The same conclusion is obtained after the analysis of the X-ray morphology of the galaxy done by Hou et al. (2021).

\subsection{NGC4469 (VCC1190)}

NGC 4469 is an edge-on galaxy classified as a barred lenticular galaxy (see Table \ref{gal}), with a typical boxy/peanut morphology (see Fig. \ref{colour_image}). The H$\alpha$ image shows 
a thin star-forming disc with filaments of ionised gas on the z-plane. The analysis of deep optical images of this galaxy revealed the presence of a tilted
outer disc which has been interpreted by Mosenkov et al. (2020) as a possible result of harassment, in agreement with the boxy morphology witnessing a
recent gravitational instability (e.g. Gadotti 2012), although gas accretion of gas-rich satellites 
occuring several Gyr ago cannot be ruled out. The X-shape of the bulge observed in this edge-on galaxy is another indication of a recent merger (Mihos et al. 1995).
The presence of an ionised gas filament without any associated stellar counterpart discovered in this work
rather suggests a ram pressure stripping event able to remove a large fraction of the cold gas from the outer regions, producing a truncated disc.
The shock-ionised gas observed in the z-plane has been probably ionised during its interaction with the surrounding intracluster medium.

\subsection{NGC4476 (VCC1250)}

The H$\alpha$ image of NGC 4476 shows a central ring-like structure with prominent HII regions similar to those observed in spiral galaxies, although here limited 
to the very inner region. The HST F475W image shows that these star-forming regions are associated with a dusty disc with an evident spiral structure.
The large H$_2$/HI molecular-to-atomic gas ratio compared to that observed in normal late-type galaxies led Lucero et al. (2005)
to conclude that the observed gas disc of NGC 4476 is the result of a ram pressure stripping event. Indeed, this perturbing mechanism has a differential effect able to remove 
more efficiently the diffuse atomic gas, which is mainly distributed on an extended disc, than the molecular gas component, mainly located in dense giant molecular clouds 
in the inner disc. Young (2002), however, ruled out any possible internal origin of the gas because its specific angular momentum 
is about three times larger than that of the stars.
Our results are consistent with the gas having an internal origin, being the remnant of a ram pressure stripping event.

\subsection{NGC4477 (VCC1253}

The H$\alpha$ image of NGC 4477 reveals an ionised gas disc with spiral structures surrounding the nucleus (Fig. \ref{Ha_image}). The same disc
is also visible in absorption in the HST F475W (Fig. \ref{HST}) and in the NGVS (Fig. \ref{Ag}) images, although the presence of dust is less spectacular than in other 
objects. This galaxy is also characterised by the presence of hot gas whose temperature decreases to $\sim$ 0.3 keV in the inner 2-3 kpc region (Li et al. 2018; Kim et al. 2019)
where the ionised gas disc is located. The hot gas is distributed along a spiral structure (Kim et al. 2019) 
with an asymmetric halo probably formed by the gas stripped during the interaction with the surrounding 
ICM during the galaxy motion on the plane of the sky (Li et al. 2018). The X-ray observations seem to indicate that the galaxy is suffering a feedback process due to 
the central AGN which is able to produce two small cavities in the hot gas. The same observations suggests that the cold gas in the centre is accreted gas (Li et al. 2018).
Plateau de Bure interferometric (PdBI) observations suggest that the molecular gas is located on a ring-like structure  with kinematical properties similar to those of the ionised gas 
derived from the SAURON observations (Sarzi et al. 2006, Crocker et al. 2011). The difference between the stellar kinematics position angle and the ionised gas
position angle is of $\sim$ 30$^o$, which might indicate an external origin of the gas (Sarzi et al. 2006, Crocker et al. 2011).

\subsection{NGC4526 (VCC1535)}

The H$\alpha$ image shows a high surface brightness disc of ionised gas extending up to $\simeq$ 1.45 kpc (Fig. \ref{Ha_image}). 
The same disc is observed in absorption in the HST F475W (Fig. \ref{HST}) and NGVS (Fig. \ref{Ag}) images, and contains giant molecular
clouds with properties similar to those observed in other late-type galaxies (Fig. \ref{ALMA}; Utomo et al. 2015).
The galaxy also has hot gas (0.3 keV) confined within the inner $\simeq$ 3 kpc,
with a strong X-ray point source in the centre, although not identified as a known AGN (Kim et al. 2019). \textit{Herschel}/SPIRE spectroscopy indicates that
the gas in the disc has a temperature of $T$ $\simeq$ 15-20 K and an electron density of $n_e$ $\simeq$ 1.6 cm$^{-2}$ (Lapham \& Young 2019), thus significantly lower than the one
derived using the [SII] doublet ratio (Fig. \ref{SII}) or the size of the HII regions (Fig. \ref{HII}). The stellar kinematics derived using the SAURON instrument 
(Sarzi et  al. 2006, Krajnovic et al. 2011) or with the spectroscopy of globular cluster in the outer regions (Foster et al. 2016) 
indicates that the galaxy is a fast rotator with a double maxima velocity profile and a regular gas kinematics, kinematical features quite frequent in early-type galaxies.

\subsection{NGC4552 (M89, VCC1632)}

NGC 4552 (M89) is characterised by diffuse filaments of ionised gas (Fig. \ref{Ha_image}). Some ionised gas in emission was
detected by Macchetto et al. (1996).  The H$\alpha$ filamentary structure has a morphology similar to the one observed in X-rays (Machacek et al. 2006a,
Kraft et al. 2017, Kim et al. 2019). These filaments are not detected in absorption in the HST F555W (Fig. \ref{HST}) nor in the NGVS images.
The hot gas temperature varies radially, with a hot core, a decrease of temperature down to $T$ $\sim$ 0.4 keV at 4-5 kpc, and an
increase in the outer regions to $T$ $\sim$ 0.6 keV (Kim et al. 2019). The asymmetric distribution of the hot gas indicates that the galaxy is 
suffering a ram pressure stripping while traveling within the cluster. The morphology of the hot gas of the galaxy at the interface with the surrounding
ICM is due to the viscosity and the Kelvin-Helmholtz instabilities created in these gas layers (Roediger et al. 2015). NGC 4552 is a slow rotator (Emsellem et al. 2011)
with the gas kinematics strongly decoupled from the stellar kinematics in the inner 320 pc (Sarzi et al. 2006), and a non-regular rotation (Krajnovic et al. 2011). The deep optical NGVS image
of the galaxy shown in Fig. \ref{colour_image} indicates the presence of shells and tidal tails, suggesting that this object underwent a strong gravitational perturbation (Penoyre et al. 2017, Pop et al. 2018).
Although the galaxy is cleary suffering a ram pressure stripping event during its jurney within the cluster (Machacek et al. 2006a), the ionised gas filaments probably have
an external origin being gas accreted during the merging event and ionised by the shocks produced by the nuclear outflow (Machacek et al. 2006b).

\end{appendix}

\end{document}